\documentclass[11pt]{article}

\usepackage[margin=1in]{geometry}
\usepackage{microtype}
\usepackage{times}    
\usepackage{amsmath,amsfonts,amssymb}
\usepackage{graphicx}
\usepackage{booktabs}
\usepackage[hidelinks]{hyperref}
\usepackage[style=ieee,backend=biber]{biblatex}
\usepackage{pifont}
\usepackage{tabularx} 
\usepackage{float}
\usepackage{hyperref}
\usepackage{xurl}
\usepackage{adjustbox}
\title{Time Series Foundation Models\\
       for Multivariate Financial Time Series Forecasting}
\author{Ben Asher Marconi \\ Imperial College London \\ \texttt{ben.marconi21@imperial.ac.uk}}
\date{July 2025}

\begin{document}
\maketitle
\begin{abstract}
Financial time series forecasting presents significant challenges due to complex nonlinear relationships, temporal dependencies, feature interdependencies and limited data availability—particularly for tasks involving low-frequency data, newly listed instruments, or emerging market assets. Time Series Foundation Models (TSFMs) offer a promising solution through pretraining on diverse time series corpora followed by task specific adaptation. This study evaluates two TSFMs (Tiny Time Mixers (TTM) and Chronos) across three financial forecasting tasks: US 10 year Treasury yield changes, EUR/USD volatility, and equity spread prediction.

Results demonstrate that TTM exhibits strong transferability. When fine-tuning both the pretrained version of TTM and an untrained model with the same architecture, the pretrained version achieved 25–50\% better performance when fine-tuned on limited data and 15–30\% improvements even when fine-tuned on lengthier datasets. Notably, TTM's zero-shot performance outperformed naive benchmarks in volatility forecasting and equity spread prediction, with the latter demonstrating that TSFMs can surpass traditional benchmark models without fine-tuning. The pretrained model consistently required 3–10 fewer years of data to achieve comparable performance levels compared to the untrained model, demonstrating significant sample efficiency gains. However, while TTM outperformed naive baselines, traditional specialised models matched or exceeded its performance in two of three tasks, suggesting TSFMs prioritise breadth over task specific optimisation.

These findings indicate that TSFMs, though still nascent, offer substantial promise for financial forecasting—particularly in noisy, data constrained tasks—but achieving competitive performance likely requires domain specific pretraining and architectural refinements tailored to financial time series characteristics.

\end{abstract}
\section{Introduction}
\label{intro}

\subsection{Motivation and Scope}

Whether predicting the evolution of a sovereign yield curve, estimating key macroeconomic indicators for policy formulation, or forecasting cross-asset volatilities for risk management, financial time series forecasting presents a notoriously formidable challenge. Such series are characterised by low signal to noise ratios, non-stationarity, nonlinearity and frequent structural breaks, compounded by uneven data availability across instruments and markets~\cite{ratnadip2014, praveen2025}. Furthermore, financial time series are influenced by exogenous events such as political factors, government policies and geographical events~\cite{praveen2025}.
Nevertheless, these forecasts can be immensely valuable. They enable policymakers, financial institutions, and investors to anticipate market movements and proactively manage the risks inherent in fluctuating asset prices, interest rates, and exchange rates~\cite{makridakis1994}. Through accurate forecasts, organisations and investors can calibrate monetary and fiscal policies, optimise portfolio allocations, design hedging strategies, and set appropriate capital reserves in advance. Moreover, forecasts support pricing of derivative instruments, guide corporate treasury decisions, and inform macroprudential oversight, thereby helping to stabilise markets and enhance economic welfare in the face of uncertainty and exogenous shocks.
Traditional approaches to financial forecasting have relied heavily on econometric models and statistical methods, which whilst theoretically grounded, often struggle to capture the complex, nonlinear relationships inherent in financial markets. The emergence of machine learning techniques has offered new possibilities, yet these methods, particularly deep neural networks, typically require substantial amounts of data for effective training—a significant limitation given the relatively short history of many financial instruments and the challenges of obtaining high frequency data across diverse markets.
The recent development of foundation models represents a paradigm shift in machine learning, demonstrating remarkable capabilities across various domains through pretraining on large, diverse datasets followed by task specific adaptation. Time Series Foundation Models (TSFMs) extend this concept to temporal data, offering the potential to leverage patterns learned from broad time series corpora for specific forecasting applications. This approach is particularly promising for financial forecasting, where the scarcity of task specific data has long been a limiting factor.
This study evaluates the effectiveness of TSFMs in financial forecasting contexts by examining two distinct models: Tiny Time Mixers (TTM), a compact architecture designed specifically for time series applications, and Chronos, which tokenises time series values through scaling and quantisation into a fixed vocabulary before training transformer based language model architectures on these tokenised sequences. These models are selected to provide representative coverage of the different architectural approaches currently available in the TSFM landscape, as detailed in the literature review.
The evaluation encompasses three diverse financial forecasting tasks that capture different aspects of market dynamics and statistical properties: forecasting US 10 year Treasury yield changes 21 days ahead (a low signal to noise ratio change series), predicting EUR/USD realised volatility 21 days ahead (an autocorrelated series with persistent clustering), and forecasting the spread between equity indices at multiple horizons (a mean reverting series that oscillates around zero). These tasks are chosen to span different asset classes, statistical properties, and forecasting challenges, providing a comprehensive assessment of TSFM capabilities across varied financial contexts with fundamentally different underlying data generating processes.
The models are evaluated using standard regression metrics and compared against traditional benchmark approaches including linear models, tree based methods, and sequential models such as LSTMs. Performance is assessed both in zero shot mode—without any task specific training—and following fine tuning on historical data. This dual evaluation framework allows for assessment of both the immediate transferability of pretrained knowledge and the models' capacity to adapt to specific financial forecasting challenges.
Through this comprehensive evaluation, the study aims to establish whether TSFMs represent a meaningful advancement for financial forecasting applications, particularly in scenarios where data limitations have traditionally constrained model performance. The findings contribute to the growing understanding of foundation model applications in finance whilst providing practical guidance for practitioners considering these emerging technologies.

\subsection{Report Structure}

This report is structured as follows:

\begin{itemize}
    \item \textbf{Introduction} (Chapter~\ref{intro}, this chapter): introduces the motivation, objectives, and context of the project, outlining the forecasting challenges inherent to financial time series and the potential role of time series foundation models (TSFMs).
    
    \item \textbf{Background} (Chapter~\ref{ch2}): provides relevant background on financial time series forecasting, transfer learning, and foundation models. It also presents a review of prior work and an overview of the TSFMs evaluated in this study.
    
    \item \textbf{Forecasting Tasks - Overview and Data Processing} (Chapter~\ref{imp}): describes the three forecasting tasks used to benchmark model performance, detailing the problem formulation, data sources, and feature and target construction for each.
    
    \item \textbf{Experimental Setup} (Chapter~\ref{exp}): outlines the training and evaluation procedures, including model configurations, baseline comparisons, and the two core evaluation frameworks: the sample-efficiency probe and the transfer-gain test.
    
    \item \textbf{Results} (Chapter~\ref{results}): presents experimental results across all tasks, evaluating model accuracy, transferability, and training efficiency. Comparisons with naive and classical baselines are also included.
    
    \item \textbf{Conclusion} (Chapter~\ref{Conclusions}): concludes the report and suggests potential directions for future research and development in this area.
\end{itemize}

\section{Background}
\label{ch2}
\subsection{A History of Time Series Forecasting}
\label{sec:tsf_hist}
Due to the broad benefits of time series forecasting with applications ranging from finance~\cite{praveen2025} and economics~\cite{bai2008} to weather prediction~\cite{gneiting2005} and energy consumption prediction \cite{chou2018}, forecasting techniques have greatly evolved over time, from classical methods to current state of the art deep learning approaches~\cite{kolambe2024}. Reflecting this wide applicability, forecasting methodologies have evolved from early linear statistical models such as autoregressive integrated moving‐average (ARIMA) and exponential smoothing, to more flexible tree based machine‐learning techniques like random forests and gradient boosting. In recent years, the field has witnessed a paradigm shift with the advent of deep‐learning architectures (e.g., Recurrent Neural Networks (RNNs), Temporal Convolutional Networks, and Transformers) that can capture complex, nonlinear dependencies and long‐range temporal patterns~\cite{kolambe2024}. 

Deep learning methods emerged to address the shortcomings of classical statistical models when applied to real‐world time series. Traditional techniques like ARIMA assume linearity and stationarity, yet time‐series information typically comprises both time‐step dependencies and correlations between temporal variables–features that often manifest as complex, nonlinear interactions \cite{song2024}. By stacking successive nonlinear transformations and, in transformer models, using attention mechanisms that weight information from distant time steps, deep architectures can learn these temporal dependencies and variable correlations directly from data, thereby capturing the full informational structure of a time series \cite{song2024}.

Building on the capacity of deep networks to model nonlinear dynamics, researchers have explored a variety of specialised architectures for time‐series forecasting. Early efforts applied auto-encoders and stacked auto-encoders to learn compact representations of temporal sequences, directly mapping those latent features to future values \cite{Lv2014}. Convolutional neural networks (CNNs) soon followed, exploiting their local receptive fields to capture short‐term patterns in financial and economic data \cite{Gudelek2017,Markova2022}. Temporal Convolutional Network (TCN) introduced dilated causal convolutions, enabling larger receptive fields without deepening the network \cite{Bai2018}.

Recurrent neural networks (RNNs), which unlike the previous models are suited by design to handle sequential inputs, were also developed for time series forecasting tasks \cite{Shi2015,Dey2017}. However, standard RNNs struggle to effectively capture long-term dependencies, as gradient-based learning algorithms face an increasingly difficult problem as the duration of the dependencies to be captured increases~\cite{bengio1994}. Pascanu \emph{et al.} (2013) also demonstrate that the performance of recurrent neural networks decreases as sequence length increases, primarily due to the vanishing and exploding gradient problems~\cite{pascanu2013}. Gated architectures such as Long Short-Term Memory (LSTM) and Gated Recurrent Units (GRU) partially alleviated these issues by incorporating memory cells and gating mechanisms \cite{Shi2015,Dey2017}, yet their performance still degrades over very long horizons due to error accumulation \cite{Tang2022,Fan2019}.

More recently, the transformer model \cite{Vaswani2017}, with its encoder–decoder design and self‐attention mechanism, has revolutionised natural language processing \cite{Devlin2018} and computer vision \cite{Dosovitskiy2020}, and is now being adapted for time series \cite{Li2019,Wu2020,xu2022,Zhou2021}. By directly computing pairwise interactions across all time steps, transformers mitigate the recurrent error accumulation that plagues RNNs and TCNs. However, Zeng \emph{et al.} (2022) demonstrate that even with positional encodings, the permutation-invariant nature of self-attention can still incur temporal information loss \cite{Zeng2022}.

\begin{figure}
    \centering
    \includegraphics[width=1\linewidth]{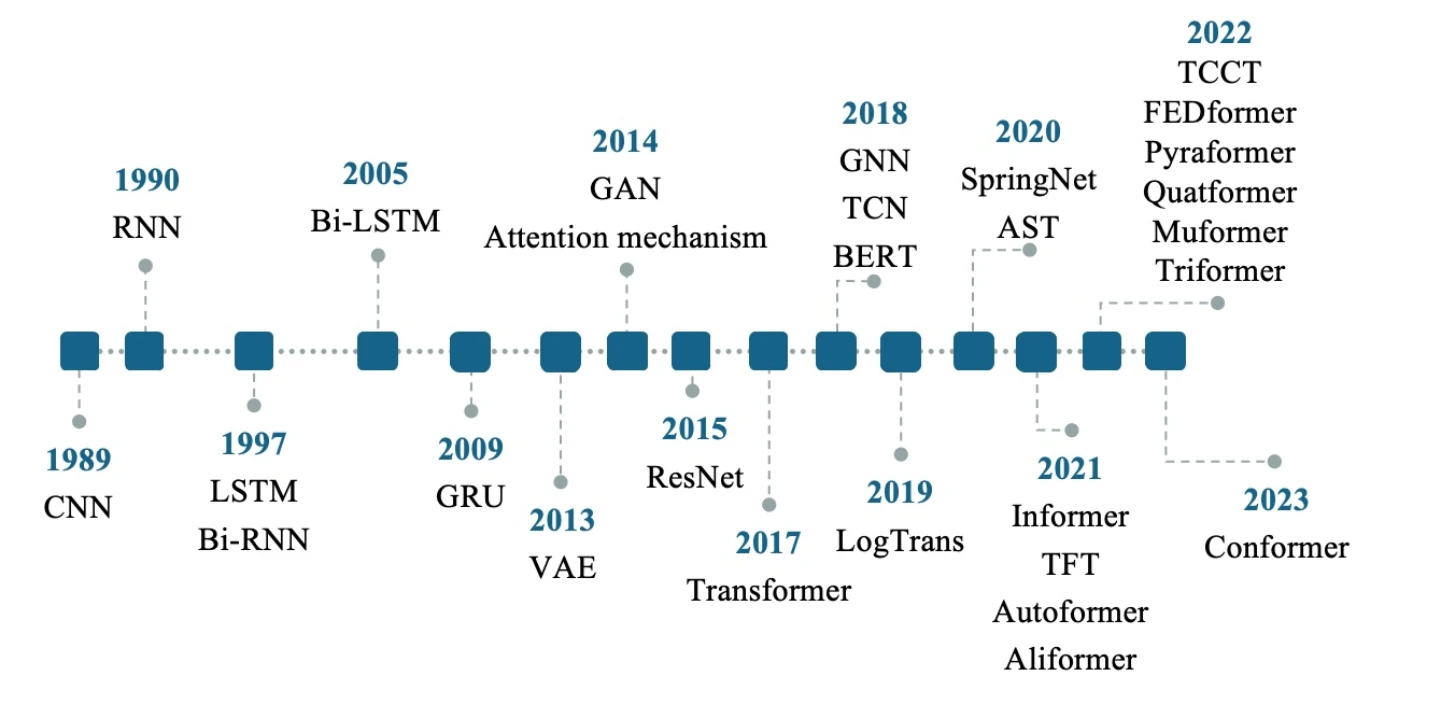}
    \caption{The Development History of Time Series Forecasting Deep Learning models, from \cite{Su2025}}
    \label{fig:tsf-development}
\end{figure}

Nevertheless, transformer architectures offer several distinct advantages for long-term time-series forecasting, and these are discussed in  Section \ref{ss:backbone}.

A timeline of the development of deep learning models can be seen in Figure \ref{fig:tsf-development}.

However, given that transformers and other deep neural network models can have from a few million up to billions of parameters, their use in daily frequency financial time series is constrained, since even a 20 year period consists of around 5000 data points–not nearly enough to train a model with a parameter count several orders of magnitude higher than that. This problem is exacerbated in certain circumstances, such as in financial instruments that have recently been listed or when dealing with macroeconomic data from emerging markets.  
Several separate also challenges arise in this context:
\begin{enumerate}
  \item Explainability: Black‐box architectures struggle to offer transparent justifications for their predictions, a critical requirement in high‐stakes applications \cite{Zhao2023}. In some industries such as energy, regulators go as far as demanding that the model decisions are based on reasonable and understandable factors \cite{ye2024, Reddy2023}.
  \item Data sparsity: Many traditional time series—such as economic indicators—are published only monthly or quarterly, resulting in even fewer training examples. This scarcity hampers the effective training of deep learning models; in most cases, available datasets remain insufficient for learning high-quality models \cite{Liu2021}.
  \item Multimodal learning: Integrating data from different modalities (for example price series and textual news) requires synchronising inputs observed at disparate intervals \cite{Tadas2017}.
  \item Transferability: Models trained on one domain may fail to generalise when seasonality and trend patterns differ across markets or instruments \cite{Shi2022}.
\end{enumerate}

Meanwhile, the field of natural language processing has witnessed the remarkable success of large language models (LLMs). By pretraining massive transformer backbones on vast amounts of unlabelled text and then fine‐tuning on comparatively small annotated datasets, these models have demonstrated powerful few‐shot transfer capabilities across diverse tasks.

Such models fall under the broader category of \textbf{Foundation Models} (FMs), which leverage a two‐stage paradigm: first, a single, general‐purpose model is pretrained on massive unlabelled corpora to learn rich, task‐agnostic representations; then, this pretrained model is fine‐tuned with limited supervision on specific downstream tasks.  Foundation models have achieved success across a wide range of AI disciplines, delivering state-of-the-art results in image understanding (e.g., CLIP, DALL·E) \cite{Dosovitskiy2020}, language processing (e.g., BERT, GPT) \cite{Devlin2018, radford2019}, and graph-based reasoning \cite{Sun2023}.

Inspired by these results and in an attempt to resolve the issues discussed above, researchers have recently begun developing \textbf{Time Series Foundation Models} (TSFMs) \cite{ye2024}. The core idea is to pre-train a large deep-learning‐based network on broad, diverse time‐series datasets so that it acquires general knowledge of common temporal patterns—such as trends, seasonality and cross‐series relationships.  Once pretrained, the model can be adapted to a particular downstream forecasting task by fine‐tuning only a small portion or all of its parameters, allowing it to leverage its broad temporal understanding while remaining robust even when the available data are limited \cite{ye2024,hu2021}.  

Such models would be ideal for many financial time series forecasting tasks, particularly where data exhibits nonlinearity, non-stationarity, cross series dependencies and long-term time step dependencies. In such settings, deep learning architectures can capture these intricate interactions, yet training them from scratch requires more data than is typically available. TSFMs models have the potential to address this gap by using pretrained temporal and multivariate representations that enable robust forecasting even when task specific observations are limited.

However, so far foundation models have seen limited applications to financial time series forecasting tasks. Xie \emph{et al.} (2023) have applied LLMs for zero shot forecasting in multimodal stock moving predictions, leveraging data from tweets and historical price data \cite{Xie2023}. They found that the results were subpar compared to benchmarks such as linear regression. These results are not discouraging since the researchers did not use a TSFM specifically but rather an LLM, and the authors also did not fine tune the model, suggesting that further training/fine-tuning could have lead to better results \cite{Xie2023}. Yu \emph{et al.} (2023) and Islam \emph{et al.} (2024) fine-tuned LLMs for stock market forecasting and obtain results that are superior to benchmarks such as ARMA and GARCH \cite{Yu2023} and show the potential for using LLMs when dealing with scarce financial datasets \cite{Islam2024}. However, these two papers also fail to use TSFMs specifically. Fu \emph{et al.} (2024) fine-tuned TimesFM \cite{timesfm}, a TSFM developed by Google Research, to forecast stock prices \cite{Fu2024}. The researchers performed mock trading for the fine-tuned TimesFM model in various financial markets and showed that it outperformed various benchmarks in terms of returns, Sharpe ratio, max drawdown and trading costs. However, one limitation of this use case is that TimesFM is a univariate model \cite{timesfm}, unable to model cross feature dependencies that are highly relevant in many financial forecasting tasks. Finally, Huynh \emph{et al.} (2024) use a fine-tuned version of Moirai \cite{moirai}, a TSFM developed by Salesforce, for anomaly detection in Vietnamese Financial Markets \cite{Huynh2024}. Their results indicate that the fine-tuned model significantly improves forecast accuracy and successfully identifies critical anomalies in the VN30-Index and demonstrates promising outcomes compared to traditional benchmarks such as Bollinger Bands and Moving Average Crossovers (MAC) \cite{Huynh2024}. The research gap this paper aims to address is visualised in Table \ref{tab:existing_literature}. 

\begin{table}[htbp]
\centering
\caption{Overview of Foundation Model Applications in Financial Time Series Forecasting}
\label{tab:existing_literature}
\begin{adjustbox}{max width=\textwidth}
\begin{tabular}{lccccc}
\toprule
\textbf{Paper} & \textbf{Used TSFM} & \textbf{Multivariate} & \textbf{Fine-tuned} & \textbf{Regression} & \textbf{Multiple FMs}\\
\midrule
Xie et al.\ (2023) \cite{Xie2023}     & \ding{55} & \checkmark & \ding{55} & \checkmark & \ding{55} \\
Yu et al.\ (2023) \cite{Yu2023}       & \ding{55} & \ding{55}  & \checkmark & \checkmark & \checkmark \\
Islam et al.\ (2024) \cite{Islam2024} & \ding{55} & \ding{55}  & \checkmark & \checkmark & \checkmark\\
Fu et al.\ (2024) \cite{Fu2024}       & \checkmark & \ding{55} & \checkmark & \checkmark & \ding{55} \\
Huynh et al.\ (2024) \cite{Huynh2024} & \checkmark & \checkmark & \checkmark & \ding{55} & \ding{55} \\
\midrule
\textbf{This paper}                   & \checkmark & \checkmark & \checkmark & \checkmark & \checkmark \\
\bottomrule
\end{tabular}
\end{adjustbox}
\end{table}

\subsection{Transfer Learning}
As noted in Section \ref{sec:tsf_hist}, a lot of financial time series forecasting are constrained by two related issues: the inherent scarcity of daily-frequency data (twenty years of daily prices yields only a few thousand observations) and the need to capture highly nonlinear, cross-feature and temporal dependencies. Traditional remedies for data scarcity include augmentation methods such as noise injection or time-series warping~\cite{wen2020}, synthetic oversampling techniques like SMOTE~\cite{huang2020}, adversarial generation via GANs~\cite{fathy2021}, and CNN-based feature extraction on limited samples~\cite{jeribi2024}. While each can expand or enrich a dataset, none confers the broad, domain-spanning pattern knowledge required for robust forecasting.  

By contrast, \textbf{transfer learning}~\cite{alstouhi2016} offers a fundamentally different solution: pretrained foundation models can be fine-tuned on small, task-specific financial datasets, enabling zero- or few-shot adaptation despite limited observations.  
Transfer learning has the following definition: "Given a source domain $\mathcal{D}_S$ and learning task $\mathcal{T}_S$, a target domain $\mathcal{D}_T$ and learning task $\mathcal{T}_T$, transfer learning aims to help improve the learning of the target predictive function $f_T(\cdot)$ in $\mathcal{D}_T$ using the knowledge in $\mathcal{D}_S$ and $\mathcal{T}_S$, where $\mathcal{D}_S \neq \mathcal{D}_T$, or $\mathcal{T}_S \neq \mathcal{T}_T$." \cite{pan2010}, A domain in this case is defined as a pair $\mathcal{D} = \{ X, P(X) \}$, where \(X\) is a feature space and \(P(X)\) is a marginal probability distribution. Therefore, from the transfer learning definition, it follows that if $\mathcal{D}_S \neq \mathcal{D}_T$, either $\mathcal{X}_S \neq \mathcal{X}_T \text{ or } P_S(X) \neq P_T(X)$. Additionally, a task $\mathcal{T} $ can be defined as a pair $ \mathcal{T} = \{ \mathcal{Y}, P(Y \mid X) \} $ , where $\mathcal{Y}$ is a label space and $P(Y \mid X)$  is a probabilistic viewpoint of the function \(f(x)\) mapping a feature space \(X\) to the corresponding label space $\mathcal{Y}$ . Thus, from the definition of transfer learning, it follows that either $\mathcal{Y}_S \neq \mathcal{Y}_T \text{ or } P(Y_S \mid X_S) \neq P(Y_T \mid X_T)$ \cite{pan2010}. Different combinations of differences in source and target domains and source and target tasks lead to the rise of different transfer learning types, which are described in Table \ref{tab:learning_settings}. 

\begin{table}[ht]
\centering
\caption{Types of Transfer Learning and Their Characteristics (from \cite{pan2010})}
\label{tab:learning_settings}
\begin{tabular}{lcc}
\toprule
\textbf{Learning Type} & \textbf{Domain Alignment} & \textbf{Task Alignment} \\
\midrule
Traditional Machine Learning     & Same                     & Same                      \\
Inductive Transfer Learning      & Same or Different        & Different but Related     \\
Transductive Transfer Learning   & Different but Related    & Same                      \\
Unsupervised Transfer Learning   & Different but Related    & Different but Related     \\
\bottomrule
\end{tabular}
\end{table}

Each type of transfer learning requires a different approach in "how" the transfer learning is done. Four of the main approaches are as follows.
\begin{enumerate}
    \item Instance-based transfer learning: 
    This approach focuses on re-weighting or selecting relevant instances from the source domain to improve learning in the target domain. The idea is to identify and use data points from the source domain that are most similar or relevant to the target domain while down-weighting or ignoring irrelevant ones. Algorithms like TrAdaBoost fall into this category \cite{pan2010, alstouhi2016}.

    \item Feature-representation transfer: 
    This method aims to learn a shared feature space or representation that is effective across both the source and target domains. The goal is to map both domains into a common feature space where the differences between the domains are minimised. Subspace-based, transformation-based, and construction-based are among the main feature-based domain adaptation methods \cite{farahani2021, pan2010}. One example of such a domain adaptation technique is adversarial learning \cite{zhang2023}.

    \item Parameter transfer: 
    In this approach, models trained on the source domain share parameters or priors with the target domain model. The source domain provides knowledge in the form of model parameters, which are then fine-tuned or adapted for the target task \cite{pan2010}. Fine-tuning foundation models is a common example of parameter transfer.

    \item Relational Knowledge transfer: 
    This type of transfer learning focuses on transfer learning within relational domains, where the assumption is that the relationships among data in the source and target domains are similar. In this context, the transferable knowledge lies in the relationships among the data rather than the data itself \cite{pan2010}.

\end{enumerate}

Table \ref{tab:transfer_learning_approaches} shows what approach can be used depending on what transfer learning setting one is dealing with. 

\begin{table}[ht]
\centering
\caption{Transfer Learning Approaches by Setting (from \cite{pan2010})}
\label{tab:transfer_learning_approaches}
\begin{tabular}{lccc}
\toprule
\textbf{Approach} & \textbf{Inductive} & \textbf{Transductive} & \textbf{Unsupervised} \\
\midrule
Instance Transfer               & \checkmark & \checkmark &               \\
Feature Representation Transfer & \checkmark & \checkmark & \checkmark    \\
Parameter Transfer              & \checkmark &            &               \\
Relational Knowledge Transfer   & \checkmark &            &               \\
\bottomrule
\end{tabular}
\end{table}

When fine-tuning foundation models, the transfer learning type is Inductive Transfer Learning, given that the source task (forecasting in a variety of different industries and settings) and target tasks (forecasting treasury yields or equity volatilities) are different but related. The transfer learning approach used is parameter transfer.

\subsection{Preliminaries on Time Series Foundation Models}
Foundation models are large neural networks pretrained on massive unlabelled datasets using self-supervised objectives to learn domain-agnostic representations.  After this pretraining phase, either all parameters or only a small subset of their parameters (or an added task-specific head) is fine-tuned on labelled data, enabling powerful few-shot adaptation across a wide range of downstream tasks \cite{ye2024}.

Time series data consist of observations indexed by time, represented either as a univariate sequence \(\mathbf{x}=\{x_t\}_{t=1}^T\), \(x_t\in\mathbb{R}\), or as a multivariate sequence \(\mathbf{X}=\{\mathbf{x}_t\}_{t=1}^T\), \(\mathbf{x}_t\in\mathbb{R}^N\).  Effective analysis must capture both autocorrelation (temporal dependencies within each channel) and cross-correlation (interactions across channels).  

Time Series Foundation Models (TSFMs) extend the foundation-model paradigm to time series data by first pretraining a deep-learning based backbone (typically a transformer) on broad, heterogeneous corpora of time series so as to learn general patterns of trend, seasonality, noise and inter-series relationships.  In the fine-tuning stage, this pretrained network can be adapted to specific time series task using several methods, as described in Section \ref{sec:fine-tune}, thus leveraging broad temporal knowledge even when task-specific datasets are limited. 

While  time series tasks include classification of entire sequences, detection of anomalies, imputation of missing values, and forecasting of future values, in this paper we focus on the task of forecasting.  Time series forecasting aims to predict the next \(H\) future values given a lookback window of length \(L\) \cite{ye2024}.  Formally, let
\[
\mathbf{X}_{L}=\{\mathbf{x}_{1},\mathbf{x}_{2},\dots,\mathbf{x}_{L}\},\quad \mathbf{x}_{t}\in\mathbb{R}^N,
\]
and define a learnable forecasting function
\[
f:\mathbb{R}^{N\times L}\longrightarrow\mathbb{R}^{N\times H},
\]
so that
\[
\{\hat{\mathbf{x}}_{L+1},\hat{\mathbf{x}}_{L+2},\dots,\hat{\mathbf{x}}_{L+H}\}
=f(\mathbf{X}_{L}).
\]
Depending on the value of \(H\), one distinguishes short-term from long-term forecasting, and when data are extremely scarce, few-shot or zero-shot forecasting pipelines can be employed using pretrained TSFMs to generalise with minimal additional training \cite{ye2024}.  

\subsection{Types of TSFMs}
Time Series Foundation Models differ along several key properties, including how they are pretrained, their underlying architecture, and their input handling. This section surveys these distinctions to clarify the design space and capabilities of modern TSFMs.
\subsubsection{Pretrained-from-Scratch vs.\ LLM-Adapted Models}
\begin{figure}
    \label{fig:LLM-adapted}
    \centering
    \includegraphics[width=1\linewidth]{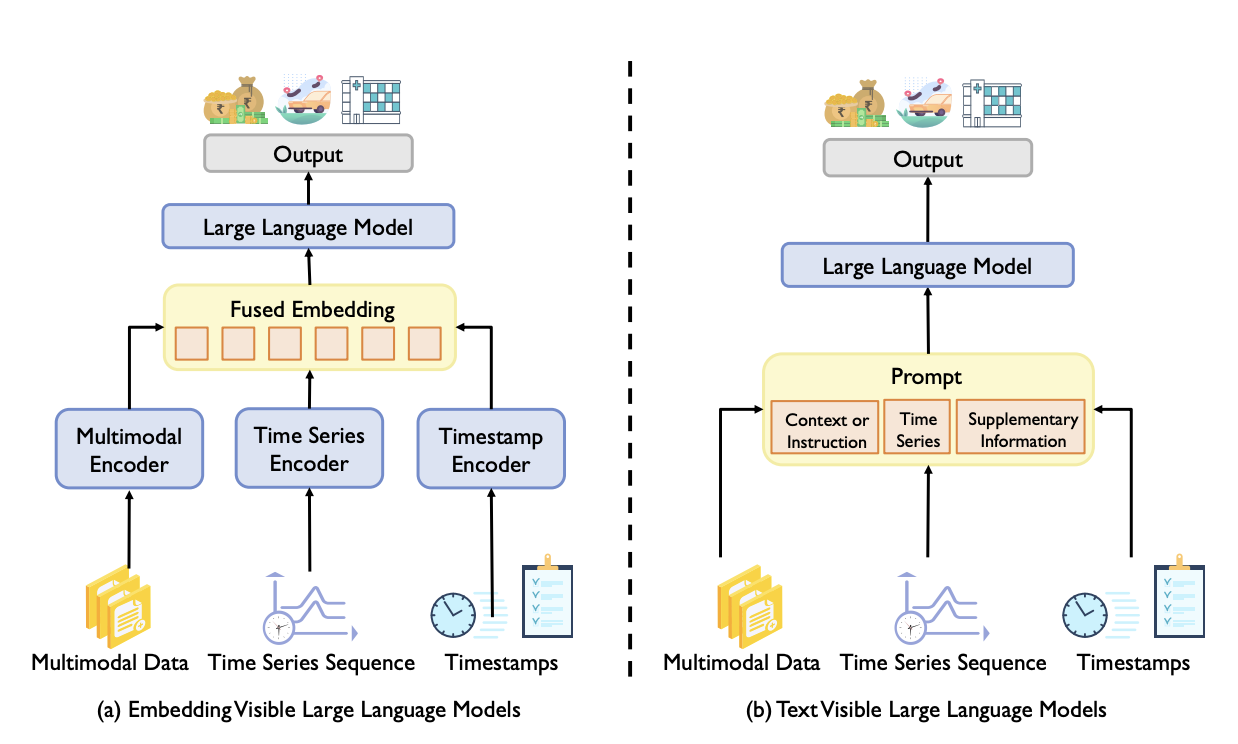}
    \caption{Types of LLM Adaptation to Time Series Methods, from \cite{ye2024}}
    \label{fig:enter-label}
\end{figure}

Ye \emph{et al.}\ \cite{ye2024} categorise Time Series Foundation Models into two main classes.  The first class comprises models \textbf{pretrained from scratch} on large collections of time series data: here the entire transformer backbone is trained with time series data.  Pioneering examples include TimesFM \cite{timesfm} and TimeGPT \cite{timegpt}. These models, however, face unique challenges such as the relatively small size of available time‐series data compared to text and the lack of language generation capabilities.
The second class consists of \textbf{LLM‐adapted} foundation models. One approach in adapting LLMs to time series tasks, known as embedding-visible adaptation, maps time series inputs into vectorised time series representations that can be passed as inputs to an LLM, often by slicing the input into fixed‐length “patches” or by applying a learned encoder, and these embeddings are presented directly to the LLM as if they were word tokens \cite{ye2024}.  A second adaptation method, known as the text visible adaptation, leaves the LLM entirely unchanged and instead reformulates time series tasks as textual prompts.  Numerical values or summary statistics are serialised into natural‐language sentences or CSV‐style strings, combined with task instructions (for example “Forecast the next 10 days of closing prices: …”),  and passed to the LLM in its original text‐processing mode.  This “pre‐train, prompt and predict” workflow uses in‐context learning and obviates any model‐architecture changes \cite{ye2024}. The difference between the two approaches in adapting LLMs into TSFMs is shown in Figure \ref{fig:LLM-adapted}. 

\subsubsection{Backbone Type}
\label{ss:backbone}
Additionally, TSFMs can be distinguished by their \textbf{backbone}. A TSFM's backbone is a scalable deep learning model which forms the basis of the TSFM \cite{ye2024}. Yeh \emph{et al.} (2023) \cite{yeh2023} explored four different neural network architectures as the backbone model for foundation models, and these are explained below.

\begin{enumerate}
  \item \textbf{LSTM:}  A classic type of recurrent neural network with gated memory cells \cite{Hochreiter1997}.
  \item \textbf{GRU:}  The Gated Recurrent Unit is another type of recurrent neural network that uses two gates: an update gate and a reset gate—to control information flow \cite{Dey2017}. GRUs have fewer parameters than LSTMs and often train faster, while still capturing long‐range dependencies in sequential data.
  \item \textbf{ResNet:}  A residual convolutional neural network backbone adapted from computer vision \cite{he2016}. Fawaz \emph{et al.} (2019) have demonstrated that ResNet is one of the strongest models for time series classification \cite{Fawaz2019}.
  \item \textbf{Transformer:}  Introduced by Vaswani \emph{et al.} \cite{Vaswani2017}, the transformer replaces recurrence with a self‐attention mechanism that computes pairwise interactions across all input positions.  Given query, key and value matrices \(Q\), \(K\) and \(V\), scaled dot‐product attention is defined as 
  \begin{equation}
    \mathrm{Attention}(Q,K,V)=\mathrm{softmax}\!\bigl(\tfrac{QK^{\!\top}}{\sqrt{d_k}}\bigr)V
  \end{equation}
  which allows each time step to attend to every other, capturing long‐range dependencies in a single layer.  Multiple such attention “heads” run in parallel, their outputs concatenated and passed through linear projections and feed‐forward networks. 
  
Transformer architectures offer several distinct advantages for long-term time-series forecasting. First, the self-attention mechanism provides global context by allowing each output time step to attend directly to all past inputs \cite{Vaswani2017, Lim2021}, thereby capturing long-range dependencies and multivariate interactions without the error accumulation inherent in recurrent update rules \cite{Su2025}. Second, unlike RNNs or TCNs, transformers process entire sequences in parallel, yielding substantially faster training and inference on modern hardware accelerators \cite{Vaswani2017}. Third, the attention operation is highly flexible: it can be made sparse or structured (e.g.\ ProbSparse or LogSparse attention), combined with frequency-domain blocks (as in FEDformer \cite{Zhou2022}), or hybridised with decomposition modules (as in Autoformer \cite{Wu2021}), enabling practitioners to tailor the trade-off between expressivity and computational cost \cite{Su2025}. Additionally, by attending not only over time but also across input features, transformers can explicitly model complex relationships between variables in multivariate series. Finally, empirical benchmarks demonstrate that transformer-based models achieve state-of-the-art accuracy on long-horizon forecasting tasks, consistently outperforming classical statistical methods and earlier deep-learning approaches when sufficient data are available \cite{Su2025}. 

Figure~\ref{fig:transformer} illustrates a single transformer block with its multi‐head attention and feed‐forward blocks, along with the residual connections and positional encodings that inject sequence order information.

    \begin{figure}
        \centering
        \includegraphics[width=0.5\linewidth]{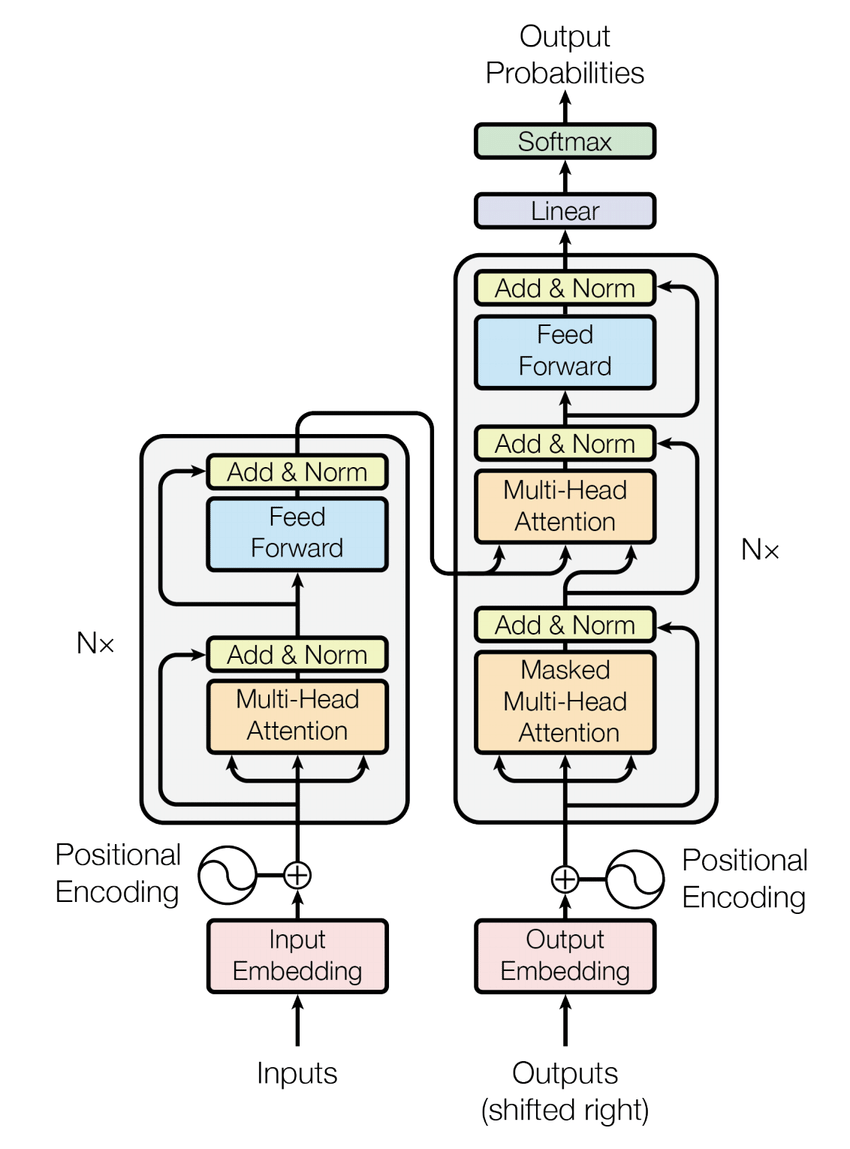}
        \caption{Transformer Architecture, from \cite{Vaswani2017}}
        \label{fig:transformer}
    \end{figure}

Transformers come in three architectural variants. The original encoder--decoder model uses an encoder to transform the entire input sequence into a latent representation and a decoder to generate each output element autoregressively, making it ideal for multi-step forecasting where the input and output horizons differ. Encoder--only models, such as BERT \cite{Devlin2018}, build deep contextual embeddings of a fixed input and excel at tasks like representation learning or anomaly detection; they can also be extended to forecasting by attaching a simple prediction head. Decoder--only models, exemplified by the GPT family \cite{radford2019}, generate sequences one step at a time and are well suited to univariate next-value prediction and few-shot forecasting scenarios, where the model must flexibly adapt to new series with minimal additional training \cite{ye2024}. In the time-series domain, encoder--decoder transformers are most commonly applied to multi-horizon forecasting tasks, since their separate encoder and decoder stacks naturally handle differing input and output lengths and can model complex conditional dependencies across long prediction windows.

\end{enumerate}

In their study, Yeh \emph{et al.} (2023) \cite{yeh2023} found that although self-supervised pretraining improved all four backbones, the Transformer consistently yielded the strongest downstream performance. This has been the motivation for the majority of TSFMs to use a transformer backbone.

\subsubsection{Channel Setting}
TSFMs can also differ in their \textbf{channel setting}. In the channel‐independence configuration, each feature series is treated separately: a univariate TSFM is applied independently to each channel, often by first vectorising or “patching” the series (for example, grouping adjacent time steps into fixed‐length patches as in PatchTST \cite{nie2023} or constructing lag‐vector tokens at varying intervals as in Lag‐Llama \cite{Rasul2023}). This approach simplifies model design, since the same backbone can be reused across any number of series, but it may miss cross‐series interactions.

By contrast, the channel‐mixing configuration ingests all channels simultaneously, fusing them into a joint embedding space so that the TSFM can learn interdependencies directly.  For example, GTT \cite{Feng2024} reshapes multivariate inputs by folding channel dimensions into the batch and then applies spatio-temporal attention to capture both temporal patterns and variable correlations.  While channel‐mixing models can exploit rich cross‐feature relationships critical in finance, they require a fixed or explicitly handled channel layout and can be more complex to pre-train and fine‐tune.  

\subsection{A Comparison of Available TSFMs}
Table \ref{tab:differet_TSFMs} gives an overview of some of the most prominent TSFMs, characterised by the features outlined in this section. The following section \ref{TSFM_selection} explains the models that will be used in this paper in further detail.

\begin{table}[ht]
\label{tab:differet_TSFMs}
\caption{Characteristics of prominent Time Series Foundation Models. “Mode” refers to transformer architecture; “Init. Source” indicates whether the model was pretrained from scratch on time‐series data or adapted from a large language model. 
$^{\dagger}$TTM does not use a transformer architecture—it is based on lightweight MLP-based mixing rather than a transformer.}
\centering
\begin{tabular}{lcccl}
\toprule
\textbf{Model}        & \textbf{Params} & \textbf{Mode}         & \textbf{Channels} & \textbf{Init. Source}           \\
\midrule
TimesFM \cite{timesfm}               & 225M            & Decoder-only          & Uni.             & Scratch                 \\
MOIRAI (Small) \cite{moirai}& 13.8M& Encoder-only          & Multi.             & Scratch                 \\
TimeGPT \cite{timegpt}               & –               & Encoder–decoder       & Uni.             & Scratch                 \\
PatchTST \cite{nie2023}        & 15M             & Encoder-only          & Multi.           & Scratch                 \\
Lag-Llama \cite{Rasul2023} & 2.45M               & Decoder-only          & Uni.             & Scratch                 \\
GTT \cite{Feng2024}                   & 57M             & Encoder-only          & Multi.           & Scratch                 \\
Tiny Time Mixers (TTM) \cite{vijay2024}& 0.805M& Decoder-only$^{\dagger}$& Multi.             & Scratch                 \\
Chronos-Bolt (Small) \cite{chronos}& 47.7M& Encoder–decoder& Multi.           & Adapted from LLM               \\
 MOMENT-Small \cite{moment}& 37.9M& Encoder-only& Multi.&Adapted from LLM\\
\bottomrule
\end{tabular}

\end{table}

\subsection{TSFMs Selected}
\label{TSFM_selection}
Two TSFMs were selected to cover the main categories of model design, namely Chronos-Bolt (Small) \cite{chronos}, and Tiny Time Mixers (TTM) \cite{vijay2024}. Chronos represents both an encoder–decoder architecture and is also a TSFM adapted from an LLM. TTM, on the other hand, is a TSFM trained from scratch for time series forecasting and also does not have a transformer backbone. It has a decoder-only architecture.  Given that in this paper only multivariate financial forecasting applications are explored, both these models are multivariate. While they differ in parameter size, they are both small enough to train efficiently on limited hardware.

\subsubsection{Tiny Time Mixers (TTM)}

\begin{figure}
    \centering
    \includegraphics[width=1\linewidth]{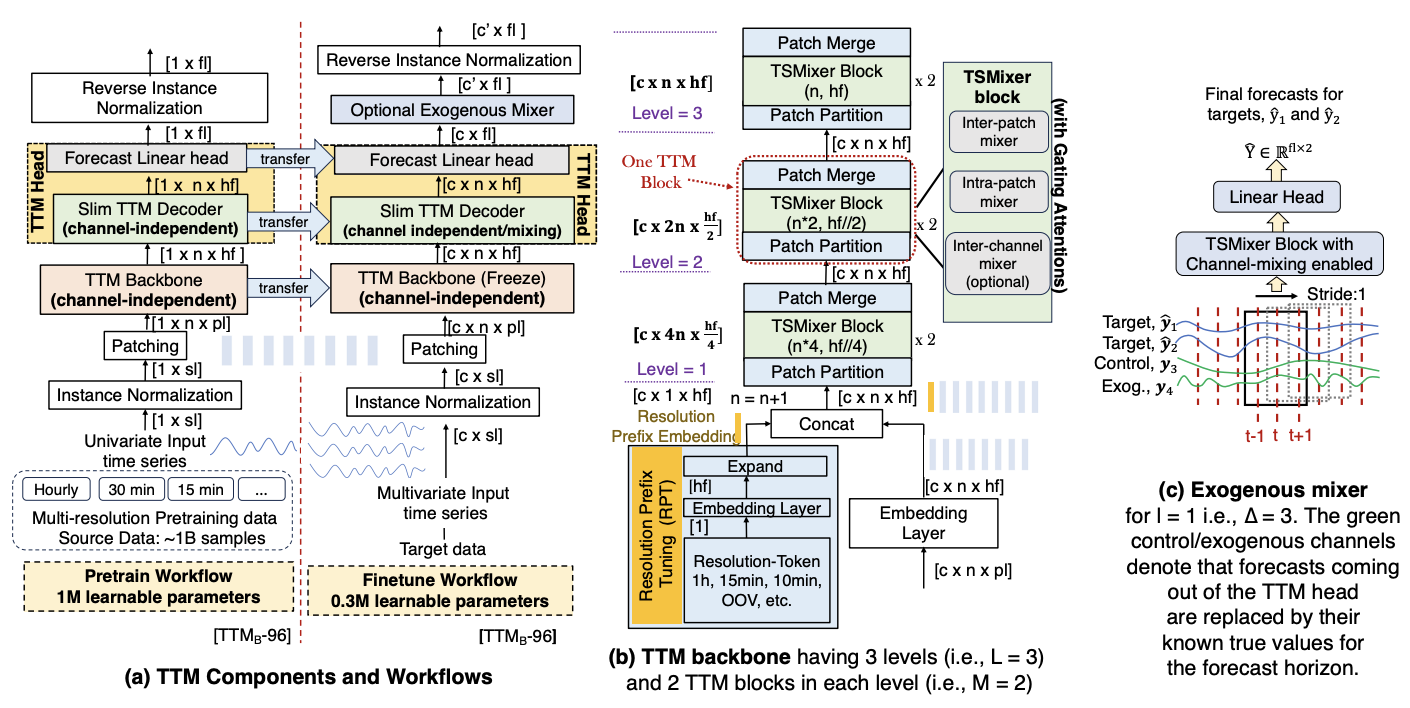}
    \caption{Architecture of Tiny Time Mixers (TTM), a TSFM Developed by IBM, from \cite{vijay2024}}
    \label{fig:ttm_architecture}
\end{figure}

Tiny Time Mixers (TTM), developed by IBM, is a parameter-efficient foundation model for time series forecasting that avoids the use of a transformer. Instead, it is built on top of TSMixer, a lightweight feedforward architecture that replaces self-attention with multi-layer perceptrons (MLPs). As shown in Figure~\ref{fig:ttm_architecture}, TSMixer alternates between intra-patch, inter-patch, and optional inter-channel MLP blocks to capture dependencies along different dimensions. This design greatly reduces computational complexity while retaining the capacity to model temporal structure, enabling fast training and inference.

As shown in Figure~\ref{fig:ttm_architecture}a, TTM follows a two-stage workflow: a univariate pretraining phase with a channel-independent backbone, and a multivariate fine-tuning phase using a lightweight decoder and head. To support generalisation across series with different temporal resolutions, the model introduces several innovations: (i) \textit{Resolution Prefix Tuning} encodes sampling frequency via a learnable token (Figure~\ref{fig:ttm_architecture}b); (ii) \textit{Adaptive Patching} allows different layers to operate on different patch sizes across levels; and (iii) \textit{Diverse Resolution Sampling} expands the training set with subsampled versions of the original time series.

The architecture is hierarchical, comprising multiple levels of patch partitioning and merging, where each level applies two TSMixer blocks. Each block mixes across time, patches, and optionally channels, using inter- and intra-patch MLPs (Figure~\ref{fig:ttm_architecture}b, right). During fine-tuning, an optional \textit{Exogenous Mixer} can be used to inject true exogenous values during decoding, which improves multivariate forecasting (Figure~\ref{fig:ttm_architecture}c).

The forecasting process in TTM operates by taking the input time series and dividing it into fixed-length patches, which are then processed by alternating intra-patch and inter-patch MLP mixer blocks to learn hierarchical representations that capture both short-term dynamics and long-term patterns. During inference, the model leverages these learned representations to directly output the entire forecast horizon in a single forward pass, optimising across all prediction steps with a joint loss. This direct multi-step approach eliminates the recursive feedback loop of autoregressive methods, preventing accumulation of errors over successive predictions. As a result, TTM is particularly well-suited for modelling series with weak or no autocorrelation at long forecast horizons, such as financial returns.

Despite its small size of approximately 1 million parameters in the full model, TTM is trained on over one billion time series samples drawn from a wide range of real-world domains. Particularly noteworthy is the inclusion of Bitcoin price data, which may enhance transferability to financial forecasting tasks due to its closer alignment with the target domain. A full list of pretraining datasets is provided in Appendix \ref{appendix:ttm_datasets}.

\subsubsection{Chronos}

Chronos is a pretrained, probabilistic time series foundation model developed by Amazon introduced as part of the Chronos framework, which repurposes large language models for forecasting tasks \cite{chronos}. Unlike traditional time series models that operate on continuous real values, Chronos-Bolt first discretises the input space using a fixed vocabulary of quantised bins, allowing the use of transformer-based language models with minimal architectural modifications.

\begin{figure}
    \centering
    \includegraphics[width=1\linewidth]{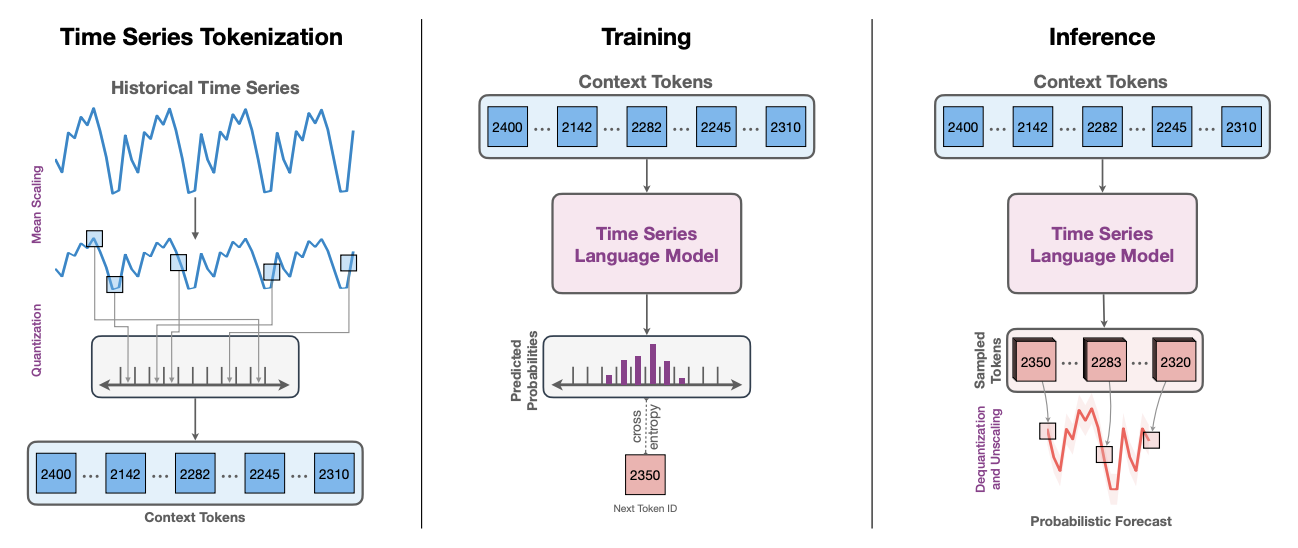}
    \caption{Chronos Forecasting Process. From left to right: time series tokenisation via scaling and quantisation; training via language modelling on token sequences; and inference via autoregressive sampling and dequantisation. Image from \cite{chronos}.}
    \label{fig:chronos_diagram}
\end{figure}

As illustrated in Figure~\ref{fig:chronos_diagram}, the input historical time series is first normalised via mean scaling. The normalised values are then quantised into a finite set of discrete tokens using uniform binning. This yields a tokenised sequence analogous to word tokens in NLP, which can be processed by a standard language model.
Chronos adopts the encoder-decoder T5 architecture, pretrained from scratch on a large corpus of publicly available time series data, as well as synthetic series generated using Gaussian processes. This broad pretraining corpus encourages generalisation across domains and improves zero-shot forecasting capabilities. Notably, no time-series-specific features (such as calendar information or lag variables) are used during training.

The model is trained using categorical cross-entropy loss over token IDs. This corresponds to modelling the time series as a sequence of categorical variables, thereby enabling regression via classification. Despite the categorical nature of the objective, the model can express complex and potentially multimodal output distributions, a key advantage for probabilistic forecasting.

Chronos is pre-trained on a variety of datasets across several domains, although it is important to note that these spanned only energy, nature, transport and web datasets, and the exclusion of any finance related datasets is noteworthy. The specific list of datasets used during pre-training is provided in Appendix \ref{appendix:chronos_datasets}.

During inference, Chronos operates autoregressively. Given a sequence of context tokens, it predicts the next token's distribution and samples from it to generate the next step. This sampled token is then appended to the input, and the process is repeated for the desired forecast horizon. The sampled tokens are then mapped back to real values using the dequantisation function and unscaled to produce the final forecast. This procedure allows multiple forecast trajectories to be generated, enabling probabilistic forecasts.

While Chronos models may vary in size and architecture, this study uses \textit{Chronos-Bolt-Small}, a lightweight yet effective variant to the one introduced in the original work. The Chronos-Bolt series is significantly more efficient than earlier versions: models in this family achieve 5\% lower error, run up to 250 times faster, and require 20 times less memory than the original Chronos models of equivalent size \cite{chronos_t5_small_hf}.

\subsection{Fine-Tuning Methods}
\label{sec:fine-tune}
Once a model is selected, there are several ways to fine-tune it, with different fine-tuning methodologies offering different trade-offs between computational efficiency, accuracy and generalisation capabilities. The most common fine-tuning methods are explored in this section. 

\subsubsection{Full Fine-Tuning}
\begin{figure}
    \centering
    \includegraphics[width=0.75\linewidth]{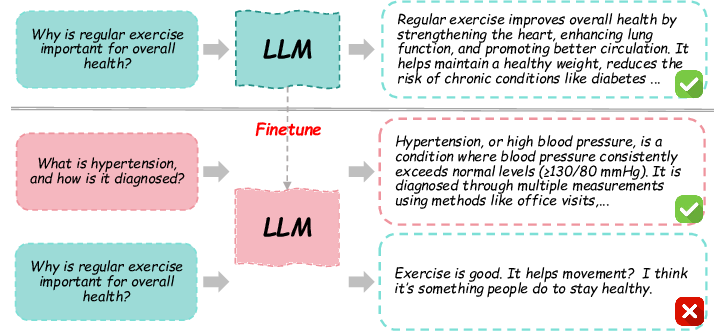}
    \caption{Illustration of Catastrophic Forgetting, where a Foundation Model "Forgets" the Answer to a Previously known Question, from \cite{Song2025}}
    \label{fig:catastrophic}
\end{figure}

Full fine‐tuning updates every weight in the pretrained model, allowing maximal flexibility to adapt to the target task \cite{Lv2024}. However, this approach incurs substantial computational and memory costs, since gradients and optimiser states must be stored for all parameters, and training times can become prohibitive for large transformer backbones. In addition, full fine‐tuning carries a heightened risk of catastrophic forgetting, whereby task-agnostic knowledge acquired during pretraining is overwritten by task-specific updates unless careful regularisation or very low learning rates are employed \cite{Song2025}. Figure \ref{fig:catastrophic} illustrates an example of catastrophic forgetting in an LLM. Consequently, while full fine-tuning can yield strong in-domain performance, practitioners often seek lighter-touch alternatives that adapt to the target task more effectively. 

\subsubsection{No Fine-Tuning}
In the no fine-tuning paradigm, the pretrained model is used directly (often via an API) without any updates to its parameters. This zero-shot approach relies entirely on the model’s existing pattern-recognition and reasoning abilities to tackle downstream time-series tasks. Although it avoids the computational and infrastructure overhead of training, it can lead to poor performance in the target task. For example, Xie \emph{et al.} (2023) apply zero-shot inference of GPT-4 to multimodal stock movement forecasting, yet report results below those of simple benchmarks \cite{Xie2023}. Another study also showed that such off-the-shelf inference can yield underwhelming results, sometimes falling short of even simple statistical baselines \cite{Yu2023}.  Moreover, repeated API calls, especially for long input sequences, can incur significant latency and monetary cost, making no-tuning methods less practical for high-frequency forecasting applications.  

\subsubsection{Parameter-Efficient Fine-Tuning}

Parameter-Efficient Fine-Tuning (PEFT) adapts large pretrained models by updating only a small fraction of their parameters, drastically reducing training cost and memory usage while preserving most of the backbone’s representations.  Broadly speaking, PEFT methods fall into three categories, which can be visualised in Figure \ref{fig:peft} \cite{Lialin2023}. These are as follows:

\begin{figure}
    \centering
    \includegraphics[width=0.75\linewidth]{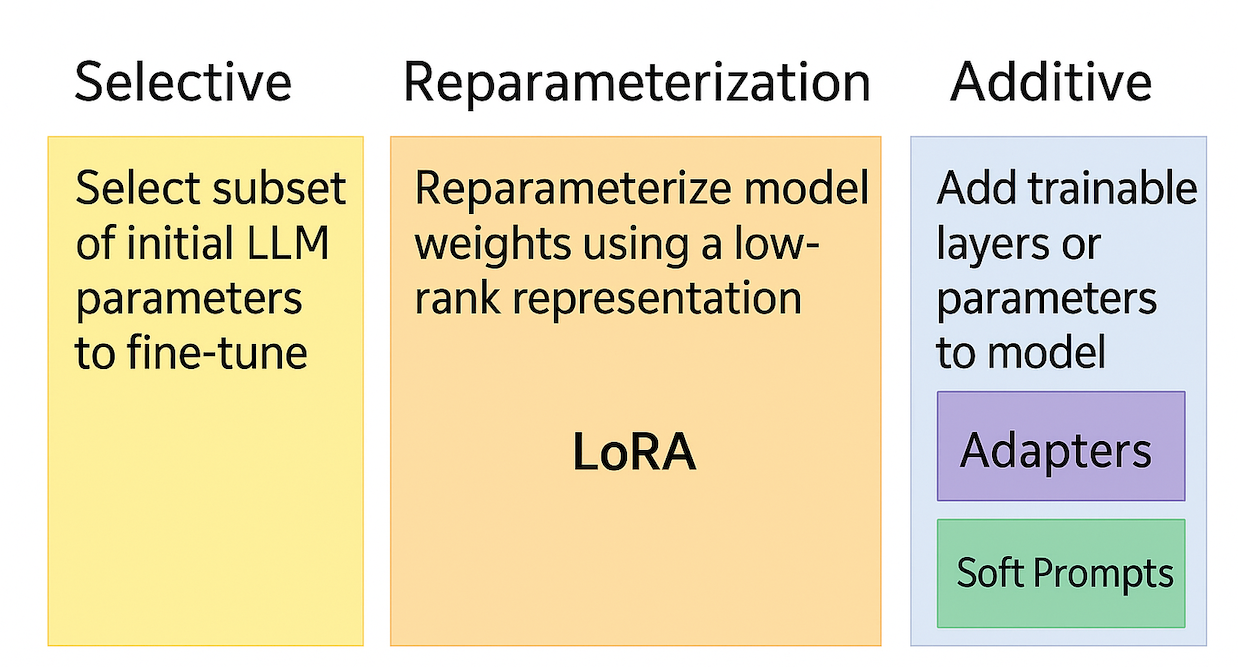}
    \caption{Types of Parameter Efficient Fine-Tuning (PEFT) Methods, from \cite{Adik2023}}
    \label{fig:peft}
\end{figure}

\textbf{Selective Fine-Tuning Methods:}  In this fine-tuning method, only a subset of the original model’s weights is unfrozen and fine-tuned on the target data, while the rest remain fixed \cite{Donahue2013}. Most often, the whole TSFM backbone is frozen while the rest of the parameters, such as the final prediction head or a few top layers, are updated. Other selective PEFT methods include fine-tuning only the bias terms within the model \cite{Zaken2021} or only updating the parameters of the LayerNorm components present in the attention module of the model architecture \cite{Zhao2023LN}. By limiting updates to just these modules, selective PEFT slashes gradient storage and computation, and also mitigates catastrophic forgetting of pretrained knowledge.

\textbf{Reparameterisation Methods:} Reparameterisation-based methods in PEFT aim to reduce the number of trainable parameters by constraining updates to low-rank subspaces. The idea is grounded in the observation that deep neural networks often have low intrinsic dimensionality, allowing them to be effectively adapted using low-rank structure \cite{Maddox2020, Li2018}.

Aghajanyan \emph{et al.} (2020) provided empirical evidence that fine-tuning can be performed effectively within low-rank subspaces of the full parameter space \cite{Aghajanyan2020}. Their results showed that the number of trainable dimensions required for good performance is significantly smaller than the total number of model parameters, particularly for larger or more extensively pre-trained models. This supports the broader motivation for methods that reparameterise model updates using compact, low-rank formulations.

One of the most widely adopted methods in this category is Low-Rank Adaptation (LoRA) \cite{hu2021}, which expresses the weight update $\Delta W$ as:
\[
\Delta W = W^{\text{down}} W^{\text{up}},
\]
where $W^{\text{down}} \in \mathbb{R}^{d \times r}$ and $W^{\text{up}} \in \mathbb{R}^{r \times k}$, with $r \ll \min(d, k)$. By freezing the original model weights and training only the low-rank components, LoRA enables parameter-efficient adaptation while maintaining strong downstream performance \cite{hu2021, Lialin2023}.

Recent methods have extended this idea through alternative reparameterisations, such as Kronecker product decompositions \cite{Edalati2022}, which further reduce the number of trainable parameters while preserving expressiveness.

\textbf{Additive Fine-Tuning Methods:}  
Additive methods extend pre-trained models by introducing new trainable parameters, typically in the form of lightweight modules or embeddings, while keeping the original model weights frozen. This approach is among the most widely explored in the PEFT literature due to its flexibility and modularity \cite{Lialin2023}.

Two major branches of additive methods have emerged: \textbf{adapter-based} techniques and \textbf{soft prompting}.

Adapters introduce small bottleneck-style feedforward layers into the Transformer architecture, typically inserted between the output of sub-layers and the residual connections \cite{houlsby2019}. These layers are initialised from scratch and trained while the main model parameters remain fixed. Adapters have gained popularity due to their efficiency and the ease with which they can be composed or swapped for multi-task settings \cite{Pfeiffer2020}.

Numerous variations have been proposed, including modifying the placement of adapter modules \cite{Zhu2021} or combining them with pruning techniques \cite{He2022, Lialin2023}.

Soft Prompting techniques aim to guide the model’s output by modifying its input rather than its weights. Unlike traditional textual prompts, soft prompts are continuous trainable embeddings prepended to the model’s input sequence \cite{Liu2021, Liu2021b}. These embeddings are optimised via gradient descent, shifting the adaptation task from discrete prompt engineering to a differentiable learning problem \cite{Lialin2023}.

\begin{figure}[H]
    \centering
    \includegraphics[width=1\linewidth]{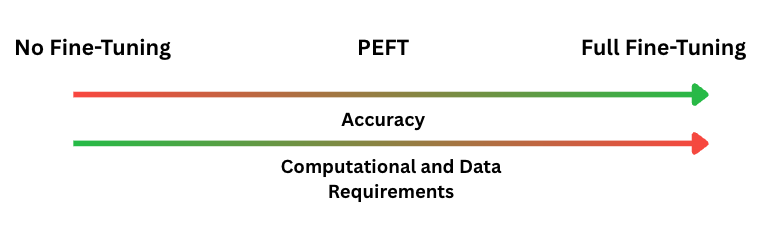}
    \caption{Trade-offs When Using Different Fine-Tuning Methods}
    \label{fig:trade-off}
\end{figure}

Each fine-tuning strategy presents a different tradeoff between computational cost, data efficiency, and task performance. Full fine-tuning offers maximal flexibility and often yields the highest accuracy, but it requires significant compute and storage, particularly for large models. It also requires significantly more data, which is the original problem TSFMs were meant to overcome. At the other end of the spectrum, no fine-tuning leverages pretrained representations in a zero-shot or few-shot manner, eliminating training costs but at the expense of performance. PEFT methods, however, strike a practical middle ground by enabling adaptation with a fraction of the parameters and compute, while preserving much of the predictive power of full fine-tuning. Further detail of what fine-tuning techniques are used in this paper are provided in Section \ref{exp}. Figure \ref{fig:trade-off} illustrates this trade-off. 

\subsection{Transferability Assessment of Pretrained TSFMs}
\label{sec:transferability_evaluation}
Transferability refers to the ability of a model or representation learned on a source task or domain to improve performance on a target task. In the context of deep transfer learning, estimating transferability helps decide whether and how to leverage pretrained models for new tasks. Xue \emph{et al.}~\cite{Xue2024} propose a taxonomy of four broad families of transferability estimation methods, three of which are applicable for regression tasks:

\begin{enumerate}
\item Task Relatedness: assesses how similar the source and target tasks are. Some approaches look at how well theoretical learning bounds hold across domains, while others use statistical measures to compare the distributions of data from each task. The basic assumption is that the more similar the tasks, the more transferable the model will be.
\item Source–Target Comparison Methods: a pretrained model is applied to target-domain inputs, and its outputs are compared to target labels. Metrics such as negative conditional entropy or Log Expected Empirical Prediction (LEEP) treat the model’s predictions as soft labels and quantify their informativeness about the true labels. Since these techniques require no further model retraining, they offer rapid, deployment-friendly estimates of transferability. This method, however, is only suitable for classification tasks.
\item Representation Analysis: examines how well source-trained features align with target-task requirements. Xue \emph{et al.} (2024) \cite{Xue2024} describe two sub-approaches: (i) \textit{Task-representation methods}, which train a shared encoder on multiple tasks and assess few-shot performance when plugging encoded features into a new head; and (ii) \textit{Pretrained-model representation methods}, which rank models by correlating their source-task and target-task performance or by analysing statistics of extracted embeddings (e.g.\ Fisher information). High alignment or correlation suggests that pretrained features are well suited to the target.
\item Optimal Transport: measure the distance between the source and target distributions by solving an optimal transport problem. These approaches assume that closer distributions in a geometric sense are more likely to transfer well, and they often quantify this alignment through entropically regularised transport costs. Although conceptually appealing, they can be computationally intensive and typically require consistent feature spaces or aligned embeddings \cite{Xue2024}.
\end{enumerate}

\subsubsection{Challenges with Black-Box Pretrained TSFMs}
\label{sec:transferability}
In practice, when using pretrained time-series foundation models (TSFMs) via high-level APIs like AutoGluon, users often lack visibility into source tasks, model weights, or hidden-layer activations. Further, some predictors emit only point forecasts rather than full probabilistic distributions. These constraints render many transferability estimators infeasible: without internal representations, representation analysis and optimal-transport methods cannot be applied directly; without predictive distributions, conditional-entropy metrics cannot be computed; and theoretical task relatedness bounds may be too loose in the absence of source-domain details.

To address this, we adopt two purely empirical evaluation strategies:

\subsubsection{Transfer-Gain Test}

An alternative, more balanced evaluation compares the impact of pretraining under both zero-shot and fine-tuned regimes. Specifically, we propose two paired experiments on the same data split.

For zero-shot, we initialise two instances of the TSFM: one initialised with random weights (untrained) and one with pretrained weights. Without any further training, apply each model in zero-shot mode to the target validation set, obtaining errors $\mathrm{Err}_{\mathrm{untrained,ZS}}$ and $\mathrm{Err}_{\mathrm{pretrained,ZS}}$. The zero-shot transfer gain is defined as

\begin{equation}
\Delta_{\mathrm{ZS}} = \frac{\mathrm{Err}_{\mathrm{untrained,ZS}} - \mathrm{Err}_{\mathrm{pretrained,ZS}}}{\mathrm{Err}_{\mathrm{untrained,ZS}}}.
\end{equation}

A positive $\Delta_{\mathrm{ZS}}$ indicates that pretrained weights confer immediate predictive advantage without any task-specific adaptation.

Next, fine-tune both models—starting from random and from pretrained weights—on the training split, yielding errors $\mathrm{Err}_{\mathrm{untrained,FT}}$ and $\mathrm{Err}_{\mathrm{pretrained,FT}}$ respectively. We quantify fine-tuning transfer gain as

\begin{equation}
\Delta_{\mathrm{FT}} = \frac{\mathrm{Err}_{\mathrm{untrained,FT}} - \mathrm{Err}_{\mathrm{pretrained,FT}}}{\mathrm{Err}_{\mathrm{untrained,FT}}}.
\end{equation}

A positive $\Delta_{\mathrm{FT}}$ demonstrates that pretrained initialisation leads to improved learning even after adaptation.

By examining both $\Delta_{\mathrm{ZS}}$ and $\Delta_{\mathrm{FT}}$, one can distinguish the relative contributions of pretrained priors versus task-specific fine-tuning to overall performance. A decision threshold (e.g. $\Delta>0.1$) can be applied to each gain to assess whether the pretrained model’s knowledge is transferable.

\subsubsection{Sample-Efficiency Probe}
One clear indicator of effective transfer is improved sample efficiency. To probe this, train both the untrained and pretrained variants on incremental fractions of the available training data (e.g.\ 10\%, 20\%, …, 100\%), recording the test error at each fraction. Plotting these learning curves reveals whether the pretrained model outperforms scratch training in low-data regimes. If the pretrained model reaches a given error level with significantly fewer samples, this provides strong evidence of relevant transfer.

Collectively, the Transfer-Gain Test and Sample-Efficiency Probe offer a practical, black-box-compatible method for assessing the transferability of TSFMs to financial time series forecasting applications.

\section{Forecasting Tasks - Overview and Data Processing}
\label{imp}
This chapter outlines the three forecasting tasks used to evaluate how well Time Series Foundation Models transfer to financial domains .These tasks were selected to span a range of asset classes, temporal dynamics, and modelling challenges. While each task is formulated as a multivariate regression problem, the tasks differ in structure and characteristics.

\subsection{Overview of Forecasting Tasks}
The three forecasting tasks in this study were chosen to reflect a range of financial modelling challenges across asset classes. Each task varies in its target complexity, signal-to-noise ratio (SNR), data availability, and relevance of multivariate structure, thereby enabling a thorough evaluation of model generalisation. They are as follows.

\begin{enumerate}
    \item \textbf{Bond Yield Forecasting:} A macroeconomic task involving prediction of the changes in US 10-year Treasury yields 21 business days ahead.
    \item \textbf{Foreign Exchange (FX) Volatility Forecasting:} A risk-oriented task that requires forecasting realised FX volatility 21 business days ahead. Realised volatility exhibits strong autocorrelation, which provides a relatively higher signal-to-noise ratio than raw returns.
    \item \textbf{Equity Spread Forecasting:} A cross-sectional relative-value task using mid-frequency co-integration signals between country equity indices. The goal is to forecast mean-reverting spreads. 
\end{enumerate}

Together, these tasks allow us to test TSFMs in settings that demand different combinations of temporal, cross-feature, and structural generalisation.

\subsection{Task 1: Bond Yield Forecasting}
\label{task1}
Forecasting the 10-year U.S. Treasury yield 21 business days ahead is a well-established but challenging problem in the macro-finance literature. The target is commonly defined as the percentage change or first difference in the 10-year yield over the forecast horizon rather than its raw level, in order to mitigate issues of non-stationarity and scale dependence~\cite{hall1992, duffee2013}. Conventional approaches for this task have included autoregressive models such as VARs, macro-factor regressions, and structural yield curve models (e.g., dynamic Nelson–Siegel or no-arbitrage term structure models) that combine yield data with macroeconomic indicators~\cite{ang2003}. More recent studies have explored non-linear models such as gradient boosting and deep learning architectures (e.g., LSTMs and GANs) to better capture the complex dependencies and interactions between interest rates and macro drivers~\cite{ping2024, walia2025}.

Despite the breadth of modelling approaches, yield forecasting remains difficult due to the low signal-to-noise ratio inherent in short-term rate movements. These movements are often driven by a mix of macroeconomic releases, policy expectations and other latent factors, making most high-frequency changes appear close to unpredictable noise. Given the complexity and nonlinearity of the yield forecasting problem, along with the limited availability of macro-financial data (especially in emerging markets), this task is well aligned with the strengths of TSFMs.

\subsubsection{Data Sources}
For this forecasting task, the data used consist of three main categories: macroeconomic indicators, interest rate data, and technical indicators derived from interest rate movements. All datasets were aligned at daily frequency and span from January 2005 to January 2025. Interest rate data were sourced from the US Department of the Treasury \cite{treasury_yield_curve}, including 1-month, 2-year, 5-year, and 10-year maturity yields. Macroeconomic and financial variables such as M2 money supply and credit spreads were retrieved from the Federal Reserve Economic Data (FRED) database maintained by the Federal Reserve Bank of St. Louis \cite{fred2025}. A range of macro indicators such as GDP growth, employment trends, and inflation expectations were sourced from Macrosynergy \cite{macrosynergy_jpmaqs}. Finally, market-based series such as the VIX, MOVE index, and S\&P 500 returns were obtained from Yahoo Finance's Python based API.

\subsection{Features Overview}

The features used in this task fall into three broad categories: macroeconomic indicators, market-based sentiment variables, and technical signals derived from yield dynamics. Macroeconomic indicators include traditional growth metrics such as quarterly real GDP (seasonally adjusted annualised rate) and an “intuitive” GDP proxy that captures smoothed year-on-year changes. Labour market conditions are represented through employment growth and the unemployment rate, both smoothed using a 3-month moving average to reduce noise. Inflation is captured via headline CPI and market-implied inflation expectations at one-year and five-year horizons, providing forward-looking information on the expected price level path.

Sentiment and risk appetite are reflected by volatility indices such as the VIX (equity market) and MOVE (fixed income), which are log-differenced and standardised using rolling z-scores to normalise their typically skewed distributions. Money supply growth, proxied by daily changes in M2, acts as a high-frequency liquidity measure, while S\&P 500 returns are smoothed via an exponentially weighted moving average to capture equity trends.

Finally, technical indicators derived from the 10-year Treasury yield include short-term momentum, mean reversion signals (Bollinger Band position), and an EWMA-based measure of yield volatility. Additionally, interest rate series at multiple maturities (1M, 2Y, 10Y) and the 5Y–2Y spread were differenced to ensure stationarity and preserve dynamic information. All features were aligned at a daily frequency and tested for stationarity using the Augmented Dickey–Fuller test, which they passed at the 5\% significance level following appropriate transformations. Full details on each variable are provided in Table~\ref{tab:bond_features}.

\begin{table}
\centering
\small
\caption{
Features used in the bond yield forecasting task. \textbf{QoQ} = quarter-on-quarter. \textbf{SAAR} = seasonally adjusted annualised rate. \textbf{YoY} = year-on-year. \textbf{3MMA} = 3-month moving average. \textbf{SA} = seasonally adjusted. \textbf{EWMA} = exponentially weighted moving average. All features were aligned at daily frequency. Interest rate series refer to constant maturity Treasury yields. Ticker names correspond to Macrosynergy JPMAQS \cite{macrosynergy_jpmaqs} identifiers.
}
\label{tab:bond_features}
\begin{tabularx}{\textwidth}{>{\bfseries\raggedright\arraybackslash}X >{\centering\arraybackslash}X >{\centering\arraybackslash}p{3cm} >{\centering\arraybackslash}p{2.2cm}}
\toprule
\textbf{Feature} & \textbf{Notes} & \textbf{Transformation} & \textbf{Source} \\
\midrule
Real GDP growth (QoQ, SAAR) & Ticker: {\footnotesize\texttt{RGDP\_SA\_P1Q1QL1AR}} & Rolling standardisation & Macrosynergy \\
\midrule
Intuitive GDP growth (\% YoY, 3MMA) & Ticker: {\footnotesize\texttt{INTRGDP\_NSA\_P1M1ML12\_3MMA}} & Rolling standardisation & Macrosynergy \\
\midrule
Employment growth (\% YoY, 3MMA) & Ticker: {\footnotesize\texttt{EMPL\_NSA\_P1M1ML12\_3MMA}} & Rolling standardisation & Macrosynergy \\
\midrule
Unemployment rate (SA, 3MMA) & Ticker: {\footnotesize\texttt{UNEMPLRATE\_SA\_3MMA}} & Rolling standardisation & Macrosynergy \\
\midrule
Change in CPI over past 3 months & Ticker: {\footnotesize\texttt{CPIH\_SA\_P1M1ML12\_D1M1ML3}} & Rolling standardisation & Macrosynergy \\
\midrule
Market-implied 1Y inflation expectation & Ticker: {\footnotesize\texttt{IMPINFM1Y\_NSA}}, from Inflation Swaps & Rolling standardisation & Macrosynergy \\
\midrule
Market-implied 5Y inflation expectation & Ticker: {\footnotesize\texttt{IMPINFM5Y\_NSA}}, from Inflation Swaps & Rolling standardisation & Macrosynergy \\
\midrule
Money supply (M2) & – & Daily percent change & FRED \\
\midrule
VIX & Volatility index on S\&P 500 & Rolling z-score of log-diff (20d) & Yahoo Finance \\
\midrule
MOVE & Volatility index on US Treasuries & Rolling z-score of log-diff (20d) & Yahoo Finance \\
\midrule
Equity returns & Daily S\&P 500 returns & EWMA (Half-life of 5 days) & Yahoo Finance \\
\midrule
Interest rates (1M, 2Y, 10Y, 5Y--2Y) & Constant maturity Treasury yields and 5Y--2Y slope & First difference & US Treasury \\
\midrule
Various Technical indicators & Short-term trend, Bollinger band position, and EMA-based volatility from 10Y yield & Various & Derived from US Treasury\\ 
\bottomrule
\end{tabularx}
\end{table}

\subsection{Target Construction}

The target variable for this task is the percentage change in the 10-year US Treasury yield. Since the forecasting model is designed for multi-horizon prediction, the target includes future returns over a sequence of horizons rather than a single fixed step.

To construct the target, the 10-year yield series was first converted into daily percentage changes. For each forecast horizon $h$, the target at time $t$ was defined as the cumulative percentage change from $t$ to $t + h$:

\[
y_{t,h} = \left( \prod_{i=1}^{h} \left(1 + \frac{r_{t+i} - r_{t+i-1}}{r_{t+i-1}} \right) \right) - 1,
\]

where \( r_t \) denotes the 10-year yield at time \( t \), and \( h \in \{1, 2, \dots, H\} \) indexes the forecast horizons.

\subsection{Task 2: Foreign Exchange Volatility Forecasting}
\label{task2}
Volatility forecasting refers to the task of predicting the future variability of asset returns, typically expressed as the standard deviation or variance of returns over a given horizon. It plays a central role in financial applications, informing risk management, derivative pricing, and portfolio construction~\cite{Poon2003}. Unlike returns, which are notoriously noisy and often resemble white noise, volatility tends to be persistent and mean-reverting. This phenomenon, known as volatility clustering~\cite{Bollerslev1994}, implies that periods of high volatility are likely to be followed by similar periods, and vice versa. As a result, volatility exhibits a higher signal-to-noise ratio than returns, making it more amenable to statistical forecasting~\cite{Poon2003, Andersen2003}. Numerous studies have shown that while returns may be nearly unpredictable, meaningful forecasts of volatility can still be achieved.

A wide range of approaches have been developed to model and forecast volatility. On the econometric side, the foundational Autoregressive Conditional Heteroskedasticity (ARCH) model introduced by Engle~\cite{Bollerslev1994} and its generalisation, GARCH, model the current variance as a function of past variances and shocks.  To address long-memory features in volatility series, the Heterogeneous Autoregressive (HAR) model~\cite{Corsi2008} uses realised volatilities at multiple time horizons as predictors, capturing slow-decaying autocorrelations while retaining a simple linear structure.

These models benefit from theoretical grounding and interpretability, but may struggle with structural breaks, omitted exogenous drivers, or highly nonlinear dynamics. More recently, machine learning (ML) approaches such as neural networks, tree-based ensembles, and hybrid models have been applied to volatility forecasting. These methods can flexibly model non-linearities and interactions. For instance, deep learning models like LSTMs have shown promise in uncovering temporal patterns in volatility that traditional models might miss, with some studies reporting improvements over GARCH-family models~\cite{Bucci2020}. However, ML methods are data-hungry, risk overfitting, and often lack transparency. A study~\cite{Hansen2001} finds that their out-of-sample gains are sometimes below that of traditional benchmarks such as GARCH(1, 1), although this study was published in 2001, before many techniques used today were introduced.

The FX market presents unique challenges for volatility forecasting. It operates continuously, is sensitive to macroeconomic news and central bank policy, and can shift volatility regimes abruptly. Scheduled announcements, such as inflation reports or central bank meetings, are can also cause jumps in volatility. Furthermore, high-frequency FX data are affected by microstructure noise and intraday seasonality, complicating the estimation of realised volatility~\cite{Andersen1998}.

Despite these challenges, FX volatility remains forecastable to a useful degree. Even basic GARCH models can explain a large portion of next-day variance, with some studies suggesting they approach the limits of predictability~\cite{Andersen1998}. In this study, we focus on forecasting the realised volatility of the EUR/USD exchange rate—the most liquid FX pair globally.

\subsection{Data Sources}
For this second forecasting task, the dataset includes volatility measures, global risk sentiment indicators, and macro-financial variables across currency areas. All data were sampled at daily frequency and span the period from January 2005 to January 2025. Spot EUR/USD FX prices were sourced from Yahoo Finance, from which multiple short- and long-horizon volatility proxies were derived. Broader risk sentiment was captured using the VIX and MOVE indices, also obtained from Yahoo Finance. To account for cross-currency monetary dynamics, the model includes yield curve information from both EUR and USD government bonds at key maturities. Yield data for the US were taken from the US Department of the Treasury \cite{treasury_yield_curve}, while EUR rates were sourced from the European Central Bank \cite{ecb2025}. The ECB yield data, which spans multiple countries in the euro area, is based on the composite euro area yield curve for government bonds issued by AAA-rated sovereigns.  Information related to  monetary policy announcement days for the ECB and Federal Reserve was obtained from Investing.com \cite{investing2025}.

\subsection{Feature Construction}

The features for this task were selected to capture various influences on FX volatility, including realised market activity, risk sentiment, and macro-financial conditions. Volatility itself was approximated using multiple estimators based on FX returns, including exponentially weighted moving average (EWMA) volatility, rolling mean absolute returns, and the Garman--Klass estimator \cite{Garman1980}. Both short-term (5-day) and longer-term (21-day) realised volatility were computed, along with their ratio, to detect relative shifts in short- versus medium-term fluctuations. Broader market sentiment was reflected through the VIX and MOVE indices, each transformed using rolling z-scores of their log-differences. Yield curve information was incorporated using EUR--USD yield spreads at 3-month, 2-year, and 10-year maturities, as well as the first principal component (PC1) of the EUR and USD yield curves, which summarise overall level shifts. All interest rate features were differenced to improve stationarity. A final categorical feature marked the dates of key monetary policy announcements by the ECB and Federal Reserve, dates in which volatility is expected to increase.

Full details on each feature, including notes, transformations and sources, are provided in Table~\ref{tab:fx_vol_features}. To ensure suitability for forecasting, all features were transformed as necessary to achieve approximate stationarity and normality. In particular, volatility-related features were transformed using a method denoted as \textbf{VolTransform}, which applies the natural logarithm followed by a rolling z-score standardisation using a half-year window. This transformation reduces heteroskedasticity and positive skew in raw volatility estimates, yielding an approximately Gaussian and stationary series that improves model performance and interpretability~\cite{Andersen2003}. All final features passed the Augmented Dickey--Fuller Stationarity test with p-values below 0.05.

\begin{table}
\centering
\small
\caption{
Features used in the FX volatility forecasting task. \textbf{EWMA} = exponentially weighted moving average. \textbf{Garman--Klass} volatility is a high--low--open--close estimator \cite{Garman1980}. Interest rate features are expressed as spreads between EUR and USD yields at corresponding maturities. \textbf{VolTransform} corresponds to taking the natural logarithm of a volatility series following by applying rolling standardisation with a window of half a year.
}
\label{tab:fx_vol_features}
\begin{tabularx}{\textwidth}{>{\bfseries\raggedright\arraybackslash}X >{\centering\arraybackslash}X >{\centering\arraybackslash}p{3cm} >{\centering\arraybackslash}p{2.2cm}}
\toprule
\textbf{Feature} & \textbf{Description} & \textbf{Transformation} & \textbf{Source} \\
\midrule
EWMA volatility & EWMA of standard deviation of log FX returns & EWMA (span=10), VolTransform & Derived, from Yahoo Finance \\
\midrule
Rolling Mean absolute returns & 5-day average absolute log return & VolTransform & Derived, from Yahoo Finance\\
\midrule
Garman--Klass volatility & High–low–open–close estimator over 5-day window & Annualised, VolTransform & Derived, from Yahoo Finance \\
\midrule
Long-term volatility & Realised volatility over 21-day window & Annualised, VolTransform & Calculated \\
\midrule
Volatility ratio (5d/21d) & Ratio of 5-day to 21-day realised volatility & None & Calculated \\
\midrule
VIX & Volatility index on S\&P 500 & Rolling z-score of log-diff (50d) & Yahoo Finance \\
\midrule
MOVE & Volatility index on US Treasuries & Rolling z-score of log-diff (50d) & Yahoo Finance \\
\midrule
EUR–USD yield spread (3M, 2Y, 10Y) & Yield differential between EUR and USD government bonds at corresponding maturities & Differenced & ECB / US Treasury \\
\midrule
USD yield curve PC1 & First principal component of USD yield curve & Differenced & Derived from US Treasury \\
\midrule
EUR yield curve PC1 & First principal component of EUR yield curve & Differenced & ECB \\
\midrule
Interest Rate Meetings & Categorical Variable indicating ECB Interest Rate Decision Days and FOMC Meetings & None & Investing.com \\
\bottomrule
\end{tabularx}
\end{table}

\subsection{Target Construction}
To construct a stable and tractable forecasting target,  the target variable is defined as the natural logarithm of the rolling standard deviation of daily FX log‐returns (i.e.\ log realised volatility).  This transformation addresses the strong positive skew and heteroskedasticity inherent in raw volatility measures, yielding an approximately Gaussian and stationary series that enhances model performance and interpretability \cite{Andersen2003}.  Such a log‐volatility target is compatible with both classical time‐series models (e.g., GARCH, HAR) and modern nonlinear methods (e.g., gradient boosting, neural networks), facilitating robust one‐month‐ahead FX volatility forecasts.

\subsection{Task 3: Equity Spread Forecasting}
\label{task3}
Pairs trading is a market-neutral strategy that exploits the temporary divergence in prices between two historically correlated assets \cite{Schizas2011}. It exploits the principles of co-movement and mean reversion: the notion that prices of related securities tend to move together over time, and that deviations from this relationship eventually correct themselves. In equity markets, such strategies have gained popularity due to their relative robustness to broad market trends and their capacity to isolate relative value. Traditionally, this technique has been applied to similar stocks or assets within the same sector, but recent work has demonstrated its effectiveness across international exchange-traded funds (ETFs), particularly those tracking country-specific equity indices \cite{Schizas2011}. 

Traditional approaches in pairs trading typically employ distance‐based methods, which select pairs according to the Euclidean distance between normalised price series \cite{Gatev2006}, and cointegration‐based techniques, which first test for a stationary long‐run relationship via Engle–Granger or Johansen procedures and then fit an Error Correction Model (ECM) or Vector Error Correction Model (VECM) to forecast the spread \cite{Zhu2024}. These forecasting models capture mean‐reversion by using the cointegration residual as an adjustment term while also accommodating short‐term autoregressive dynamics. Although effective in developed markets with ample data, such methods require sufficiently long estimation windows and tend to become less effective when history is sparse and when multiple variables are used.  TSFMs, in contrast, can capture long‐term dependencies with comparatively lower data requirements, making them potentially suitable for pairs trading in situations where data may be scarce, such as Emerging Market (EM) indices, where available price histories and macroeconomic data may be insufficient to support fully specified cointegration or multivariate factor models.

In this task, we focus on forecasting the spread between two highly correlated international equity ETFs: MSCI Australia (EWA) and MSCI Canada (EWC). Both ETFs track developed commodity-exporting economies and have been shown to exhibit strong long-term co-movement, making them an ideal candidate pair for relative value forecasting~\cite{mitchell2015pairs}. Understanding and predicting fluctuations in their spread allows us to anticipate potential convergence or divergence opportunities. By doing so, we aim not only to improve signal quality for trading strategies but also to assess the capacity of modern time series forecasting models to learn mid-frequency patterns of mean reversion in noisy financial time series.

\subsection{Data Sources}
The data used for this task spans three categories, namely fundamental data, macro data and technical indicators. All data were sampled at daily frequency and span the period from January 2005 to January 2025. The prices for EWC and EWA were sources using Yahoo Finance, from which the technical indicators were derived. Macro data was once again sourced from Macrosynergy \cite{macrosynergy_jpmaqs}. Fundamental data such as dividend yield was also calculated using data obtained through Yahoo Finance. 

\subsection{Feature Construction}
The features used for equity spread forecasting were selected to capture both the short-term technical divergences and longer-term macroeconomic drivers that may explain relative price movements between international equity indices. Following the empirical findings of \cite{Schizas2011}, who highlight the role of unemployment, earnings per share, and dividend yield in explaining the profitability of pairs trades, we include corresponding proxies where available. Macroeconomic features include growth and inflation indicators such as real GDP (quarter-on-quarter, SAAR), intuitive GDP growth (year-on-year, 3MMA), CPI momentum (3-month change in YoY inflation), employment growth, unemployment rate, and household inflation expectations. Measures of real interest rates—both backward-looking and forward-looking—at short and long maturities are also included to account for monetary conditions across countries \cite{Costa2025}. Fiscal and external balance indicators, such as government debt-to-GDP and the current account-to-GDP ratio, further capture macroeconomic divergence. In addition, technical indicators derived from the spread series—such as moving average convergence, Bollinger band position, and drawdown z-scores—were calculated following approaches outlined by \cite{smith2017cointegrated}. These are designed to detect mid-frequency reversals and identify points of stretched relative valuation. All features were aligned daily and, where appropriate, transformed via rolling standardisation or differencing to ensure stationarity. Further details on each feature, its construction, and transformation are provided in Table~\ref{tab:equity_features}.  

\begin{table}
\centering
\small
\caption{
Features used in the equity spread forecasting task. \textbf{QoQ} = quarter-on-quarter. \textbf{SAAR} = seasonally adjusted annualised rate. \textbf{YoY} = year-on-year. \textbf{3MMA} = 3-month moving average. Ticker names correspond to Macrosynergy JPMAQS identifiers. Technical indicators are derived from the EWA--EWC spread using standard methods such as Bollinger bands, moving average divergence, and drawdown z-scores.
}
\label{tab:equity_features}
\begin{tabularx}{\textwidth}{>{\bfseries\raggedright\arraybackslash}X >{\centering\arraybackslash}X >{\centering\arraybackslash}p{3cm} >{\centering\arraybackslash}p{2.2cm}}
\toprule
\textbf{Feature} & \textbf{Description} & \textbf{Transformation} & \textbf{Source} \\
\midrule
Real GDP growth (QoQ, SAAR) & Ticker: \texttt{RGDP\_SA\_P1Q1QL1AR} & Rolling standardisation & Macrosynergy \\
\midrule
Intuitive GDP growth (\% YoY, 3MMA) & Ticker: \texttt{INTRGDP\_NSA\_P1M1ML12\_3MMA} & Rolling standardisation & Macrosynergy \\
\midrule
Employment growth (\% YoY, 3MMA) & Ticker: \texttt{EMPL\_NSA\_P1M1ML12\_3MMA} & Rolling standardisation & Macrosynergy \\
\midrule
Unemployment rate (SA, 3MMA) & Ticker: \texttt{UNEMPLRATE\_SA\_3MMA} & Rolling standardisation & Macrosynergy \\
\midrule
CPI inflation momentum (3-month YoY change) & Ticker: \texttt{CPIH\_SA\_P1M1ML12\_D1M1ML3} & Rolling standardisation & Macrosynergy \\
\midrule
Household inflation expectations (1Y ahead) & Ticker: \texttt{HHINFESCORE\_NSA} & Z-score & Macrosynergy \\
\midrule
Real short-term interest rate (expectation-based) & Ticker: \texttt{RIR\_NSA} & None & Macrosynergy \\
\midrule
Real short-term interest rate (backward-looking) & Ticker: \texttt{RIRB\_NSA} & None & Macrosynergy \\
\midrule
Real 5Y IRS yield (expectation-based) & Ticker: \texttt{RYLDIRS05Y\_NSA} & None & Macrosynergy \\
\midrule
Real 5Y IRS yield (backward-looking) & Ticker: \texttt{RYLDIRS05YB\_NSA} & None & Macrosynergy \\
\midrule
Government debt to GDP (next year) & Ticker: \texttt{GGDGDPRATIONY\_NSA} & None & Macrosynergy \\
\midrule
Current account to GDP (12MMA) & Ticker: \texttt{CABGDPRATIO\_NSA\_12MMA} & None & Macrosynergy \\
\midrule
Bollinger band position & Relative location of spread in 20d Bollinger bands & Rolling z-score & Calculated \\
\midrule
Moving average divergence & Difference between short and long EMA of spread & EWMA (5--20d) & Calculated \\
\midrule
Spread drawdown & Relative drawdown from recent peak in spread & Rolling z-score & Calculated \\
\bottomrule
\end{tabularx}
\end{table}

\subsection{Target Construction}

The target variable for this task is the daily change in the spread between the log prices of the MSCI Australia ETF (EWA) and the MSCI Canada ETF (EWC), defined as the difference between their respective log closing prices. Before constructing the target, the two log price series were tested for cointegration using the Engle--Granger two-step method~\cite{Zhu2024}. This approach first estimates a long-run linear relationship between the two series via ordinary least squares (OLS), then tests the stationarity of the residuals using the Augmented Dickey--Fuller (ADF) test. The results confirmed that the log price series of EWA and EWC are cointegrated over the sample period, supporting the use of their spread as a stationary, mean-reverting target.

To ensure stationarity and consistent scaling across time, the rolling standardized log-price spread was used as the final target variable. Specifically, the target $y_t$ is defined as the difference in log prices between the two assets (EWA and EWC), normalised by the rolling mean and standard deviation over a window of length $w$:

\begin{equation}
    y_t = \frac{\log\left( P_t^{\text{EWA}} \right) - \log\left( P_t^{\text{EWC}} \right) - \mu_t^{(w)}}{\sigma_t^{(w)}}
\end{equation}

Here, $\mu_t^{(w)}$ and $\sigma_t^{(w)}$ denote the rolling mean and standard deviation of the log-price spread, computed as:

\begin{equation}
    \mu_t^{(w)} = \frac{1}{w} \sum_{i=0}^{w-1} \left[ \log\left( P_{t-i}^{\text{EWA}} \right) - \log\left( P_{t-i}^{\text{EWC}} \right) \right]
\end{equation}

\begin{equation}
    \sigma_t^{(w)} = \sqrt{ \frac{1}{w} \sum_{i=0}^{w-1} \left( \left[ \log\left( P_{t-i}^{\text{EWA}} \right) - \log\left( P_{t-i}^{\text{EWC}} \right) \right] - \mu_t^{(w)} \right)^2 }
\end{equation}

Where $\omega$ denotes the rolling window length used to compute the mean and standard deviation of the log-price spread, in this case set to $\omega=42$.
\section{Experimental Setup}
\label{exp}

To evaluate the practical utility and transferability of pretrained TSFMs in  multivariate financial forecasting tasks, we design an empirical framework aligned with the black-box constraints outlined in Section~\ref{sec:transferability}. Since pretrained models often expose only their output forecasts and not internal representations or training data provenance, we adopt two empirical evaluation strategies that remain valid in such settings: the \textit{Transfer-Gain Test} and the \textit{Sample-Efficiency Probe}.

The first experimental stage varies the amount of available training data, fine-tuning each TSFM on increasing historical windows and evaluating performance against a naive benchmark. This setup allows us to assess (i) whether the pretrained model provides consistent improvement over simple baselines and (ii) how performance scales with data availability. When a model outperforms the naive benchmark, we extend the experiment by comparing the fine-tuned TSFM to its untrained counterpart, quantifying the transfer gain through relative error reduction.

The second stage conducts a rolling forecast evaluation. For each task, we retrain the model every six months using a fixed-size trailing window and evaluate it over a fixed test period. This rolling setup mirrors a realistic deployment scenario, capturing regime shifts and testing robustness across market conditions.

Together, these experiments provide insights into transfer effectiveness, data efficiency, and time stability of TSFM-based forecasting workflows in financial applications.

\subsection{Training and Evaluation Design}
\label{sec:training}
\subsubsection{Tiny Time Mixers}

For all experiments involving Tiny Time Mixers (TTMs), a fixed random seed was used to ensure reproducibility. The context length (i.e., the number of past time steps used as input) and the forecast horizon (i.e., the number of future time steps predicted) were chosen based on validation performance for each task and are summarised in Table~\ref{tab:ttm_config}. Features were labelled as either \textit{conditional}, meaning they are observed only up to the forecast origin (e.g., lagged variables), or \textit{observable}, which includes future-known inputs (e.g., scheduled events). All features in all tasks were conditional except for central bank meeting indicators in the volatility forecasting task, which were treated as observable. The temporal resolution was set to \texttt{‘B’} (business-day frequency), and no internal scaling was applied, since all features were preprocessed and standardised during the data pipeline.

The model was trained using the \texttt{AdamW} optimiser, a variant of Adam that decouples weight decay from the gradient update, improving generalisation and stability in deep learning training\cite{adamw}. To select an appropriate learning rate, we employed a learning rate range test inspired by the method proposed in~\cite{Smith2015}. This involves gradually increasing the learning rate from a small value to a large one over several mini-batches, either linearly or exponentially. At each step, the model performs a forward and backward pass, and the corresponding loss is recorded. The optimal learning rate is then identified using the valley criterion—just before the point where the loss begins to increase rapidly. As the model is briefly trained during this procedure, its weights may be reused or reinitialised before full training. A \texttt{OneCycleLR} learning rate scheduler was used, which increases the learning rate to a peak value before gradually annealing it to a minimum. This schedule has been shown to accelerate convergence while reducing the risk of getting stuck in sharp minima. To further enhance generalisation and reduce overfitting, dropout regularisation was applied at a fixed rate of 0.2, randomly deactivating a subset of neurons during training.

Epoch size and batch size were treated as a hyperparameter and selected for each task via validation. For Task 1, the model was trained for 25 epochs; for Task 2, 50 epochs; and for Task 3, 4 epochs. The batch size was 64 for all tasks. Given the relatively small size of the TTM architecture and empirical results showing superior validation accuracy with end-to-end training, full fine-tuning of all model parameters was used over parameter-efficient alternatives. All training and evaluation used mean squared error (MSE) as the objective function and metric, enabling direct comparison across models and tasks.

\begin{table}[ht]
\centering
\caption{Model Configuration for Tiny Time Mixers Across Forecasting Tasks}
\label{tab:ttm_config}
\begin{tabular}{lcc}
\toprule
\textbf{Task} & \textbf{Context Length} & \textbf{Forecast Horizon} \\
\midrule
Bond Yield Forecasting & 90  & 30  \\
FX Volatility Forecasting & 512 & 48  \\
Equity Spread Forecasting & 512 & 96 \\
\bottomrule
\end{tabular}
\end{table}

\subsubsection{Chronos}

Chronos models were trained using the AutoGluon library, a high-level AutoML framework designed to simplify training and deployment of machine learning models across tabular, image, text, and time-series tasks. AutoGluon handles model selection, hyperparameter tuning, and ensembling automatically, making it a convenient choice for benchmarking pretrained time-series foundation models (TSFMs) such as Chronos. While AutoGluon's abstraction facilitates ease of use, it also restricts control over certain low-level parameters such as the choice of optimiser or fine-tuning strategy. The underlying Chronos paper adopts the AdamW optimiser, but the exact optimisation strategy used during fine-tuning via AutoGluon is not exposed.

For reproducibility, a fixed random seed was set across all experiments. Unlike Tiny Time Mixers (TTM), which use fixed combinations of context lengths and forecast horizons, Chronos allows the context length to be set arbitrarily, so it was selected from a broad range using validation performance. When no context length configuration outperformed a naive benchmark, the context length was set equal to that used for TTM for comparability. The forecast horizon was fixed to match the prediction interval required for each application. Table~\ref{tab:chronos_config} outlines the final context length and forecast horizon used for each task.

\begin{table}[ht]
\centering
\caption{Model Configuration for Chronos-Bolt (small) Across Forecasting Tasks}
\label{tab:chronos_config}
\begin{tabular}{lcc}
\toprule
\textbf{Task} & \textbf{Context Length} & \textbf{Forecast Horizon} \\
\midrule
Bond Yield Forecasting & 90  & 21 \\
FX Volatility Forecasting & 90  & 21 \\
Equity Spread Forecasting & 360 & 10 \\
\bottomrule
\end{tabular}
\end{table}

AutoGluon supports two types of additional features: static covariates and time-varying covariates. All features used in this study are time-varying, with the exception of central bank meeting indicators in Task~2. However, AutoGluon does not allow users to declare these as observable (i.e., known in advance at forecast time), which limits how this information can be exploited.

The model frequency was set to business days (\texttt{‘B’}), with no internal feature scaling applied since all inputs were preprocessed and standardised prior to model ingestion. For fine-tuning, a cosine learning rate scheduler was used, which gradually reduces the learning rate following a cosine decay pattern to help achieve better convergence and avoid oscillations near the optimum. Epoch size was selected based on validation performance, although the choice was somewhat arbitrary since Chronos performed similarly poorly across all tested values. The final configuration used 40 epochs for Task 1, 50 for Task 2 and 20 for Task 3. A batch size of 32 and a default dropout rate of 0.1 were used.

AutoGluon does not currently support parameter-efficient fine-tuning (PEFT) options such as LoRA, so full fine-tuning was applied. This was computationally feasible given GPU availability and relatively short training times. Models were trained using the Weighted Quantile Loss (WQL), with the median (50\% quantile) prediction extracted and evaluated using mean squared error (MSE).

\subsection{Baseline Models}
\label{sec:baseline}
To contextualise the performance of the pretrained TSFMs, a range of baseline models was assessed across tasks. These included both naive forecasters and classical statistical and machine learning models. The specific choice of baselines depended on the structure of each forecasting task and the type of target variable involved.

Two types of naive forecasters were used depending on the formulation of the target and the statistical properties of the series. For tasks where the model predicted first differences in a target variable (such as Task 1 \ref{task1}), the naive baseline simply predicted a constant value of zero at every horizon, corresponding to no expected change. For tasks where the target was a level (e.g., Task 2 \ref{task2} and Task 3 \ref{task3}), the naive forecaster used the most recently observed value of the target as the prediction at all future horizons. However, for Task 3, where the target series is a mean-reverting spread, the naive forecast varied by horizon: the last observed value was used for 5-day forecasts to capture short-term autocorrelation, while a zero forecast was used for 10-day horizons, reflecting the spread’s zero-centred mean-reverting dynamics and an estimated reversion time of approximately 14 days. These baselines serve as meaningful references due to their simplicity and surprising robustness in financial time series.

In addition to naive forecasters, a set of classical benchmark models was used in the Transfer-Gain Test. These comprised two categories. The first included standard supervised learning models such as linear regression, Ridge regression, gradient-boosted regression (GBR), and random forests. These models were trained with a shifted target (e.g., 21-day-ahead value) and produced single-step forecasts. The second category included temporal forecasting models such as Vector Autoregression (VAR) and Long-Short Term Memory (LSTM) neural network, which were trained on unshifted target sequences and produced forecasts across multiple horizons. These were only applicable to tasks that involved level forecasting (i.e., Tasks 2 and 3), as differenced targets were incompatible with their sequential formulations.

\subsection{Sample-Efficiency Probe}
\label{sec:sep}
The sample-efficiency probe is designed to test how the performance of pretrained versus randomly initialised (untrained) TSFMs evolves with increasing amounts of training data. This procedure provides a practical assessment of transferability in low-data regimes and helps identify the training window size that balances performance with data efficiency.

The procedure is as follows:

\begin{enumerate}
    \item Fix a reference date: \textbf{22 January 2021}.
    \item For each value $n \in \{2, 3, \ldots, 14\}$:
    \begin{enumerate}
        \item Use the $n$ years preceding the reference date as the training window.
        \item Use all data after the reference date as the test window, which varies by task and typically extends into late 2024 or early 2025.
        \item Train both the pretrained and untrained versions of the model, following the procedure in Section~\ref{sec:training}.
        \item Evaluate each model using mean squared error (MSE) and compare results to a naive benchmark.
    \end{enumerate}
    \item Plot the learning curves showing test MSE as a function of training window size for both model variants.
    \item Identify the \emph{elbow point}, i.e. the smallest $n$ after which additional training data yields diminishing returns, to determine the optimal training window length $k$.
    \item Use this $k$ in the subsequent rolling Transfer-Gain Test (Section~\ref{sec:tgt}).
\end{enumerate}

\subsection{Transfer-Gain Test}
\label{sec:tgt}
The transfer-gain test evaluates the predictive benefit of pretraining relative to training from scratch across multiple rolling windows. This enables a detailed assessment of transfer learning effectiveness, generalisation capability, and robustness under varying market conditions.

In this evaluation, five model variants are compared: a TSFM fine-tuned from scratch, a pretrained TSFM fine-tuned on the target task, a pretrained TSFM in zero-shot modality, an untrained TSFM in zero-shot modality, and a set of classical baseline models as described in Section~\ref{sec:baseline}.

Given the optimal training window length $k$ identified in Section~\ref{sec:sep}, the procedure is as follows:

\begin{enumerate}
    \item Iterate through evaluation dates in 6-month intervals, beginning at the first date with $k$ years of historical data available.
    \item For each evaluation date:
    \begin{enumerate}
        \item Define the training window as the $k$ years preceding the evaluation date.
        \item Define the test window as the 2 years following the evaluation date, plus the model's context length (when significant, such as 512 in Task 2).
        \item Train both pretrained and untrained model variants as in Section~\ref{sec:training}.
        \item Evaluate both models on the test set using MSE and record the percentage improvement relative to a naive forecaster.
        \item Compute both zero-shot and fine-tuning transfer gains $\Delta_{\mathrm{ZS}}$ and $\Delta_{\mathrm{FT}}$, as defined in the background (Section~\ref{sec:transferability}).
    \end{enumerate}
    \item Include classical baseline models in this evaluation to contextualise the performance of the TSFMs.
\end{enumerate}
\section{Results}
\label{results}

This chapter presents the empirical findings of the experiments described in the previous sections. We evaluate the performance of pretrained Time Series Foundation Models (TSFMs) across three financial forecasting tasks using the experimental protocols defined in Chapter \ref{exp}. Finally, we discuss the findings in Section \ref{sec:discussion}

\subsection{Task 1: Bond Yield Forecasting}
\label{sec:task1_results}
The results for Task 1, forecasting US 10-Year Treasury yields 21 days in advance, are displayed in this section for TTM, Chronos and benchmark models. Given that the target variable is a change for this task, the naive model simply predicts zero–or no change–every time. 

\subsubsection{TTM}
\begin{figure}[H]
    \centering
    \includegraphics[width=1\linewidth]{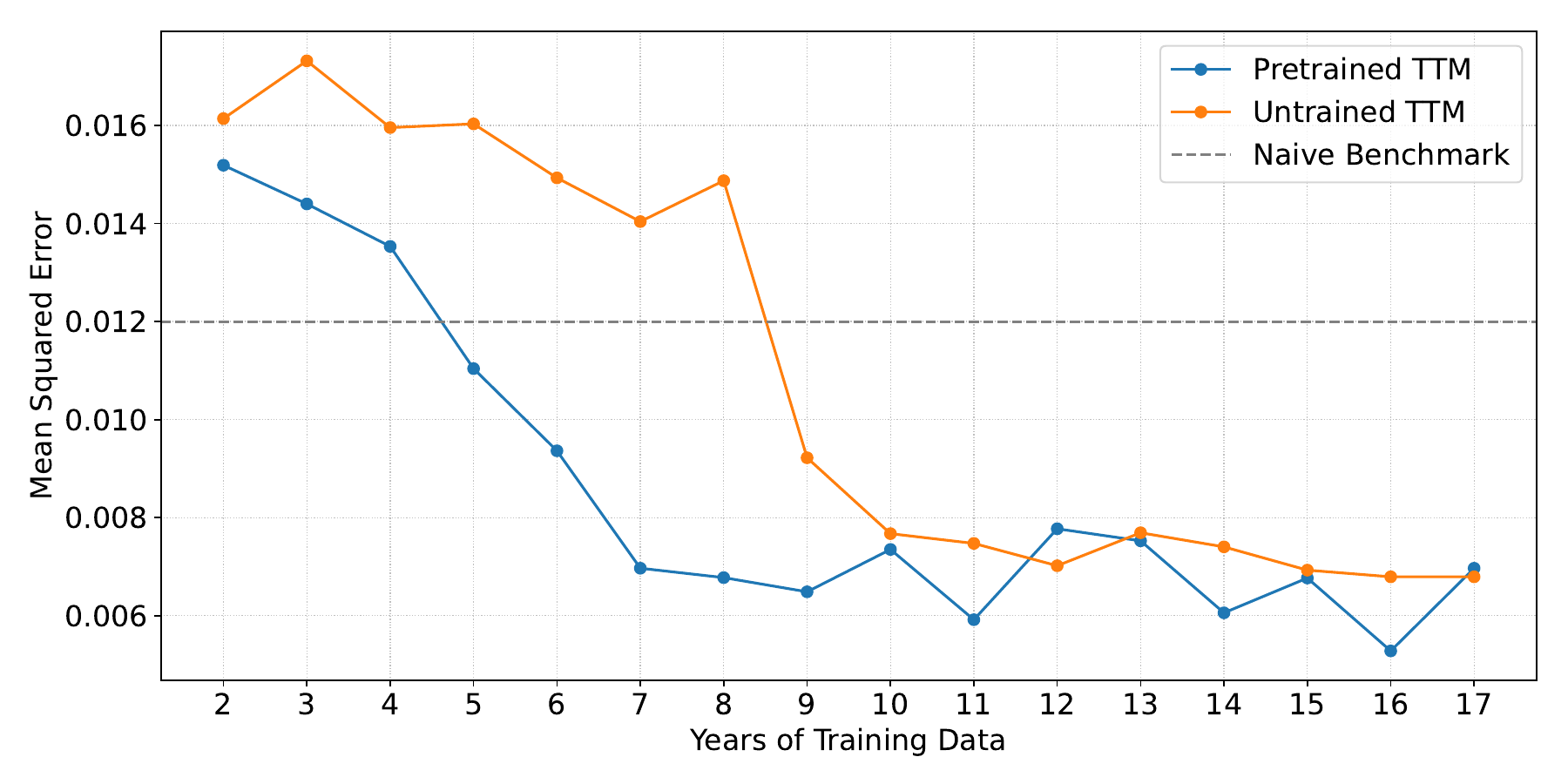}
    \caption{TTM Mean Squared Error when Fine-tuned on Different Training Dataset Sizes when Forecasting 10Y Treasury Yields 21 Business Days Ahead}
    \label{fig:ttm_expanding_yield}
\end{figure}

Figure \ref{fig:ttm_expanding_yield} shows the results of running the Sample Efficiency Probe (Section \ref{sec:sep}) using TTM for Task 1. When fine-tuning both the pretrained and untrained version of TTM, the figure makes it clear that the pretrained version needs significantly less data to learn; in fact, the curve for the untrained version looks like a lagged version of the pretrained model. \textbf{On average, the figure reveals that it takes the untrained version three more years than the pretrained version of TTM to achieve similar performance levels}. However, the plot also shows that given enough data to fine-tune the models, the untrained version achieves the same accuracy as the pretrained version.

\begin{figure}[H]
    \centering
    \includegraphics[width=1\linewidth]{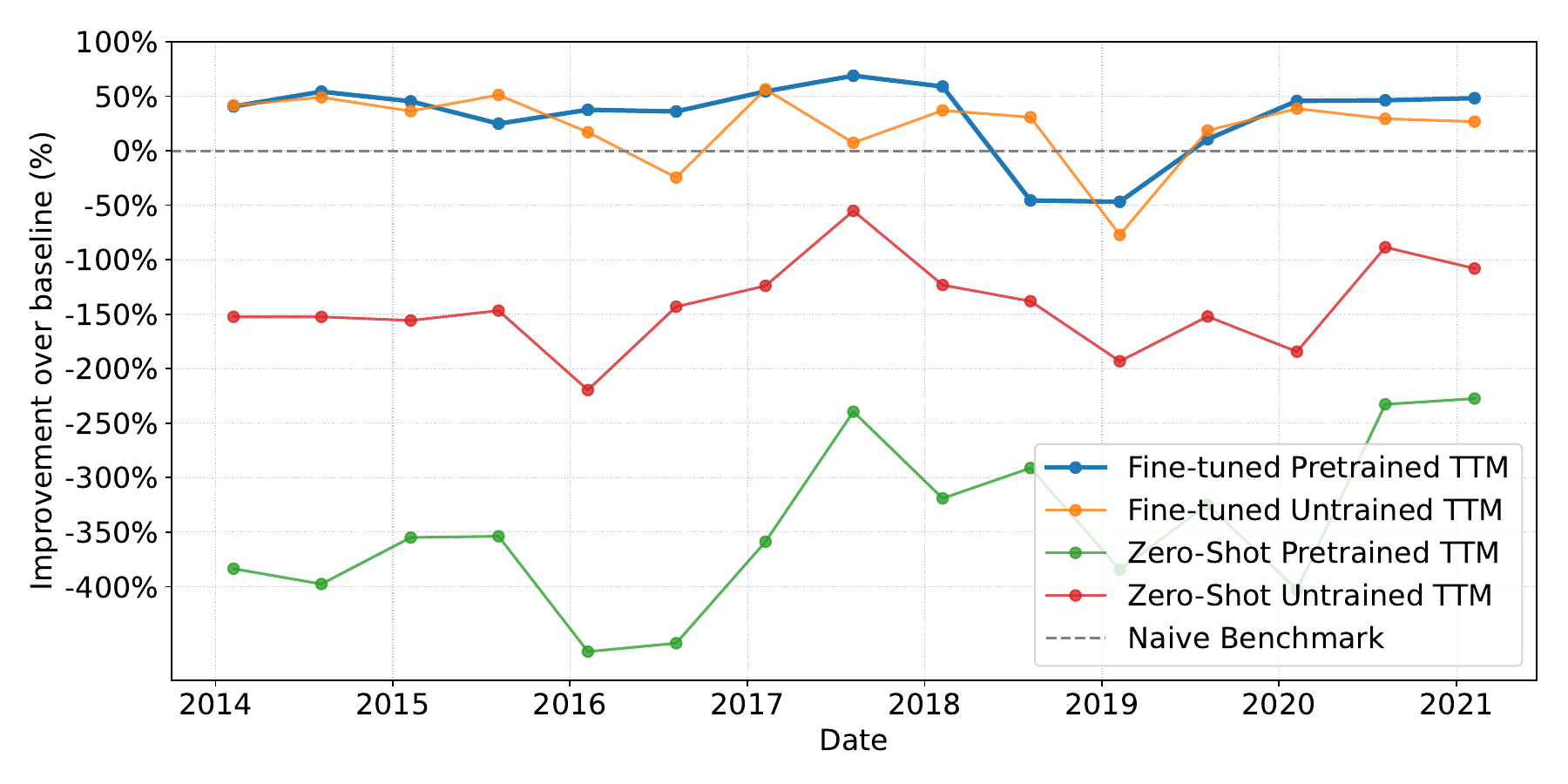}
    \caption{Rolling Performance of Different Training Settings of TTM Compared to Naive Benchmark when Forecasting 10-Year Treasury Yields 21 Business Days Ahead}
    \label{fig:ttm_rolling_yield}
\end{figure}

Figure \ref{fig:ttm_rolling_yield} shows the rolling performance of four TTM variants: the pretrained model after fine-tuning, the untrained model after fine-tuning, the pretrained zero-shot model, and the untrained zero-shot model. Both fine-tuned versions outperform the rest, with the pretrained fine-tuned TTM beating the naive benchmark by roughly 50\%. Notably, its performance dips in 2018; this is the first test window that overlaps the onset of COVID-19, indicating a regime shift driven by pandemic-era monetary policy. After that point, and before any pandemic related yield movements were included in training, the pretrained fine-tuned model’s performance returns to positive levels; this suggests it was able to incorporate those new patterns solely through its contextual understanding. Finally, the zero-shot models exhibit very poor performance, which is expected given the difficulty of this task and the absence of closely related source tasks during pretraining.

Table \ref{tab:transfer_gains_task1} shows the transfer gains (Section \ref{sec:tgt}) for TTM in Task 1. Notably, when fine-tuning data is limited (under ten years), the fine‐tune transfer gain $\Delta_{\mathrm{FT}}$ is 29.32\%—substantially higher than our 10\% threshold. With no data restrictions, however, $\Delta_{\mathrm{FT}}$ becomes statistically insignificant. Although the zero‐shot gain $\Delta_{\mathrm{ZS}}$ is negative, this figure has little interpretive value because both the pretrained and untrained TTM models in zero‐shot mode perform very poorly.

\begin{table}[h]
\centering
\caption{Transfer Gains, Task 1}
\label{tab:transfer_gains_task1}
\begin{tabular}{lc}
\toprule
\textbf{Metric} & \textbf{Gain} \\
\midrule
Fine-tune Transfer Gain (full data)      & 1.73\% \\
Fine-tune Transfer Gain (limited data)   & 29.32\% \\
Zero-shot Transfer Gain                  & -74.16\% \\
\bottomrule
\end{tabular}
\end{table}

The results of constructing a trading signal using TTM for bond yield prediction is presented in Appendix \ref{appendix:backtest} for the interested reader. 

\subsubsection{Chronos}
\begin{figure}[H]
    \centering
    \includegraphics[width=1\linewidth]{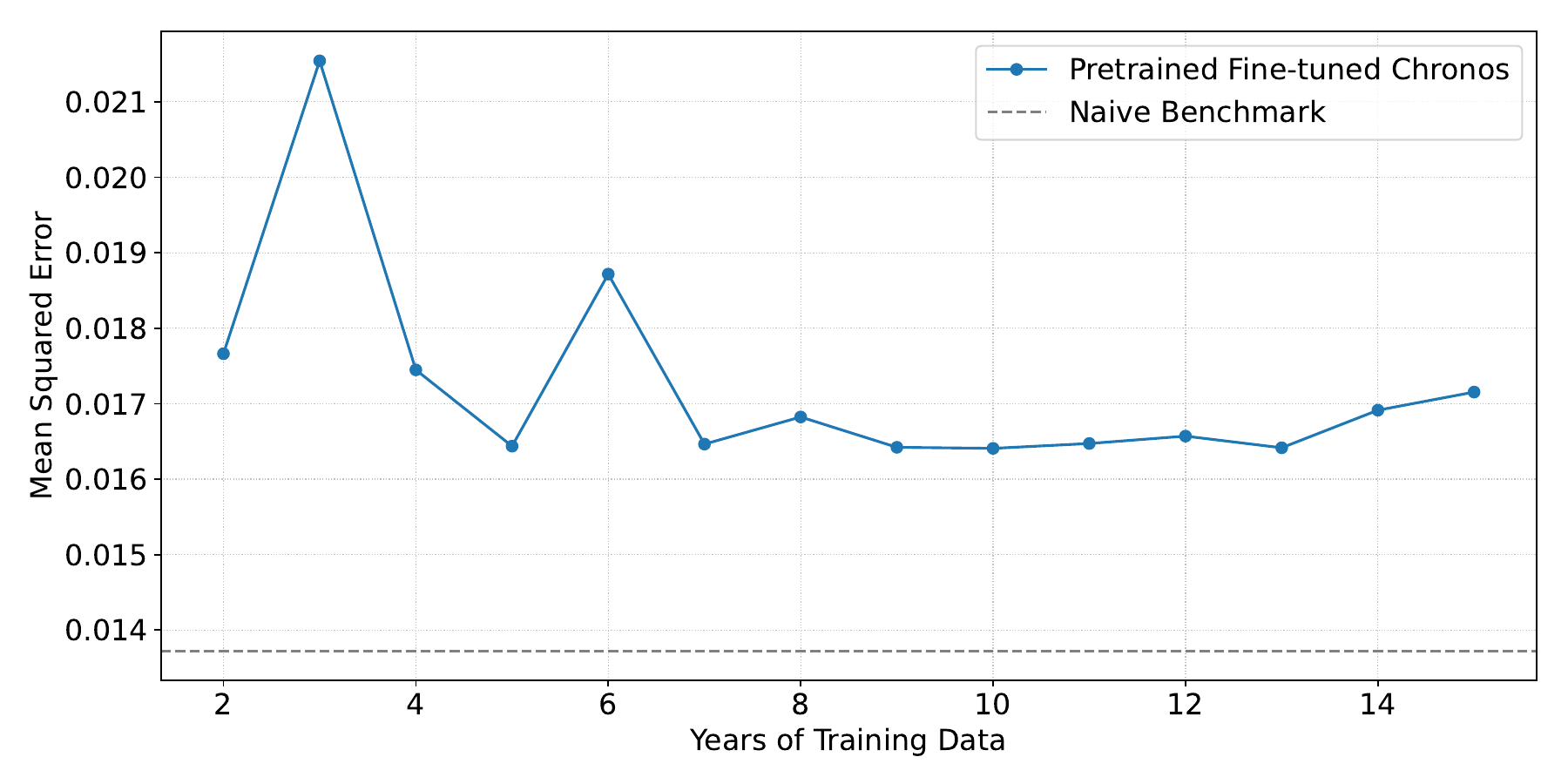}
    \caption{Chronos-Bolt (Small) Mean Squared Error when Fine-tuned on Different Training Dataset Sizes when Forecasting 10-Year Treasury Yields 21 Business Days Ahead}
    \label{fig:chronos_expanding_yield}
\end{figure}

Figure \ref{fig:chronos_expanding_yield} shows the results for Task 1 when using Chronos instead of TTM. The plot demonstrates that regardless of the volume of training data employed for model fine-tuning, performance remains substantially below the naive benchmark threshold, indicating an absence of transfer learning and adaptability for this particular task. Given these findings, we exclude Chronos from subsequent transfer gain analyses as the sample efficiency results clearly demonstrate its inability to provide meaningful knowledge transfer for this task.

\subsubsection{Benchmarks}
\begin{figure}[H]
    \centering
    \includegraphics[width=1\linewidth]{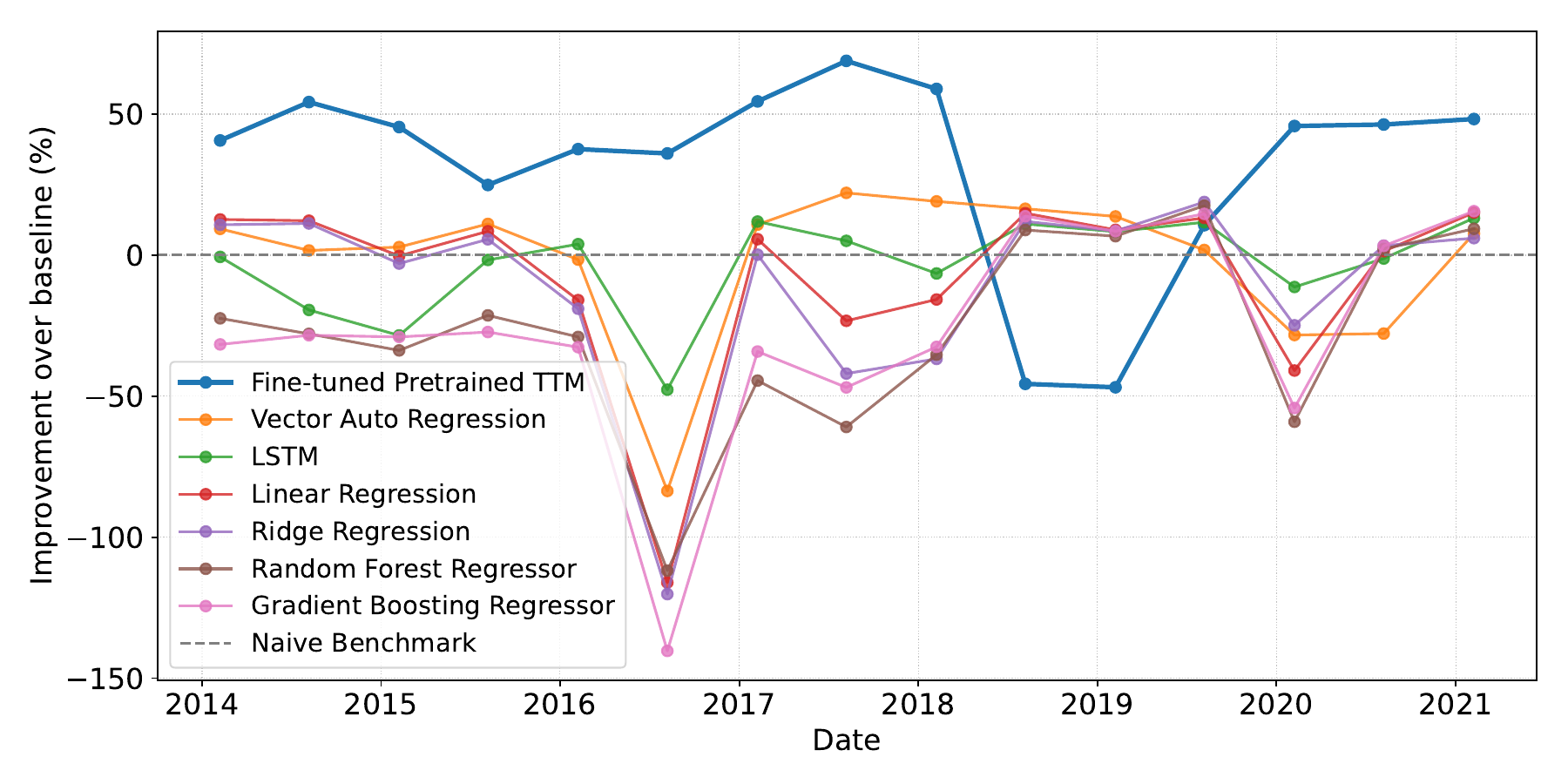}
    \caption{Rolling Performance of Different Models relative to Naive Benchmark when Forecasting 10-Year Treasury Yields 21 Business Days Ahead}
    \label{fig:benchmarks_vs_naive_yield}
\end{figure}

The performance of benchmark models on task 1 is shown in the figures below. Figure \ref{fig:benchmarks_yield} shows the MSE of these models across time, while Figure \ref{fig:benchmarks_vs_naive_yield} shows the same results relative to the naive benchmark. The plots reveal that TTM clearly outperforms all benchmark models for this task, while Chronos is the clear exception in how badly it performs. Furthermore, apart from the Vector Auto Regression (VAR) model, all other benchmark models fail to consistently beat the naive benchmark. The outperformance of TTM, LSTM and VAR compared to other models also show how sequential models are more suitable for this task compared to non-sequential ones.

\begin{figure}[H]
    \centering
    \includegraphics[width=1\linewidth]{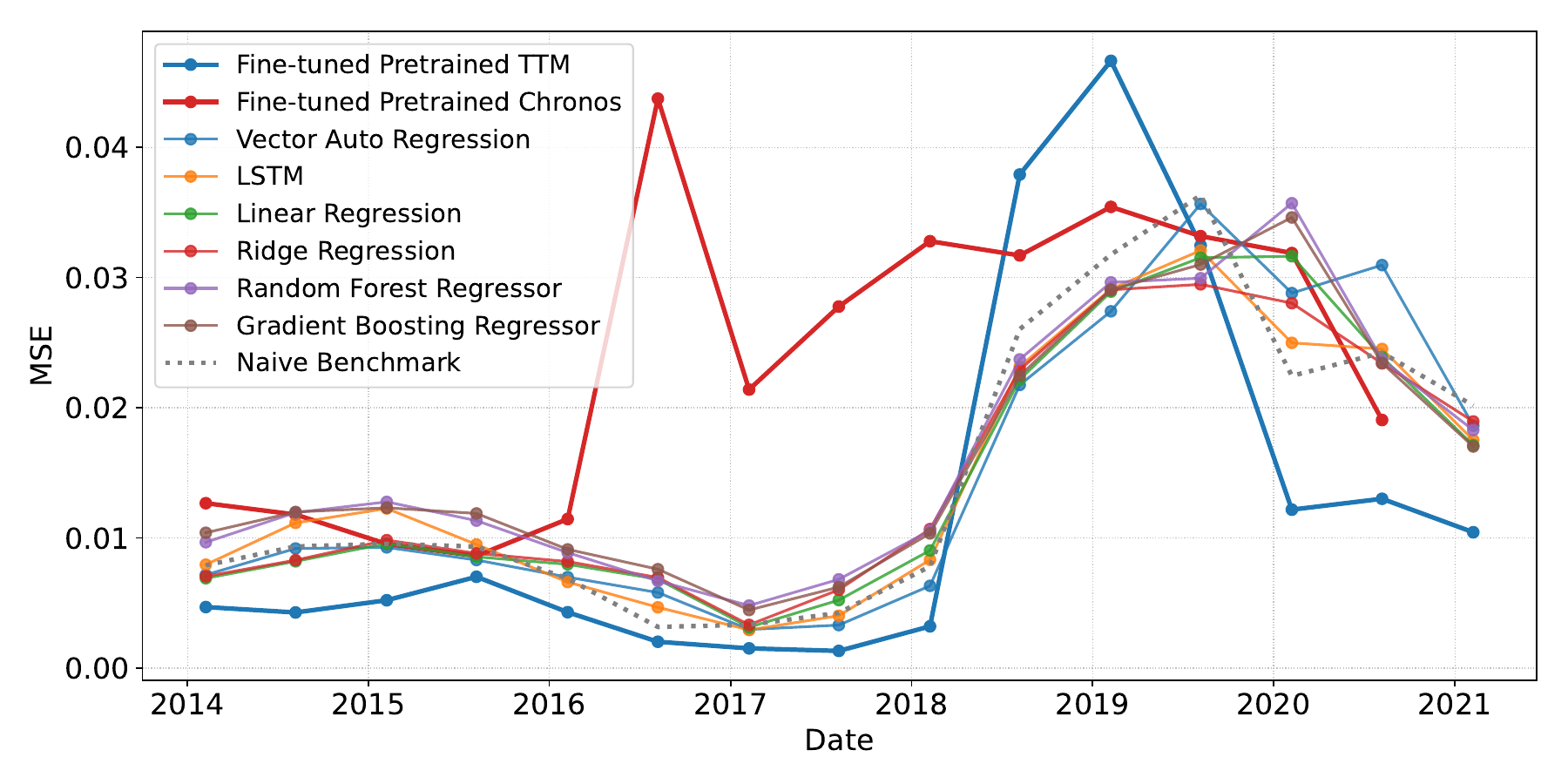}
    \caption{Rolling Performance of Different Models when Forecasting 10-Year Treasury Yields 21 Business Days Ahead}
    \label{fig:benchmarks_yield}
\end{figure}

\subsection{Task 2: Foreign Exchange Volatility Forecasting}
\label{sec:task2_results}
The results for Task 2, forecasting the EUR/USD exchange rate volatility 21 days in advance, are displayed in this section for TTM, Chronos and benchmark models. Given that the target variable is autocorrelated, the naive model simply predicts no change every time.
\subsubsection{TTM}
\begin{figure}[H]
    \centering
    \includegraphics[width=1\linewidth]{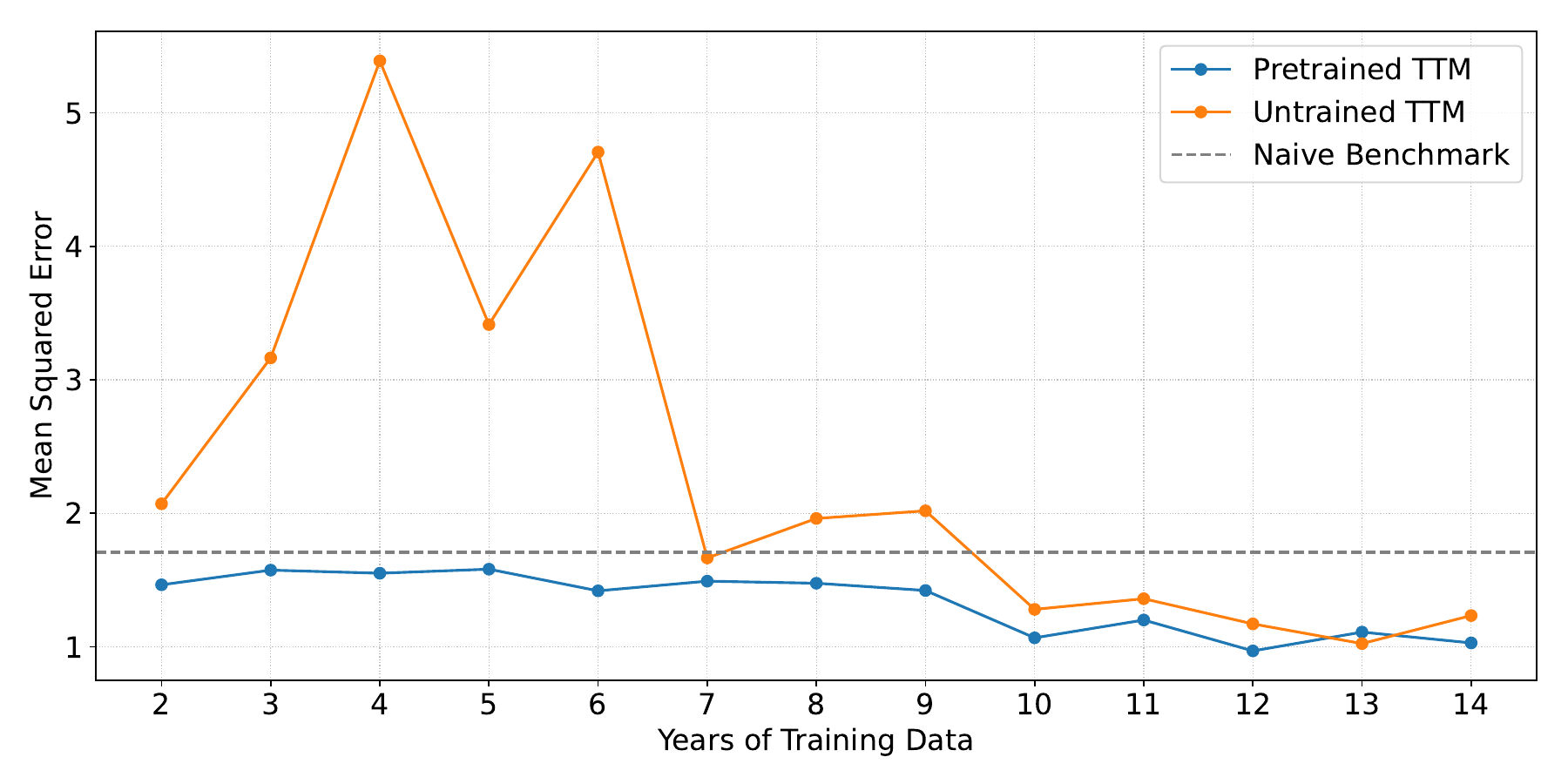}
    \caption{TTM Mean Squared Error when Fine-tuned on Different Training Dataset Sizes when Forecasting EUR/USD Realised Volatility 21 Business Days Ahead}
    \label{fig:ttm_expanding_vol}
\end{figure}

Figure \ref{fig:ttm_expanding_vol} shows the results of running the Sample Efficiency Probe (Section \ref{sec:sep}) using TTM for Task 2. Similar to the results for Task 1 (Section \ref{sec:task1_results}), the figure shows that the pretrained model needs significantly less data to learn; \textbf{while the pretrained version performs significantly better than the naive benchmark with just two years of data, the untrained model requires fine-tuning on 10 years of data before it can beat the naive benchmark}. Like in Task 1, however, the untrained model eventually reaches the performance of the pretrained model with enough fine-tuning data.

\begin{figure}[H]
    \centering
    \includegraphics[width=1\linewidth]{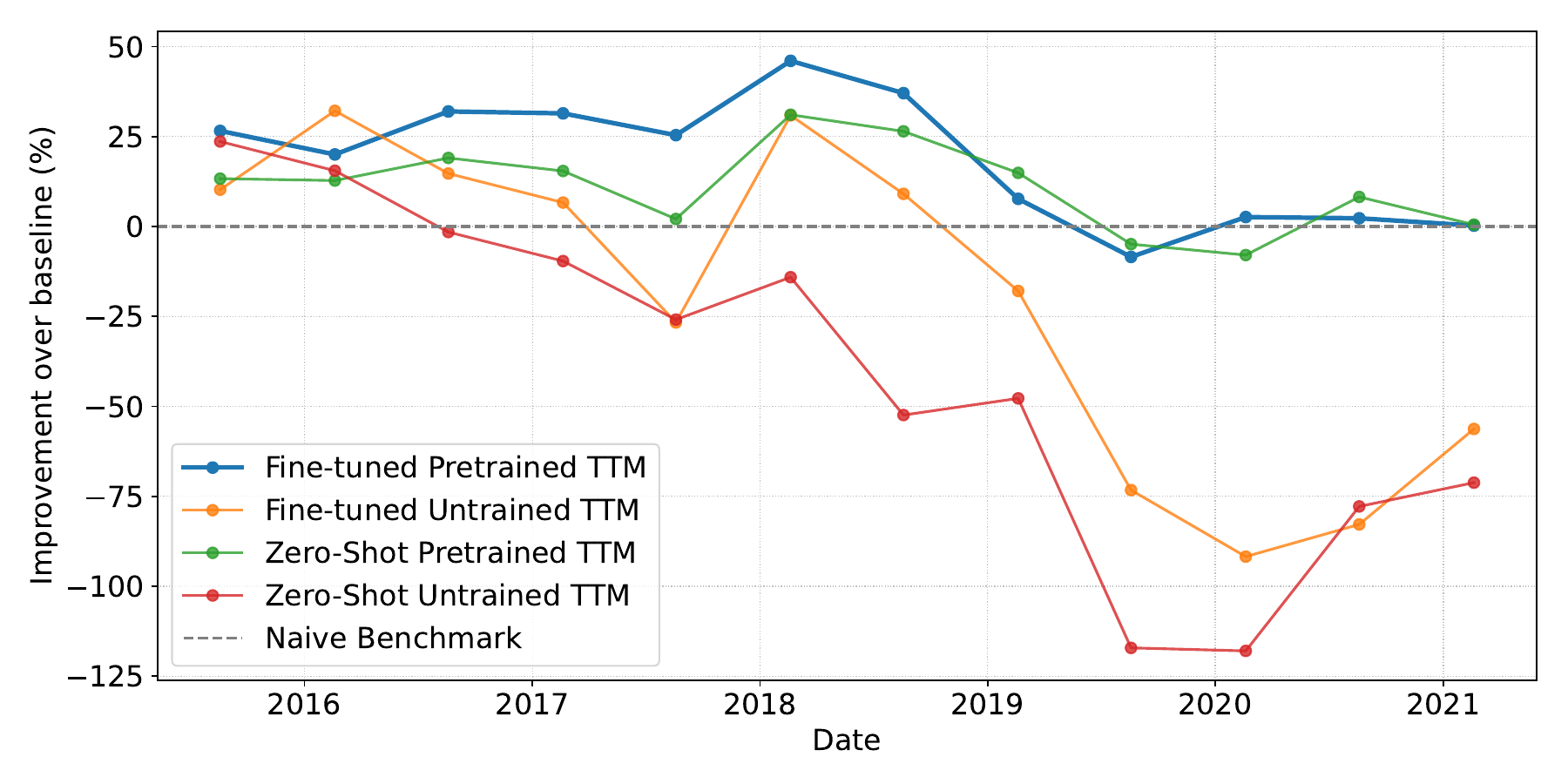}
    \caption{Rolling Performance of Different Training Settings of TTM Compared to Naive Benchmark when Forecasting EUR/USD Realised Volatility 21 Business Days Ahead}
    \label{fig:ttm_rolling_vol}
\end{figure}

Figure \ref{fig:ttm_rolling_vol} shows the rolling performance of the same four TTM variants as in Task 1. Notably, the figure reveals firstly the outperformance of the pre-trained and fine-tuned TTM compared to untrained and fine-tuned TTM, indicating a high transfer gain coefficient and secondly, that \textbf{pretrained TTM in zero-shot modality outperforms the naive benchmark consistently except for the pre-pandemic and pandemic period}. 

Table \ref{tab:transfer_gains_task2} shows the transfer gains (Section \ref{sec:tgt}) for TTM in Task 2. When fine-tuning data is limited (under ten years), the fine‐tune transfer gain $\Delta_{\mathrm{FT}}$ is 50.90\%, which is again substantially higher than our 10\% threshold. Even when using the full fine-tuning dataset, $\Delta_{\mathrm{FT}}$ is still statistically significant, with a value of 30.62\%. Furthermore, the zero‐shot gain $\Delta_{\mathrm{ZS}}$ is also substantially higher than the threshold at 34.4\%, a strong indication of the transferability of TTM's pretraining to this downstream task, and also a demonstration of how volatility forecasting is an easier task compared to forecasting yield changes.

\begin{table}[ht]
\centering
\caption{Transfer Gains, Task 2}
\label{tab:transfer_gains_task2}
\begin{tabular}{lc}
\toprule
\textbf{Metric} & \textbf{Gain} \\
\midrule
Fine-tune Transfer Gain (full data)         & 30.62\% \\
Fine-tune Transfer Gain (limited data)               & 50.90\% \\
Zero-shot Transfer Gain         & 34.40\% \\
\bottomrule
\end{tabular}
\end{table}

\subsubsection{Chronos}
\begin{figure}[H]
    \centering
    \includegraphics[width=1\linewidth]{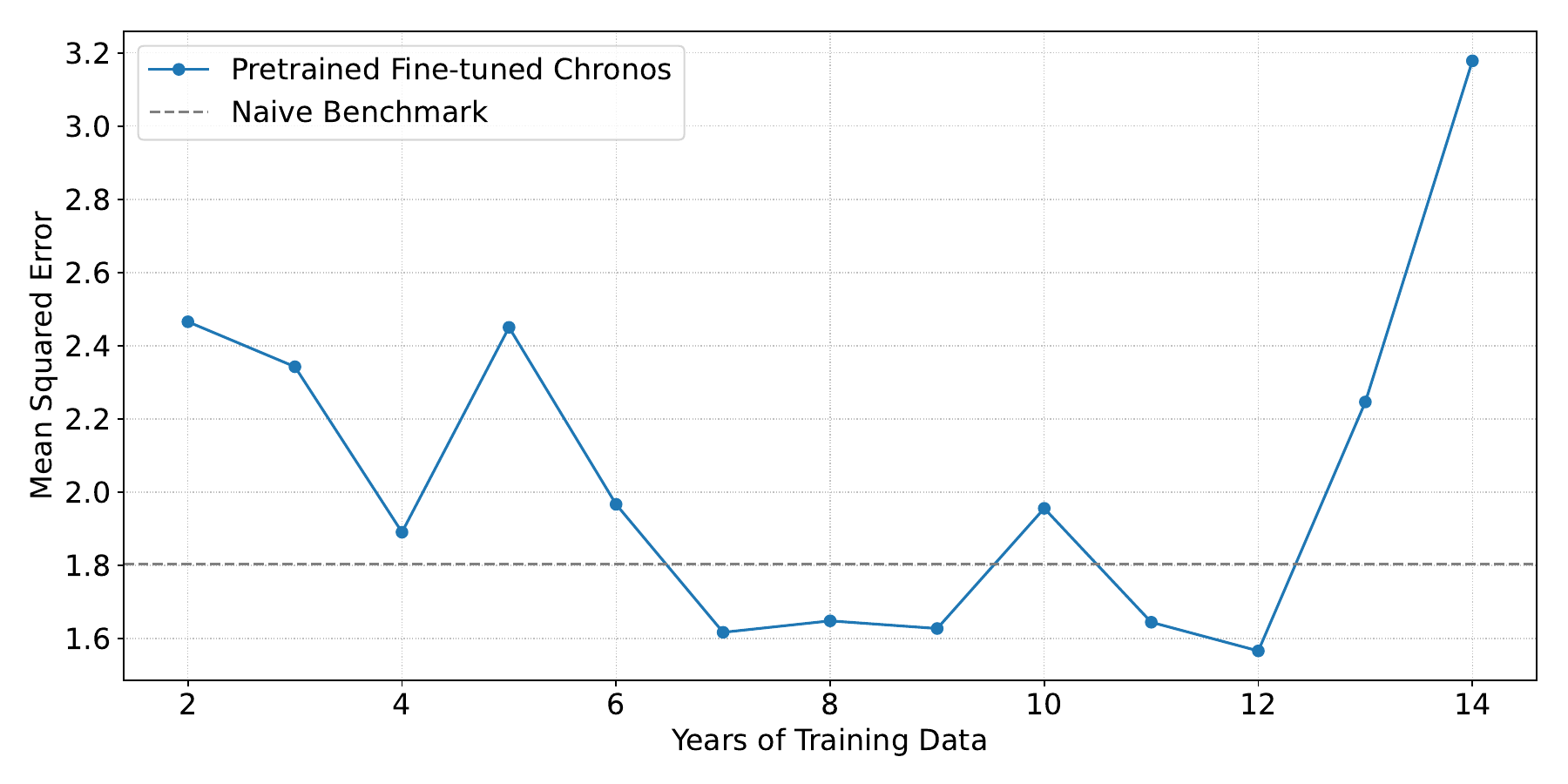}
    \caption{Chronos-Bolt (Small) Mean Squared Error when Fine-tuned on Different Training Dataset Sizes when Forecasting EUR/USD Realised Volatility 21 Business Days Ahead}
    \label{fig:chronos_expanding_vol}
\end{figure}

Figure \ref{fig:chronos_expanding_vol} shows that once again, Chronos fails to consistently surpass the naive benchmark for Task 2 regardless of the amount of training data used for model fine-tuning. Comparing to TTM, and especially to its outperformance of the naive benchmark in zero-shot modality, we conclude that there is yet again an absence of transferability for this particular task and exclude Chronos from subsequent transfer gain analyses.

\subsubsection{Benchmarks}
\begin{figure}[H]
    \centering
    \includegraphics[width=1\linewidth]{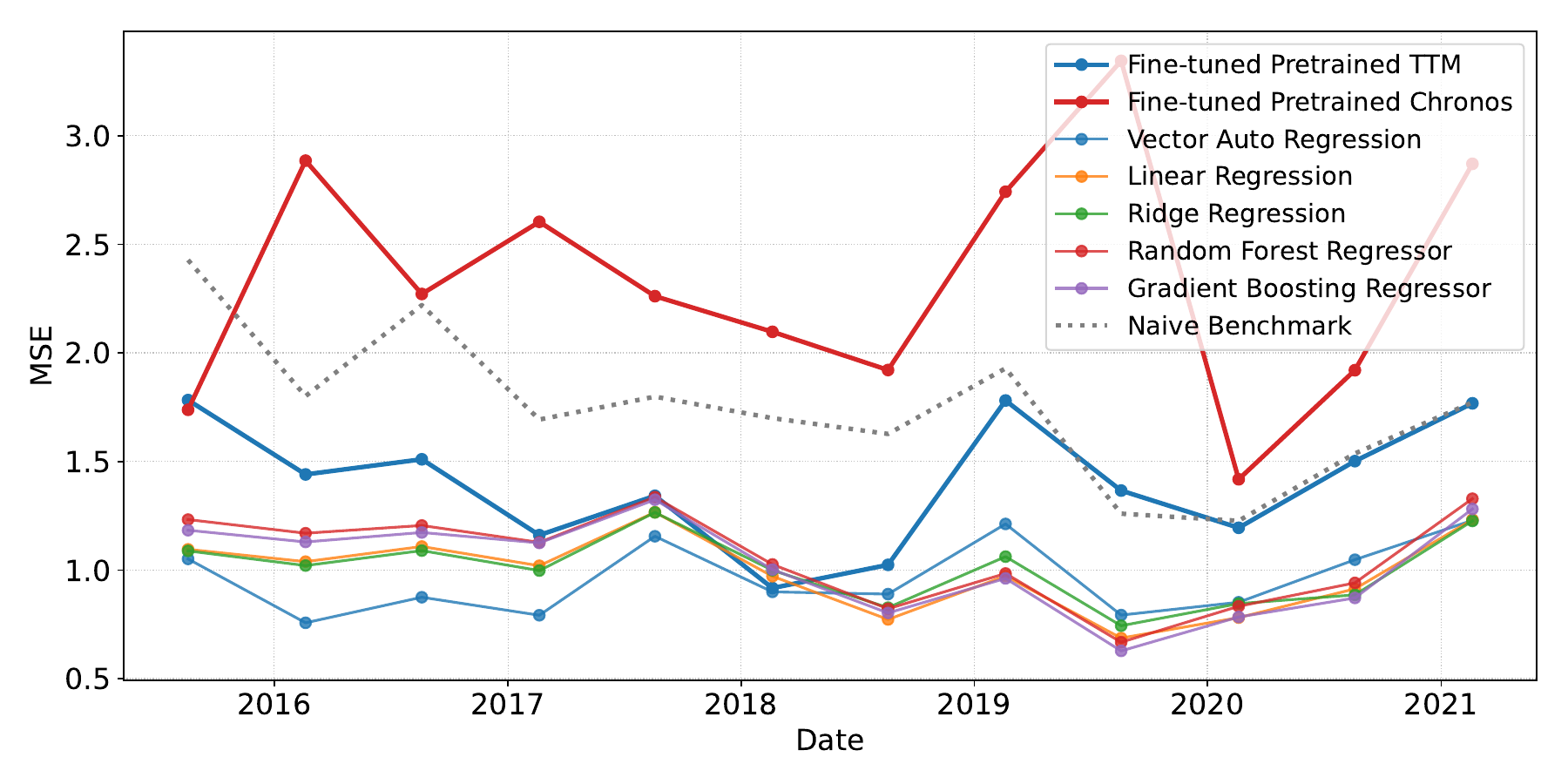}
    \caption{Rolling Performance of Different Models when Forecasting EUR/USD Realised Volatility 21 Business Days Ahead}
    \label{fig:benchmarks_vol}
\end{figure}

\begin{figure}[H]
    \centering
    \includegraphics[width=1\linewidth]{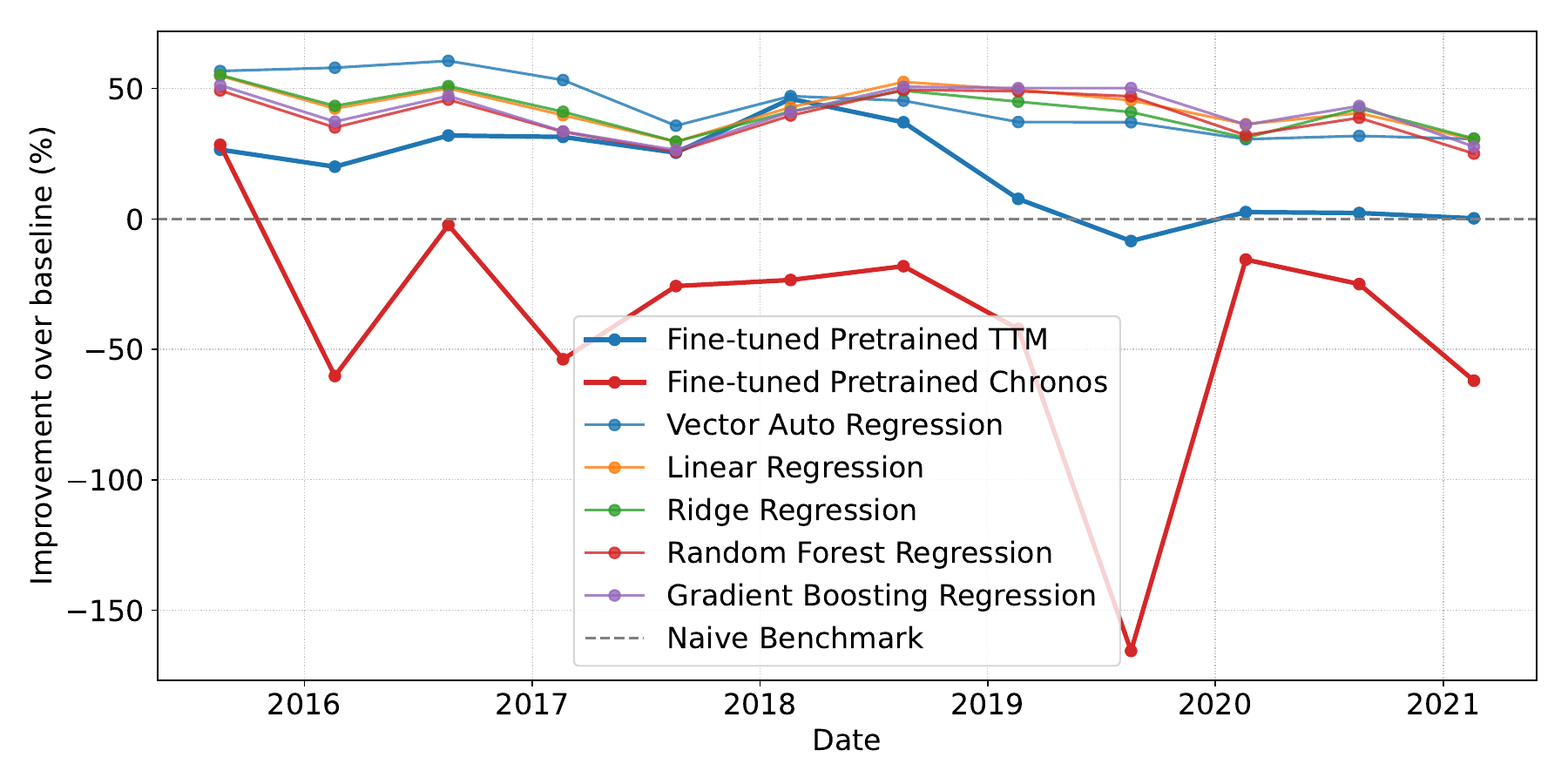}
    \caption{Rolling Performance of Different Models relative to Naive Benchmark when Forecasting EUR/USD Realised Volatility 21 Business Days Ahead}
    \label{fig:benchmarks_vs_naive_vol}
\end{figure}

The performance of benchmark models on Task 2 is illustrated in Figures \ref{fig:benchmarks_vol} and \ref{fig:benchmarks_vs_naive_vol}, which show MSE across time and performance relative to the naive benchmark, respectively. In contrast to Task 1, benchmark models consistently outperform TTM on this task, with Chronos remaining a notable exception as the only model that fails to exceed the naive benchmark. These results reinforce that volatility forecasting—a relatively straightforward task in financial prediction with high signal-to-noise ratio—benefits from simpler approaches, as complex deep neural networks tend to overfit rather than capture the underlying patterns effectively. Nevertheless, pretrained TTM showed a high transfer gain in this task compared to its untrained counterpart.

\subsection{Task 3: Equity Spread Forecasting}
The results for Task 3, forecasting the spread between MSCI Canada and MSCI Australia indices at different horizons, are presented in this section. The naive forecast selection varies by horizon based on the series' statistical properties: for 5-day forecasts, we use the previous value due to short-term autocorrelation, while for a 10-day horizon, we use zero as the naive forecast. This choice reflects the spread's mean-reverting, zero-centred nature, with an estimated mean reversion time of approximately 14 days, making zero the more appropriate long-term expectation for horizons near this threshold. 

\subsubsection{TTM}
\begin{figure}[H]
    \centering
    \includegraphics[width=1\linewidth]{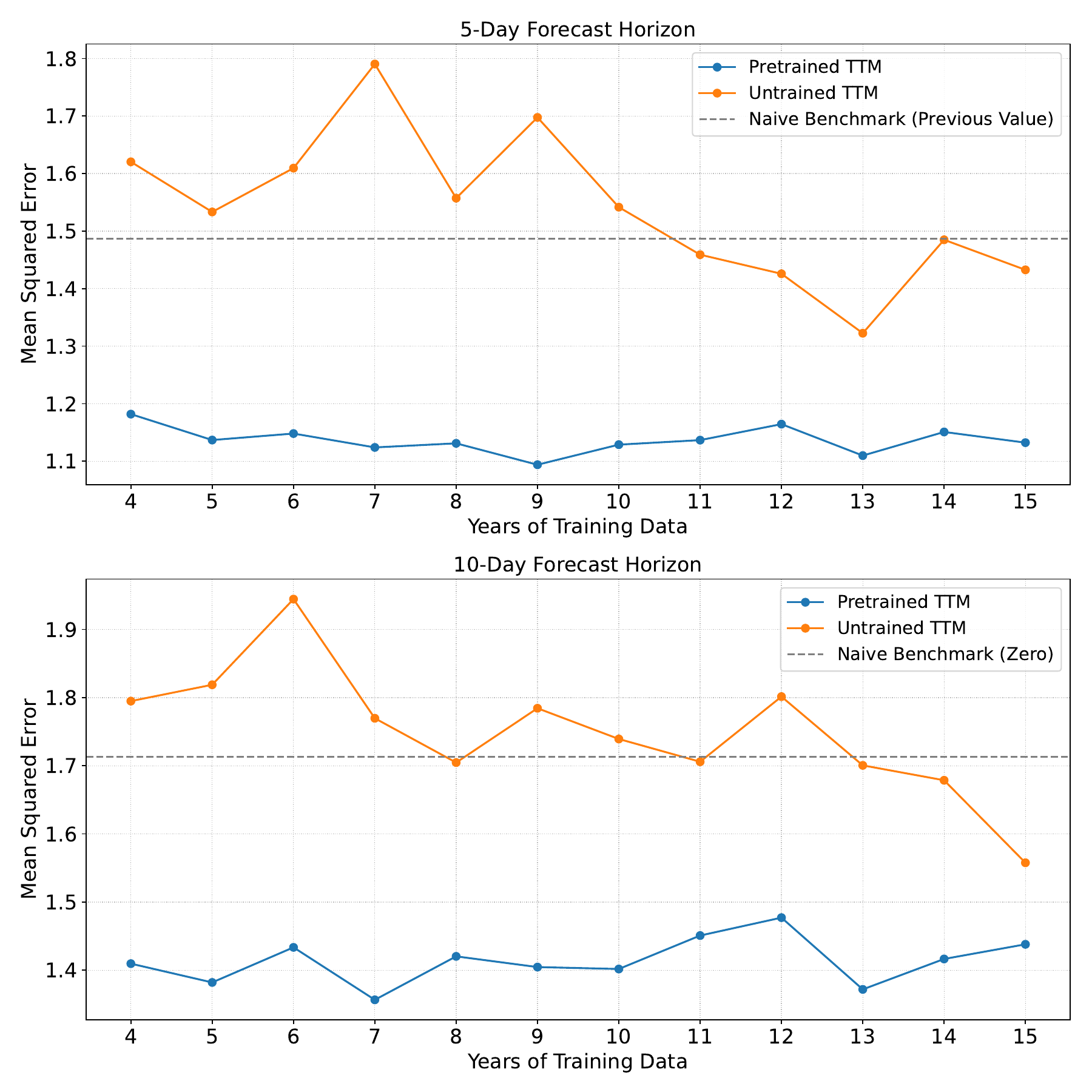}
    \caption{TTM Mean Squared Error when Fine-tuned on Different Training Dataset Sizes when Forecasting the Spread Between MSCI Australia and MSCI Canada at Different Horizons}
    \label{fig:ttm_expanding_spread}
\end{figure}

Figure \ref{fig:ttm_expanding_spread} shows the results of running the Sample Efficiency Probe (Section \ref{sec:sep}) using TTM for Task 3. Like in Tasks 1 and 2 (Sections \ref{sec:task1_results} and \ref{sec:task2_results}), the figure shows that the pretrained model needs significantly less data to learn; \textbf{while the pretrained version performs significantly better than the naive benchmark with as little as four years of data, the untrained model requires fine-tuning on 11 years of data before it can beat the naive benchmark}. Unlike in Task 1 and 2, the untrained model never reaches the performance of the pretrained model even with the full fine-tuning dataset. This robust transfer capability suggests that TTM's pretraining on diverse time series has equipped it with generalisable knowledge of mean-reverting patterns, which are common across many domains and particularly relevant for modelling financial spreads.

\begin{figure}[H]
    \centering
    \includegraphics[width=1\linewidth]{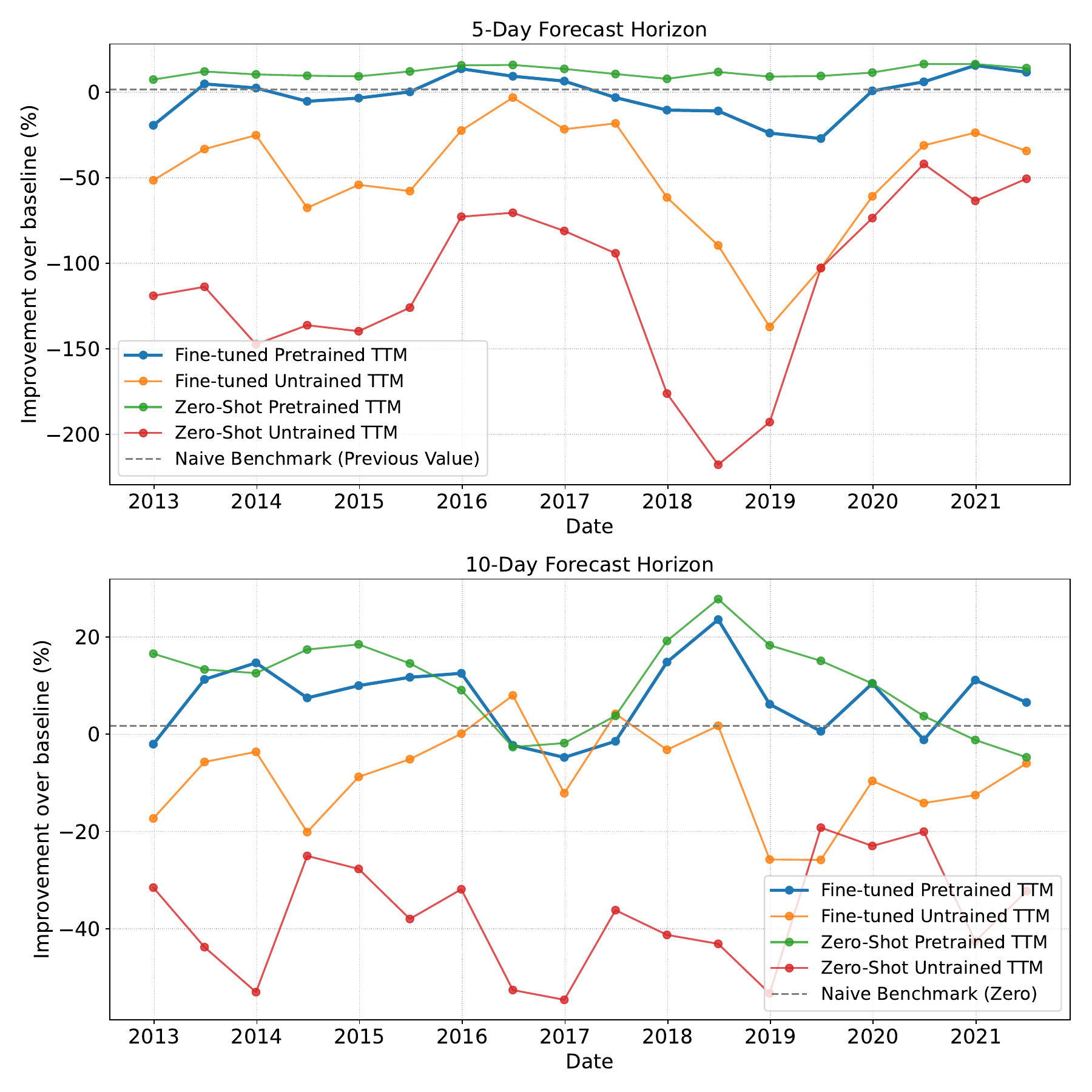}
    \caption{Rolling Performance of Different Training Settings of TTM Compared to Naive Benchmark when Forecasting the Spread Between MSCI Australia and MSCI Canada at Different Horizons}
    \label{fig:ttm_rolling_spread}
\end{figure}

Figure \ref{fig:ttm_rolling_spread} shows the rolling performance of the same four TTM variants as in Task 1 and 2 for Task 3. When predicting the spread with a forecast horizon of 5 days, the figure reveals firstly the outperformance of the pre-trained and fine-tuned TTM compared to untrained and fine-tuned TTM and secondly, that \textbf{pretrained TTM in zero-shot modality is the only model that always outperforms the naive benchmark}. Since hyperparameter tuning such as epochs and years of training data were optimised based solely on the fixed date in the Sample-Efficiency probe (January 2021), the performance at that specific date shows rough parity between zero-shot and fine-tuned pretrained models, whilst performance at other dates deteriorates for the fine-tuned variant. This suggests that the fine-tuning process may be overfitting to the validation date, and that hyperparameter choices and training procedures optimised for January 2021 do not generalise effectively to other time periods. For a forecast horizon of 10 days, we still observe an outperformance of the two pretrained models, particularly the one in zero-shot modality, although the naive benchmark is not always outperformed when using this forecasting horizon.

Table \ref{tab:transfer_gains_task3} shows the transfer gains (Section \ref{sec:tgt}) for TTM in Task 3. When fine-tuning data is limited (under ten years), the fine‐tune transfer gain $\Delta_{\mathrm{FT}}$ is significantly higher than our 10\% threshold for both forecast horizons. Even when using the full fine-tuning dataset, $\Delta_{\mathrm{FT}}$ is still statistically significant, with values of 31.68\% and 14.65\% for forecast horizons of 5 and 10 days respectively. Finally, the zero‐shot gain $\Delta_{\mathrm{ZS}}$ is also substantially higher than the 10\% threshold at 57.35\% and 34.89\%, a strong indication of the transferability of TTM's pretraining to this downstream task. Once again, this is an indication of how this mean reverting variable is easier to predict compared to forecasting yield changes and how widespread this property is across time series from various domains used in pretraining.

\begin{table}[ht]
\centering
\caption{Transfer Gains Across Forecast Horizons, Task 3}
\label{tab:transfer_gains_task3}
\begin{tabular}{lcc}

\toprule
\textbf{Metric} & \textbf{Horizon 5} & \textbf{Horizon 10} \\
\midrule
Fine-tune Transfer Gain (full data)     & 31.68\% & 14.65\% \\
Fine-tune Transfer Gain (limited data)                      & 25.36\% & 22.29\% \\
Zero-shot Transfer Gain                & 57.35\% & 34.89\% \\
\bottomrule
\end{tabular}
\end{table}

\subsubsection{Chronos}
As was this case with the first two tasks, Figure \ref{fig:chronos_expanding_vol} shows that Chronos fails to consistently surpass the naive benchmark for Task 3 regardless of the amount of training data used for model fine-tuning, particularly for a forecasting horizon 10 days. In contrast to TTM's strong performance, including its ability to outperform the naive benchmark in zero-shot settings, these results confirm Chronos' lack of effective knowledge transfer for this task, leading us to once again omit it from further transfer gain evaluations.

\begin{figure}[H]
    \centering
    \includegraphics[width=1\linewidth]{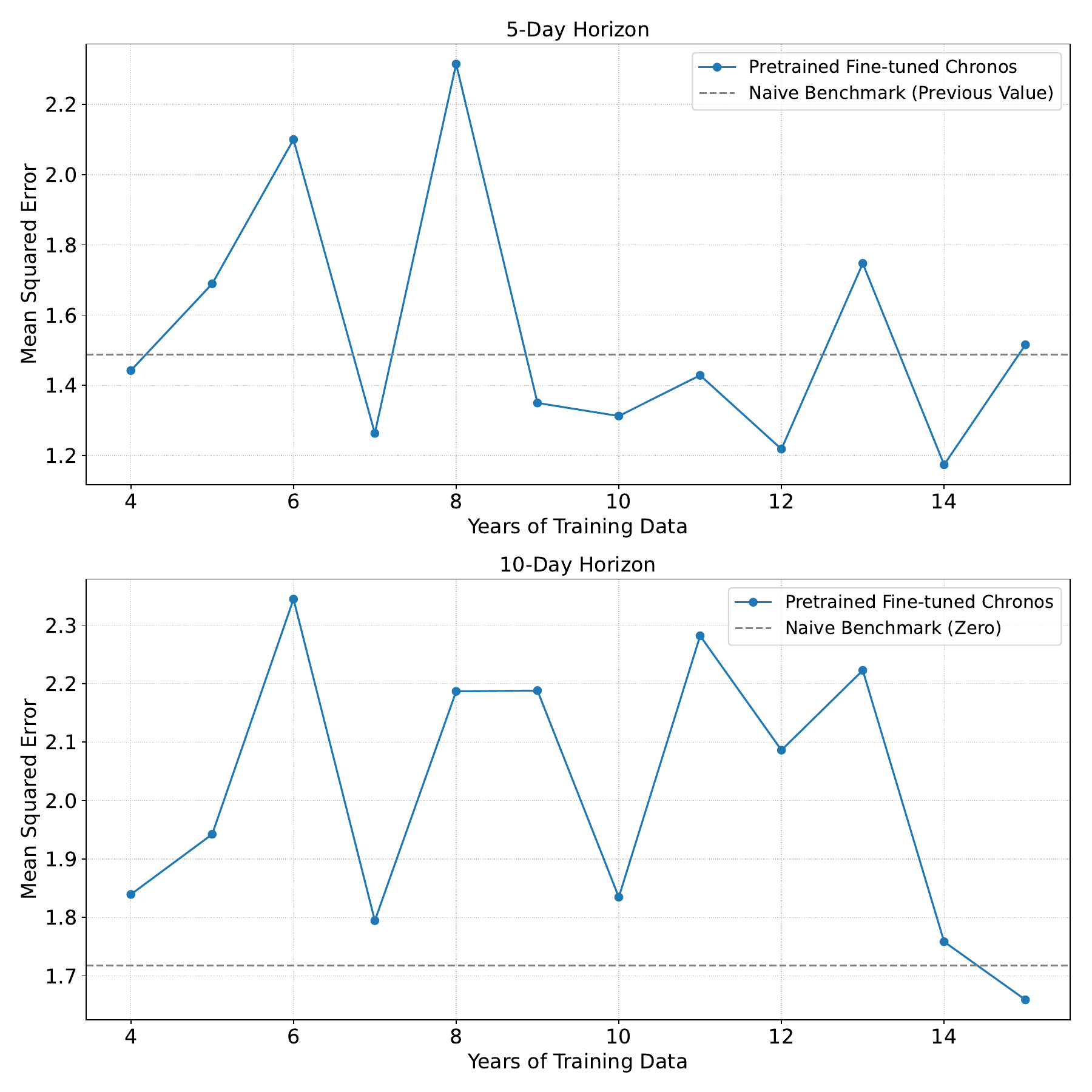}
    \caption{Chronos-Bolt (Small) Mean Squared Error when Fine-tuned on Different Training Dataset Sizes when Forecasting the Spread Between MSCI Australia and MSCI Canada at Different Horizons}
    \label{fig:chronos_expanding_spread}
\end{figure}

\subsubsection{Benchmarks}
\begin{figure}[htbp]
    \centering
    \includegraphics[width=1\linewidth]{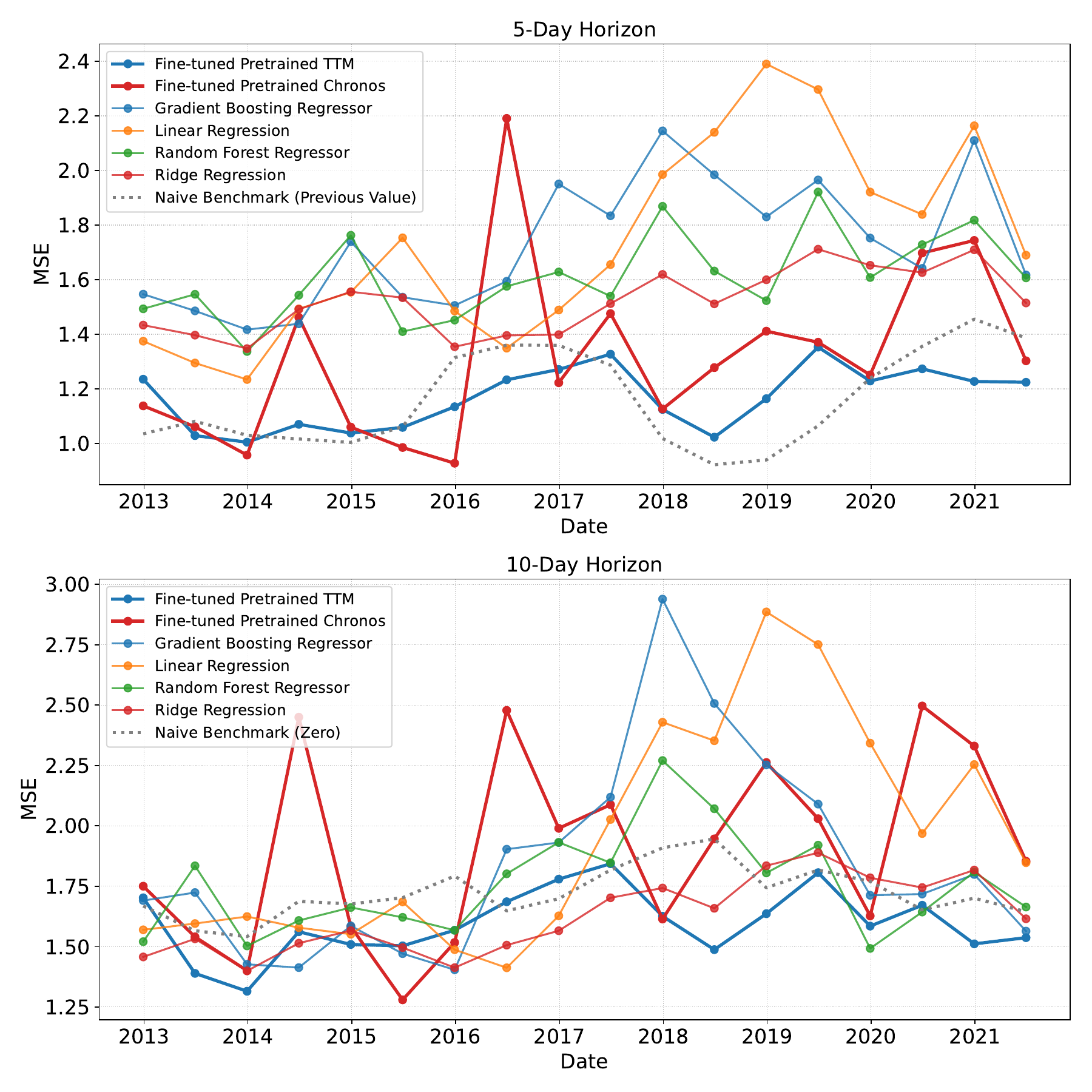}
    \caption{Rolling Performance of Different Non-Sequential Models when Forecasting the Spread Between MSCI Australia and MSCI Canada at Different Horizons}
    \label{fig:benchmarks_spread}
\end{figure}

\begin{figure}[htbp]
    \centering
    \includegraphics[width=1\linewidth]{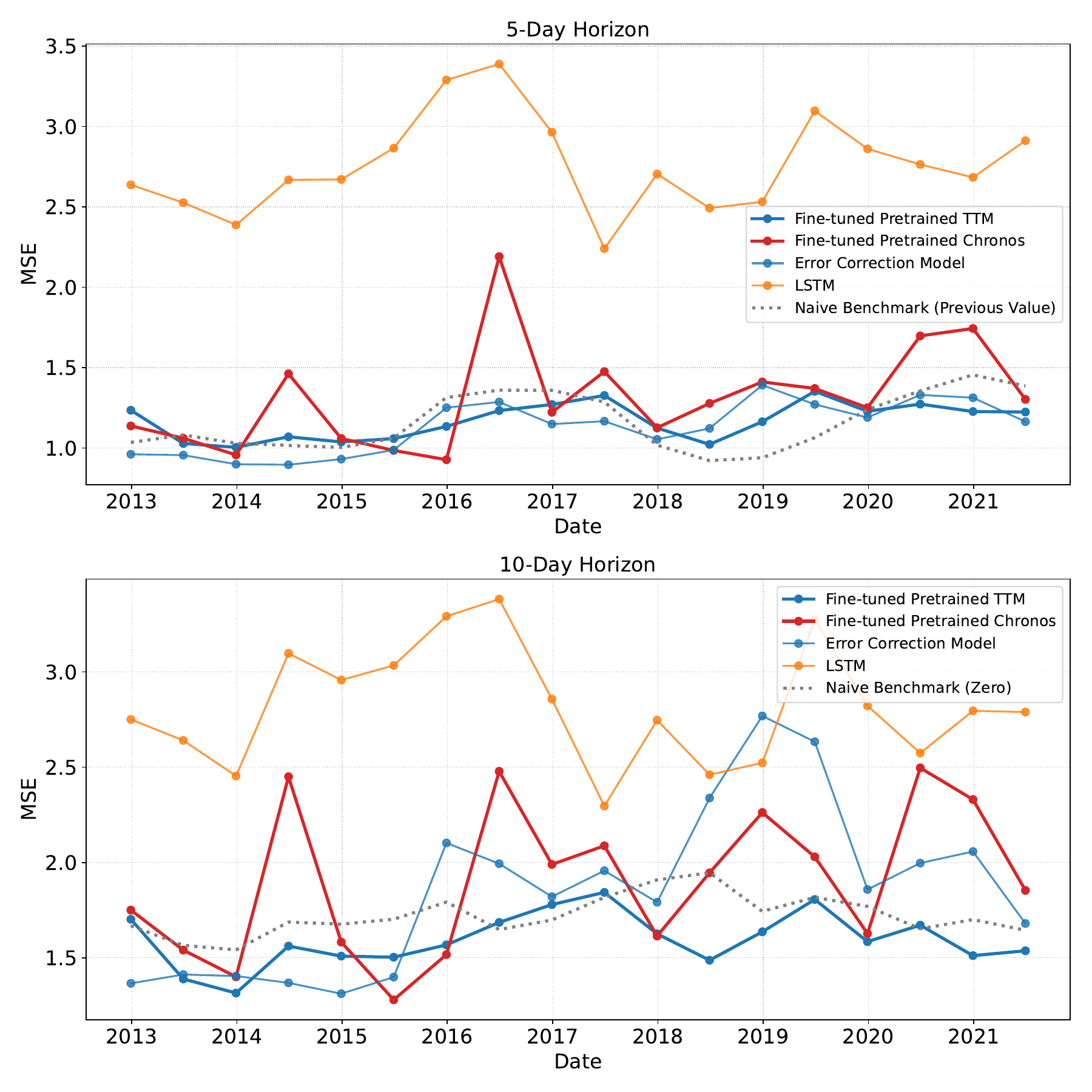}
    \caption{Rolling Performance of Different Sequential Models when Forecasting the Spread Between MSCI Australia and MSCI Canada at Different Horizons}
    \label{fig:benchmarks_spread_seq}
\end{figure}

\begin{figure}[htbp]
    \centering
    \includegraphics[width=1\linewidth]{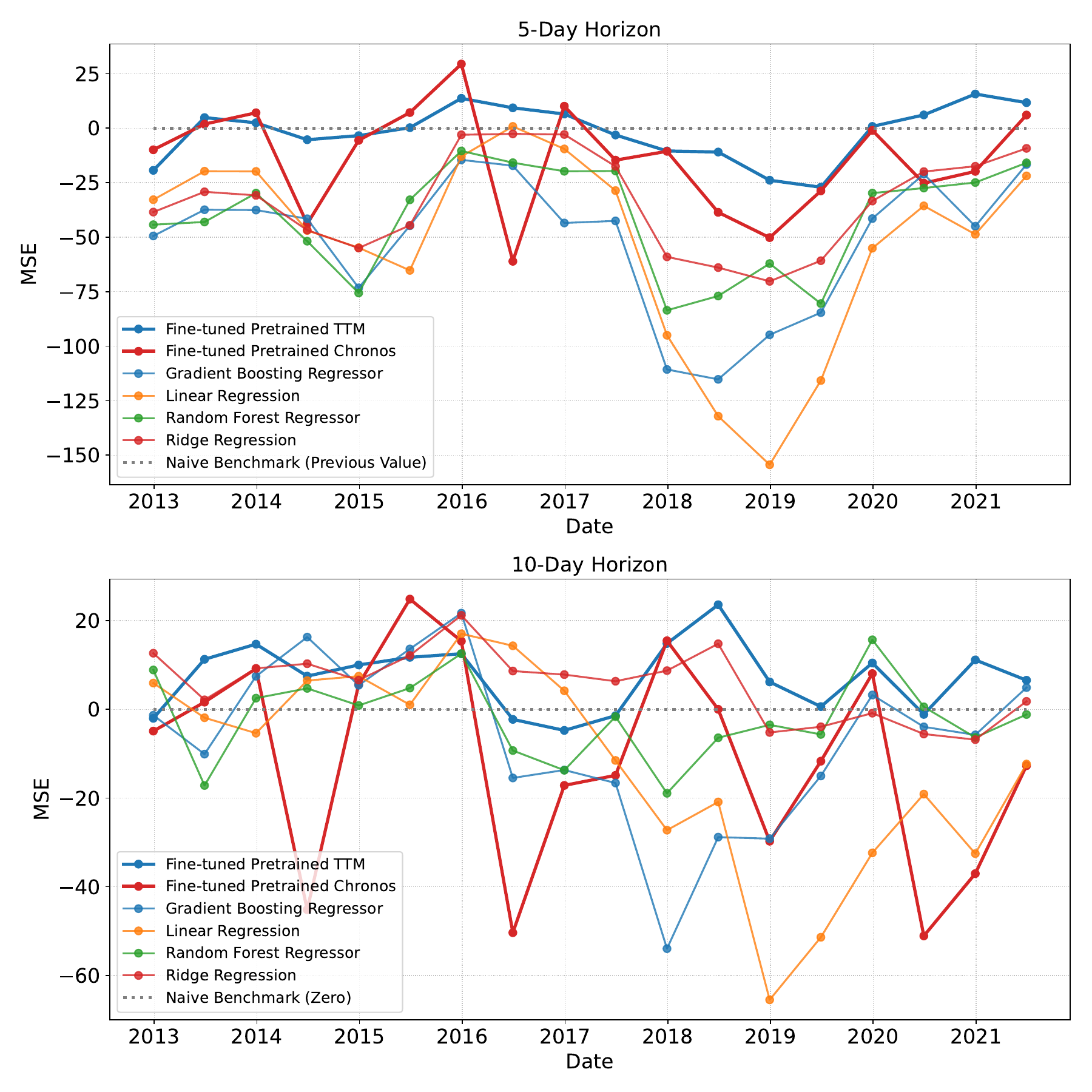}
    \caption{Rolling Performance of Different Non-Sequential Models Relative to Naive Benchmark when Forecasting the Spread Between MSCI Australia and MSCI Canada at Different Horizons}
    \label{fig:benchmarks_vs_naive_spread}
\end{figure}

\begin{figure}[htbp]
    \centering
    \includegraphics[width=1\linewidth]{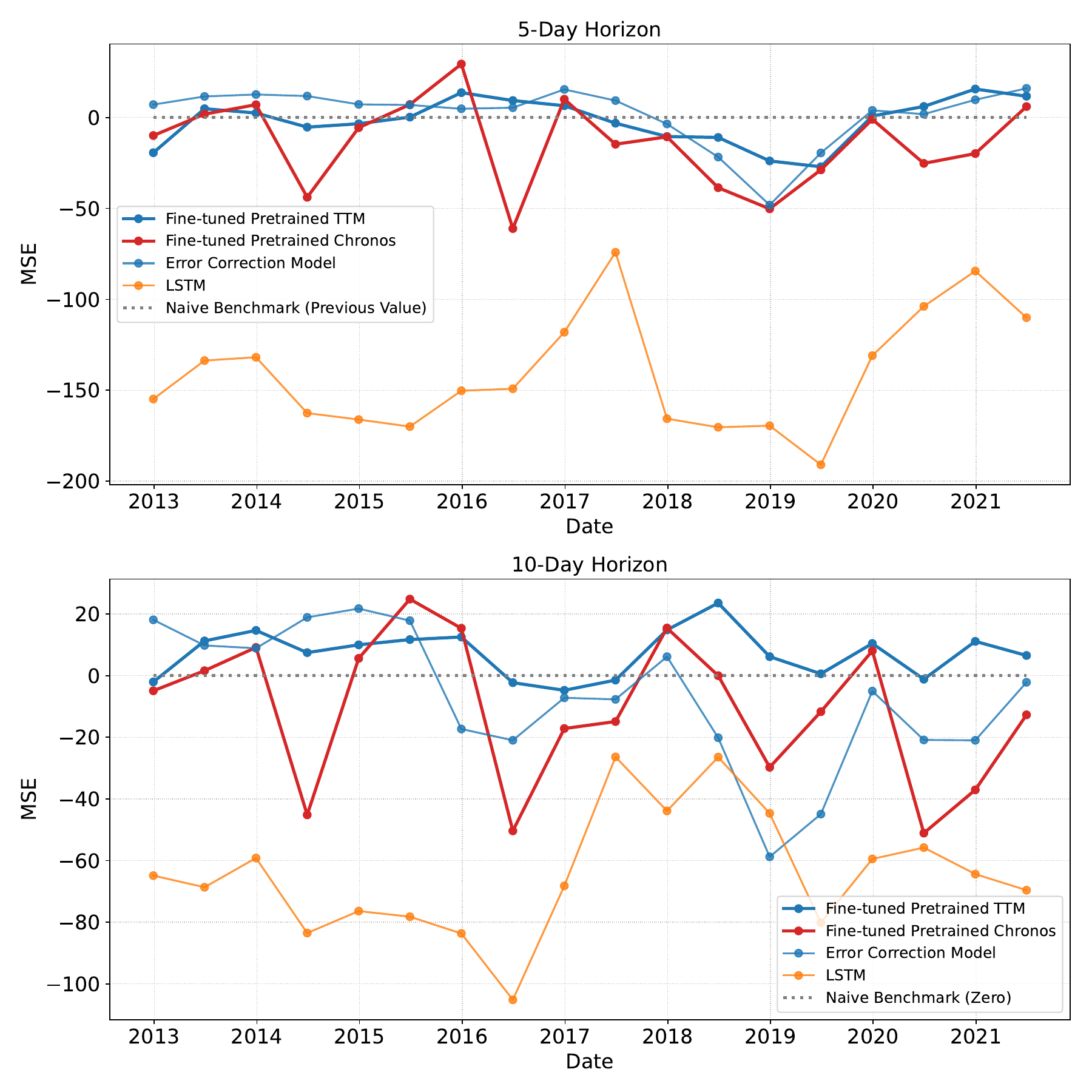}
    \caption{Rolling Performance of Different Sequential Models Relative to Naive Benchmark when Forecasting the Spread Between MSCI Australia and MSCI Canada at Different Horizons}
    \label{fig:benchmarks_vs_naive_spread_seq}
\end{figure}

The performance of benchmark models on Task 3 is shown in Figures \ref{fig:benchmarks_spread}, \ref{fig:benchmarks_spread_seq}, \ref{fig:benchmarks_vs_naive_spread} and \ref{fig:benchmarks_vs_naive_spread_seq}. For added clarity, we separate benchmark models into two groups: non-sequential models, which include Ridge regression, Linear Regression, gradient boosting regression, and random forest (RF); and sequential models, comprising the Long Short-Term Memory network (LSTM) and the Error Correction Model (ECM).
The plots reveal the inherent difficulty of outperforming the naive forecast for this task. Given the target variable's autocorrelation, beating a forecast of the previous value at short-term horizons proves challenging, while the target's mean-reverting, zero-centred nature makes it equally difficult to surpass a simple zero forecast at medium-term horizons.
Several key findings emerge from these results: First, Chronos remains a consistent outlier with poor performance across horizons. Second, while TTM ranks among the better-performing models, it still fails to always beat the naive benchmark at both horizons. Third, ECM emerges as the top performer—essentially an AR(1) model on the spread (see Appendix \ref{appendix:ecm})—which effectively exploits both the spread's autocorrelation and mean reversion through an autoregressive coefficient less than one. Finally, there is a general inability among benchmark models to consistently surpass the defined naive forecasts.

\subsection{Discussion}
\label{sec:discussion}
The empirical results across our three financial forecasting tasks reveal several technical insights about the behaviour and limitations of Time Series Foundation Models in financial applications. This section examines the specific patterns observed in our experiments and their technical implications.

\subsubsection{Sample Efficiency Patterns and Learning Dynamics}

The Sample Efficiency Probe results demonstrate consistent learning acceleration patterns for TTM across all tasks, though with notable variations in convergence behaviour. In Tasks 1 and 2, the untrained model eventually converged to similar performance levels as the pretrained version, suggesting that given sufficient data, the architectural capacity of TTM is not the limiting factor. However, Task 3 presents a different pattern—the untrained model never reaches the pretrained performance level even with the full dataset.

This divergence in Task 3 appears linked to the underlying statistical properties of the target variable. The mean-reverting nature of equity spreads creates a learning environment where early recognition of reversion patterns provides advantages throughout the training process. The pretrained model's immediate recognition of these patterns, likely acquired from similar mean-reverting series during pretraining, creates a performance gap that cannot be closed through additional data alone.

The temporal patterns in learning curves also reveal interesting dynamics. The untrained models consistently exhibit what appears to be a "discovery phase" where performance remains poor until sufficient data accumulates to identify the underlying patterns. The pretrained models bypass this discovery phase entirely, immediately leveraging transferred knowledge to achieve superior performance with minimal fine-tuning data.

\subsubsection{Rolling Performance Analysis and Regime Sensitivity}

The rolling evaluation results expose significant temporal variations in model performance that provide insights into the stability of transfer learning benefits. The performance deterioration observed during 2020-2021 across multiple models and tasks suggests that extreme market conditions-caused by the COVID-19 pandemic in this case-can overwhelm the benefits of transfer learning, at least temporarily.

Particularly notable is the recovery pattern observed in Task 1, where TTM's performance rebounds after the initial COVID-19 impact despite not having pandemic-era data in its training set. This suggests that the model's contextual understanding allows it to adapt to new regime characteristics even without explicit retraining on similar conditions.

The zero-shot performance patterns reveal task-specific transferability characteristics. Task 2's consistent zero-shot performance (except during pandemic periods) indicates that volatility patterns are sufficiently universal across domains to enable direct transfer. In contrast, Task 1's poor zero-shot performance but strong fine-tuning gains suggests that while yield forecasting requires domain-specific adaptation, the underlying temporal modelling capabilities transfer effectively.

\subsubsection{Benchmark Performance Patterns and Model Hierarchy}

The benchmark comparisons reveal a clear task-dependent model hierarchy that provides insights into when foundation models add value. In Task 1, the sequential nature of yield dynamics and the complex interaction between economic factors favour models capable of learning long-range dependencies—here TTM's architecture provides clear advantages over traditional approaches.

Task 2 presents an inverted hierarchy where simpler models outperform TTM, despite TTM showing strong transfer gains relative to its untrained version. This apparent contradiction highlights an important distinction: transfer learning success (relative to untrained performance) does not guarantee absolute performance leadership. The relatively high signal-to-noise ratio in volatility forecasting creates an environment where overfitting becomes the primary concern, favouring simpler models.

The Error Correction Model's dominance in Task 3 provides a compelling example of domain knowledge trumping architectural sophistication. The ECM's explicit incorporation of mean reversion through its autoregressive coefficient structure directly addresses the fundamental dynamics of the spread series, something that TTM must learn implicitly through its mixing operations and temporal convolutions.

\subsubsection{Transfer Gain Magnitudes and Statistical Significance}

The magnitude of transfer gains provides quantitative insights into the value of pretraining across different forecasting contexts. All three tasks demonstrate substantial fine-tuning transfer gains under data-limited conditions (25-50\%), but exhibit different patterns in zero-shot transferability. Task 1 shows strong fine-tuning transfer but negative zero-shot performance, Task 2 demonstrates both strong fine-tuning and positive zero-shot transfer, while Task 3 exhibits the strongest transfer across both dimensions. This progression correlates with the prevalence of the target statistical properties across different domains during pretraining.
The consistently higher transfer gains under data-limited conditions (25-50\%) compared to full-data scenarios (1-30\%) suggest that transfer learning's primary value lies in overcoming data scarcity rather than pushing the boundaries of achievable performance. This finding has important implications for understanding when TSFMs provide maximum value.
The statistical significance patterns also reveal the robustness of transfer effects. Tasks 2 and 3 maintain significant transfer gains even with full datasets, while Task 1's gains become statistically insignificant with abundant data. This suggests that the complexity and noise characteristics of the forecasting problem determine whether transfer benefits persist as data increases.

\subsubsection{Architectural and Pretraining Implications}

The stark performance differential between TTM and Chronos across all tasks provides insights into the importance of architectural design and pretraining strategy alignment. TTM's decoder architecture with explicit temporal modelling appears better suited to financial time series than Chronos's adapted language model approach.

The failure of Chronos to achieve meaningful transfer in any task, despite its sophisticated underlying architecture, suggests that successful TSFM development requires more than simply adapting powerful models from other domains. The tokenisation and discrete sequence modelling approaches that work well for language may be fundamentally misaligned with the continuous, temporal dynamics of financial time series.

\subsubsection{Methodological Insights}

The experimental design reveals several methodological considerations for future TSFM evaluation. The naive baseline selection proves critical—Task 3's horizon-dependent baselines (previous value for 5 days, zero for 10 days) provide more meaningful performance comparisons than uniform approaches would have.

The rolling evaluation window approach successfully captures regime-dependent performance variations that would be missed in traditional train-test splits. The ability to observe performance recovery patterns after regime shifts provides valuable insights into model robustness that static evaluation methods cannot capture.

The separation of sequential and non-sequential benchmarks in Task 3 clarifies the contribution of temporal modelling capabilities versus pure predictive power, helping to isolate the specific advantages that TSFMs bring to forecasting problems.

\section{Conclusions}
\label{Conclusions}

\subsection{Summary of Findings}
This study set out to evaluate the applicability of Time Series Foundation Models (TSFMs) to multivariate financial forecasting tasks. Through a comparative assessment of two distinct TSFM paradigms—Tiny Time Mixers (TTM), a compact model designed from the ground up for time series forecasting, and Chronos, an LLM adapted TSFM—the project investigated the effectiveness, sample efficiency, and transferability of pretrained representations across three diverse financial time series forecasting tasks.

First, the results demonstrate that pretrained TSFMs such as TTM exhibit strong potential in complex, low signal to noise ratio forecasting problems. Pretrained TTM not only required significantly less training data to reach similar accuracy compared with its untrained counterpart, but also achieved lower error after fine tuning; this shows that the network weights learned during pretraining carried useful temporal patterns that the untrained model would otherwise have to discover from scratch. Specifically, across the different tasks, \textbf{the pretrained and fine-tuned version of TTM achieved an increase in performance of 30\%, 50\% and 25\% across the three tasks under limited data availability compared to its untrained and fine-tuned counterpart}. Even with the full data set, the pretrained version outperformed the previously untrained version by 10-30\% across tasks and forecasting horizons. In addition, pretrained TTM outperformed naive forecasts without any fine tuning in two of the tasks (zero-shot mode). Furthermore, \textbf{TTM in zero-shot modality outperformed all benchmark models in Task 3}  for forecasting horizons of 5 and 10 business days ahead, which is quite significant: it shows that transfer from pretraining alone can beat a simple historical return baseline when data are scarce. This evidence supports the idea that pretraining \textbf{improves sample efficiency, accelerates convergence, and enhances generalisation, making TSFMs attractive for domains where data are expensive or limited in length or frequency}.

Second, while TTM consistently outperformed naive baselines in zero shot and fine tuned settings, it did not achieve state of the art performance overall. In fact, \textbf{traditional benchmark models matched or outperformed TTM in two of the three tasks}. This suggests that while TSFMs are broadly applicable and can be robust across diverse tasks, they may lack the specialisation required to outperform models tailored to a narrow target unless further adapted or pretrained with more relevant data. These findings are consistent with the general notion that \textbf{TSFMs aim for breadth rather than optimality on a single task}.

\textbf{Third, there is reason to believe that additional pretraining on large scale financial data could lead to improved results}. This is supported by TTM's substantially superior performance compared to Chronos, where TTM included financial data sources in its pretraining while Chronos lacked financial data in its pretraining. Future work should explore specialised versions of TSFMs pretrained on a broad set of financial series drawn from different domains—such as sovereign and corporate yields across multiple maturities, geographies, credit ratings, and equity indices. Pretraining on such a wide variety of tasks and asset classes would give the model deeper domain knowledge and a richer set of temporal patterns to draw upon, yielding a domain specific foundation model capable of superior generalisation within finance. In this sense, the current results are promising but likely suboptimal. Furthermore, pretraining from scratch would eliminate potential look-ahead bias from including correlated future data in TTM's financial pretraining dataset, as explained further in Section \ref{sec:future}. 

\textbf{Finally, preliminary comparisons suggest that TSFMs built natively for time series applications may be more suitable than TSFMs adapted from LLMs}. TTM consistently provided evidence of transferability and learned meaningful patterns, while Chronos failed to consistently outperform even naive forecasts in any task. However, this conclusion remains tentative given the limited model sample and the many confounding variables, including differences in pretraining datasets and architectures, warranting further investigation. For practitioners, these results suggest that TSFMs offer a promising middle ground between broad applicability and specialised performance, particularly valuable when developing forecasting capabilities across multiple financial instruments or when historical data is limited. Taken together, the results of this project demonstrate that TSFMs hold substantial promise for financial forecasting tasks—particularly in noisy or data constrained environments—but achieving competitive performance may require more targeted pretraining and architectural refinements tailored to the financial domain.


\subsection{Future Works}
\label{sec:future}

The promising results of this study point toward several research directions that could substantially advance the application of foundation models in financial time series forecasting. These opportunities range from addressing current methodological limitations to developing entirely new paradigms for financial AI systems.

\begin{enumerate}
    \item \textbf{Financial domain-specific pretraining}: Future research should prioritise developing TSFMs pretrained exclusively on large, diverse collections of financial time series. Pretraining from scratch would allow the use of TSFMs currently unavailable in pretrained form while dramatically improving the relevance of transferred knowledge through enhanced alignment between source and target domains. Such models would embed fundamental financial priors—including cross-asset relationships, regime changes, and market microstructure patterns—making zero-shot applications genuinely meaningful for practitioners. An intermediate approach involves taking existing pretrained TSFMs and conducting additional training on comprehensive financial datasets before task-specific fine-tuning, creating a multi-stage transfer learning pipeline. Crucially, pretraining from scratch eliminates look-ahead bias concerns that arise when using models potentially trained on future data relative to historical evaluation periods. This bias represents a significant methodological issue in financial applications, where temporal causality is paramount, and could lead to overestimating model performance in backtesting scenarios that don't reflect true out-of-sample conditions.

    \item \textbf{Task-specialised universal models}: The field should explore developing generalised models for specific financial forecasting tasks that can operate across multiple assets or instruments. For instance, a universal volatility forecasting model capable of predicting volatility for any asset class, or yield forecasting models that generalise across different issuers, credit ratings, and maturities. These models would combine the benefits of specialisation with broad applicability, potentially achieving superior performance within their domain while maintaining the flexibility to adapt to new instruments without extensive retraining. This represents a middle ground between completely general TSFMs and narrow single-asset models.

    \item \textbf{Architecture-agnostic transfer learning analysis}: While this project focused on TSFMs, the principles of parametric transfer learning extend to simpler architectures such as LSTMs, GRUs, and even traditional econometric models with learned components. Future work should systematically compare how different architectures—when pretrained on identical datasets—affect downstream performance, sample efficiency, and transferability on financial forecasting tasks. This research would provide crucial insights into whether the benefits observed stem from the scale and sophistication of foundation models or from the transfer learning paradigm itself, informing more efficient model development strategies.

    \item \textbf{Multi-modal financial foundation models}: An emerging opportunity lies in developing models that can jointly process time series data alongside textual information (earnings calls, news, regulatory filings) and alternative data sources (satellite imagery, social media sentiment). Such models could capture the complex interplay between quantitative market dynamics and qualitative information flows that drive financial markets.

    \item \textbf{Causal and interpretable TSFMs}: Financial applications demand not only accurate predictions but also interpretable explanations for regulatory and risk management purposes. Future research should explore methods for incorporating causal reasoning and explainability into TSFMs, potentially through attention mechanisms that highlight relevant historical patterns or through hybrid architectures that combine neural networks with interpretable econometric components.
\end{enumerate}

These research directions outline possible paths for advancing the use of foundation models in financial forecasting. While the results of this study are promising, they remain preliminary. Substantial work is still needed to validate these approaches across broader datasets, time periods, and use cases. Continued progress will depend on rigorous testing, careful attention to domain-specific challenges such as temporal causality and data scarcity, and deeper collaboration between financial experts and machine learning researchers. Nevertheless, the intersection of transfer learning, domain-aligned pretraining, and scalable architectures represents a worthwhile area of exploration for the future of quantitative modelling.

\subsection{Limitations}
While this study provides valuable insights into the application of TSFMs in financial forecasting, several limitations must be acknowledged that constrain the generalisability and practical applicability of the findings.
The primary reliance on Mean Squared Error (MSE) as the evaluation metric, while appropriate for assessing transferability and generalisation capabilities, does not capture the practical value that financial practitioners require. Metrics such as directional accuracy, Sharpe ratios, or trading-based performance measures would provide more meaningful assessments of real-world utility. MSE optimisation may not align with the profit maximisation or risk management objectives that drive financial decision-making, potentially overstating or understating the practical benefits of TSFMs in operational contexts.
The three financial forecasting tasks examined—while carefully selected to represent different market segments and forecasting challenges—cannot fully capture the breadth and complexity of financial time series applications. Financial markets encompass numerous asset classes, time horizons, market regimes, and structural relationships that extend far beyond the scope of this analysis. The tasks may not adequately represent the full spectrum of noise characteristics, non-stationarity patterns, or regime-switching behaviours prevalent across different financial domains, limiting the external validity of the conclusions.
The original research design intended to analyse a broader range of TSFM architectures to provide more comprehensive insights into the relative merits of different approaches. However, practical constraints significantly limited the model selection. Few models satisfied the necessary criteria of being open-source, pretrained, and capable of handling multivariate time series data. This limitation is particularly acute given the nascent state of the TSFM field, where many promising models remain proprietary or are available only in limited configurations. Consequently, the comparison between TTM and Chronos, while informative, provides insufficient evidence to support broad generalisations about the relative performance of different TSFMs.
The limited sample of models fundamentally constrains the validity of general claims about TSFMs as a class of models. With only two models examined, conclusions about the superiority of natively-designed TSFMs over LLM-adapted approaches, or broader statements about TSFM performance in financial applications, must be considered tentative. The substantial architectural and pretraining differences between TTM and Chronos introduce numerous confounding variables that complicate direct comparisons and limit the ability to isolate the factors driving performance differences.
\printbibliography[title={Bibliography},heading=bibintoc]
\appendix 
\section{Appendix}

\subsection{TTM Pretraining Datasets}
\label{appendix:ttm_datasets}

TTM models were pretrained on a wide variety of publicly available time series datasets spanning multiple domains, including energy, traffic, weather, healthcare, and finance. The specific datasets used to pretrain TTM are as follows\cite{vijay2024}:

\begin{itemize}
    \item \textbf{Australian Electricity Demand} –
    \url{https://zenodo.org/records/4659727}
    \item \textbf{Australian Weather} – \url{https://zenodo.org/records/4654822}
    \item \textbf{Bitcoin} – \url{https://zenodo.org/records/5122101}
    \item \textbf{KDD Cup 2018} – \url{https://zenodo.org/records/4656756}
    \item \textbf{London Smart Meters} – \url{https://zenodo.org/records/4656091}
    \item \textbf{Saugeen River Flow} – \url{https://zenodo.org/records/4656058}
    \item \textbf{Solar Power} – \url{https://zenodo.org/records/4656027}
    \item \textbf{Sunspots} – \url{https://zenodo.org/records/4654722}
    \item \textbf{Solar} – \url{https://zenodo.org/records/4656144}
    \item \textbf{US Births} – \url{https://zenodo.org/records/4656049}
    \item \textbf{Wind Farms Production} – \url{https://zenodo.org/records/4654858}
    \item \textbf{Wind Power} – \url{https://zenodo.org/records/4656032}
    \item \textbf{Traffic (PEMSD3, PEMSD4, PEMSD7, PEMSD8, PEMS\_BAY, LOS\_LOOP)} – \url{https://drive.google.com/drive/folders/1g5v2Gq1tkOq8XO0HDCZ9nOTtRpB6-gPe}
    \item \textbf{Covid Deaths} – \url{https://zenodo.org/records/4656009}
    \item \textbf{Covid Mobility} – \url{https://zenodo.org/records/4663809}
    \item \textbf{Extended Wikipedia Web Traffic} – \url{https://zenodo.org/records/7371038}
    \item \textbf{NN5} – \url{https://zenodo.org/records/4656117}, \url{https://zenodo.org/records/4656125}
    \item \textbf{Temperature Rain} – \url{https://zenodo.org/records/5129091}
    \item \textbf{Vehicle Trips} – \url{https://zenodo.org/records/5122537}
    \item \textbf{Kaggle Web Traffic} – \url{https://zenodo.org/records/4656075}, \url{https://zenodo.org/records/4656664}
    \item \textbf{Hierarchical Sales} – \url{https://huggingface.co/datasets/Salesforce/lotsa_data/tree/main/hierarchical_sales}
    \item \textbf{Project Tycho} – \url{https://huggingface.co/datasets/Salesforce/lotsa_data/tree/main/project_tycho}
    \item \textbf{Subseasonal} – \url{https://huggingface.co/datasets/Salesforce/lotsa_data/tree/main/subseasonal}
    \item \textbf{Subseasonal Precipitation} – \url{https://huggingface.co/datasets/Salesforce/lotsa_data/tree/main/subseasonal_precip}
    \item \textbf{Uber TLC Daily Rides} – \url{https://huggingface.co/datasets/Salesforce/lotsa_data/tree/main/uber_tlc_daily}
    \item \textbf{Wiki Rolling Traffic (GluonTS)} – \url{https://github.com/awslabs/gluonts/blob/1553651ca1fca63a16e012b8927bd9ce72b8e79e/datasets/wiki-rolling_nips.tar.gz}
    \item \textbf{CDC FluView ILINet} – \url{https://huggingface.co/datasets/Salesforce/lotsa_data/tree/main/cdc_fluview_ilinet}
    \item \textbf{CDC FluView WHO/NREVSS} – \url{https://huggingface.co/datasets/Salesforce/lotsa_data/tree/main/cdc_fluview_who_nrevss}
\end{itemize}

\subsection{Chronos Pretraining Datasets}
\label{appendix:chronos_datasets}

Chronos was pretrained on a diverse collection of real-world time series datasets across multiple domains. The specific datasets are\cite{chronos}:

\begin{itemize}
    \item \textbf{Brazilian Cities Temperature} – \url{https://www.kaggle.com/datasets/volpatto/temperature-timeseries-for-some-brazilian-cities}
    \item \textbf{Mexico City Bikes} – \url{https://ecobici.cdmx.gob.mx/en/open-data/}
    \item \textbf{Solar (5 Min.)} – \url{https://www.nrel.gov/grid/solar-power-data.html}
    \item \textbf{Solar (Hourly)} – \url{https://www.nrel.gov/grid/solar-power-data.html}
    \item \textbf{Spanish Energy and Weather} – \url{https://www.kaggle.com/datasets/nicholasjhana/energy-consumption-generation-prices-and-weather}
    \item \textbf{Taxi (Hourly)} – \url{https://github.com/mbohlkeschneider/gluon-ts/tree/mv_release/datasets}
    \item \textbf{USHCN} – \url{https://cdiac.ess-dive.lbl.gov/ftp/ushcn_daily/}
    \item \textbf{Weatherbench (Daily)} – \url{https://github.com/pangeo-data/WeatherBench}
    \item \textbf{Weatherbench (Hourly)} – \url{https://github.com/pangeo-data/WeatherBench}
    \item \textbf{Weatherbench (Weekly)} – \url{https://github.com/pangeo-data/WeatherBench}
    \item \textbf{Wiki Daily (100k)} – \url{https://wikimedia.org/api/rest_v1/}
    \item \textbf{Wind Farms (Daily)} – \url{https://zenodo.org/record/4654909}
    \item \textbf{Wind Farms (Hourly)} – \url{https://zenodo.org/record/4654909}
\end{itemize}

\subsection{Backtesting Results}
\label{appendix:backtest}

Given TTM's strong performance in forecasting 10-year US Treasury yields 21 days ahead, we implemented a rolling forecast strategy that retrains the model every six months using the most recent eight years of data. Based on these forecasts, we constructed three trading signals that determine long or short positions in the iShares 7--10 Year Treasury Bond ETF (IEF), according to predicted yield movements. Since bond prices move inversely to yields, all signals are sign-inverted to generate returns from anticipated bond price changes.
\begin{figure}[H]
    \centering
    \includegraphics[width=1\linewidth]{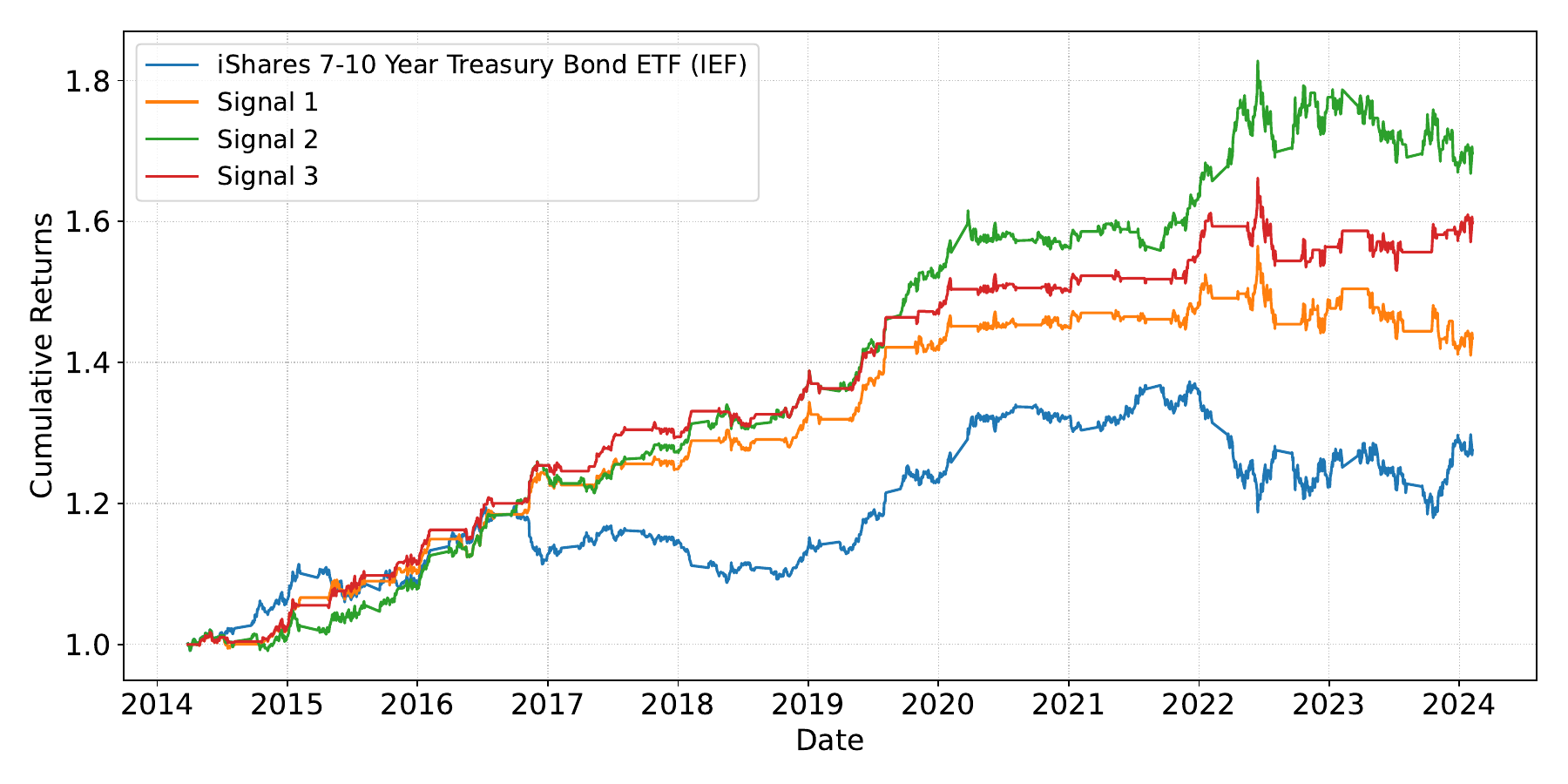}
    \caption{Cumulative Returns of Three Signals Constructed using TTM's forecasts Alongside a Benchmark ETF}
    \label{fig:backtest_yield}
\end{figure}

The three signals are defined as follows:

Signal 1 uses the 21-day rolling mean of the model’s predicted 21-day yield change. A position is entered only when the average signal magnitude exceeds a threshold of 0.01, in order to suppress noisy forecasts. The direction of the position is determined by the sign: short if yields are expected to rise, and long if they are expected to fall.

Signal 2 follows the same logic as Signal 1 but replaces the simple rolling mean with an exponentially weighted moving average (EWMA) over 21 days. This allows the signal to respond more rapidly to recent forecast changes by assigning more weight to the most recent predictions.

Signal 3 is a discrete voting mechanism based on the sign of each daily forecast. It computes a rolling sum over the past 21 predictions and enters a position only if at least 14 out of the 21 indicate the same direction. This captures sustained consensus in the model’s directional signals, rather than relying on the magnitude of the forecasts.

Each signal was applied to the daily returns of IEF, and the cumulative returns were tracked over time. The figure below shows the performance of these three strategies compared to a passive Buy and Hold benchmark. All three model-based signals exhibit substantial outperformance relative to the benchmark over the full backtest period.

Table \ref{tab:backtest-perf} reports the performance of three model-driven trading strategies against a passive Buy and Hold strategy on the iShares 7–10 Year Treasury Bond ETF (IEF). All signals outperform the benchmark in both total and annualized returns, with Signal 2 and Signal 3 delivering particularly strong results. Signal 3 exhibits the highest risk-adjusted performance, with a Sharpe ratio of 1.39 and the lowest volatility and drawdown, suggesting it achieves smoother returns through effective filtering of noisy predictions. Signal 2 offers the highest overall return but with slightly higher risk. All three signals display positive information ratios, indicating they consistently outperform the benchmark on a risk-adjusted basis.

\begin{table}[H]
\centering
\label{tab:backtest-perf}
\caption{Performance metrics of model-based trading strategies versus Buy and Hold on IEF}
\begin{tabular}{lcccc}
\toprule
\textbf{Strategy} & \textbf{CAGR} & \textbf{Sharpe} & \textbf{Max Drawdown} & \textbf{Information Ratio} \\
\midrule
Buy and Hold (Close) & 3.31\% & 0.54 & -14.09\% & -- \\
Signal 1             & 4.95\% & 0.97 & -9.92\%  & 0.17 \\
Signal 2             & 7.35\% & 1.21 & -8.75\%  & 0.38 \\
Signal 3             & 6.49\% & 1.39 & -7.91\%  & 0.35 \\
\bottomrule
\end{tabular}
\end{table}

\subsection{Error Correction Model Proof}
\label{appendix:ecm}

Let \(x_t\) and \(y_t\) be two \(I(1)\) time series that are cointegrated with cointegrating coefficient \(\beta\), so that the spread
\[
z_t \;=\; y_t \;-\;\beta\,x_t
\]
is stationary (\(I(0)\)).  The standard bivariate Error–Correction Model (ECM) takes the form
\[
\Delta y_t \;=\;\alpha\,\bigl(y_{t-1}-\beta\,x_{t-1}\bigr)\;+\;\gamma\,\Delta x_t\;+\;\varepsilon_t,
\]
where \(\alpha\) is the speed‐of‐adjustment coefficient and \(\varepsilon_t\) is a white‐noise error.

Now define the first‐difference of the spread:
\[
\Delta z_t \;=\;\Delta y_t \;-\;\beta\,\Delta x_t.
\]
Substituting the ECM expression for \(\Delta y_t\) gives
\[
\Delta z_t
=\;\alpha\,z_{t-1}
\;+\;\gamma\,\Delta x_t
\;-\;\beta\,\Delta x_t
\;+\;\varepsilon_t
=\;\alpha\,z_{t-1}
\;+\;(\gamma-\beta)\,\Delta x_t
\;+\;\varepsilon_t.
\]
If we assume that the short-run adjustment term \((\gamma-\beta)\,\Delta x_t\) is zero (or small enough to be absorbed into the noise), this simplifies to
\[
\Delta z_t = \alpha\,z_{t-1} + \varepsilon_t.
\]
But \(\Delta z_t = z_t - z_{t-1}\), so
\[
z_t - z_{t-1} = \alpha\,z_{t-1} + \varepsilon_t
\quad\Longrightarrow\quad
z_t = (1 + \alpha)\,z_{t-1} + \varepsilon_t.
\]
This is exactly the AR(1) form
\[
z_t = \phi\,z_{t-1} + \varepsilon_t,
\]
with \(\phi = 1 + \alpha\).  Hence, under these conditions, the ECM implies that the cointegrated spread \(z_t\) follows an AR(1) process.

\end{document}